\documentclass[a4paper,11pt]{article}

\usepackage[utf8]{inputenc}
\usepackage{amsmath}
\usepackage{verbatim}
\usepackage{float}
\usepackage{cite}
\usepackage{jheppub}
\usepackage{cite}
\usepackage{url}
\usepackage{slashed}
\usepackage{multirow}
\usepackage{dsfont}
\usepackage{caption}
\usepackage{subcaption}

\makeatletter
\def\maketag@@@#1{\hbox{\m@th\normalfont\normalsize#1}}
\makeatother

\def\eps{\epsilon}

\def\Tr{\mathop{\rm Tr}}
\def\PT{\text{PT}}
\def\zbar{\ensuremath{\bar{z}}}
\newcommand{\vev}[1]{\ensuremath{\langle #1 \rangle }}

\def\Im{\text{Im}}
\def\Re{\text{Re}}
\DeclareMathOperator{\sign}{sign}
\def\sym{$\mathcal{N}=4$ super Yang-Mills }
\def\sugra{$\mathcal{N}=8$ supergravity }

\newcommand{\p}[1]{(\ref{#1})}

\newcommand{\pa}{\partial}

\newcommand{\Li}{\text{Li}}

\title{Multi-Regge Limit of the Two-Loop Five-Point Amplitudes in $\mathcal{N} = 4$ Super Yang-Mills and $\mathcal{N} = 8$ Supergravity}
\author[a]{Simon Caron-Huot}
\author[b]{Dmitry Chicherin}
\author[b]{Johannes Henn}
\author[c,d]{Yang Zhang}
\author[b]{Simone Zoia}

\affiliation[a]{Department of Physics, McGill University, 3600 Rue University, Montr\'eal, QC Canada}
\affiliation[b]{Max-Planck-Institut f\"{u}r Physik, Werner-Heisenberg-Institut, D-80805 M\"{u}nchen, Germany}
\affiliation[c]{Peng Huanwu Center for Fundamental Theory, Hefei, Anhui 230026, China}
\affiliation[d]{Interdisciplinary Center for Theoretical Study, University of Science and Technology of China, Hefei, Anhui 230026, China}

\emailAdd{schuot@physics.mcgill.ca}
\emailAdd{chicheri@mpp.mpg.de}
\emailAdd{henn@mpp.mpg.de}
\emailAdd{yzhphy@ustc.edu.cn}
\emailAdd{zoia@mpp.mpg.de}

\preprint{USTC-ICTS/PCFT-20-08, MPP-2020-26}

\bigskip

\abstract{In previous work, the two-loop five-point amplitudes in $\mathcal{N}=4$ super Yang-Mills theory and $\mathcal{N}=8$ supergravity were computed at symbol level. In this paper, we compute the full functional form. The amplitudes are assembled and simplified using the analytic expressions of the two-loop pentagon integrals in the physical scattering region. We provide the explicit functional expressions, and a numerical reference point in the scattering region. We then calculate the multi-Regge limit of both amplitudes. The result is written in terms of an explicit transcendental function basis. For certain non-planar colour structures of the $\mathcal{N}=4$ super Yang-Mills amplitude, we perform an independent calculation based on the BFKL effective theory. We find perfect agreement. We comment on the analytic properties of the amplitudes.}

\begin{document}
\maketitle

\newpage

\section{Introduction}
\label{sec:Introduction}

Regge theory initially arose from the need to interpret data from high-energy experiments,
and also played a prominent role in the inception of string theory. 
Understanding the Regge or high-energy limit of scattering amplitudes and cross-sections
continues to be important both conceptually and phenomenologically.
The research aims on the one hand at describing better certain regions of phase space of collider experiments, 
and on the other hand there is hope that this limit may shed light on underlying structures of field and string theory.

In the Regge limit, highly boosted objects with a fixed transverse profile interact in an instant.
The hierarchy between longitudinal and transverse momenta allows to expand
amplitudes in powers and logarithms of a small parameter.
The general challenge is to describe
this expansion in terms of a small number of simple ingredients.
This is possible for the universal leading logarithms, which are controlled by the gluon Regge trajectory
and is closely related to light-like cusp anomalous dimension.
On the other hand, less is known about sub-leading logarithms, which have a more intricate form.  
Similarly, power-suppressed terms are conceptually much more complicated 
than their leading-power counterparts, but they can be numerically important and hence relevant for phenomenology. 
Understanding subleading power corrections in the Regge limit, but also in other kinematical limits, is an active field of research.

The Regge limit is also a useful tool to probe the structure of scattering amplitudes in quantum field theory.
Indeed, the question to what extent amplitudes are fixed by general properties and principles goes back to 
the analytic S-matrix program of the 1960's, and has seen a revival, especially in the context 
of the maximally supersymmetric Yang-Mills theory, ${\mathcal N=4}$ sYM. 
First hints of integrability in this theory can be traced to studies of this limit~\cite{Lipatov:1993yb,Faddeev:1994zg}.
More recently, it proved useful to study
multi-particle scattering amplitudes in the context of the Wilson loop / scattering amplitude duality.
Crossing symmetry for multi-particle amplitudes is nontrivial to formulate in general but it is relatively well understood
in the Regge limit,  where precise constraints, namely the absence of so-called overlapping discontinuities, 
gave early hints that a guess for the all-loop form of such amplitudes required corrections \cite{Bartels:2008ce}, and helps to constrain the form  of the latter \cite{Caron-Huot:2016owq}.
This is an example of how the Regge limit was useful in the context of the bootstrap approach to amplitudes, where an ansatz is made based on certain assumptions, 
and various conditions are being used to determine free coefficients in the ansatz~\cite{Dixon:2011pw,Caron-Huot:2019vjl}. 
Initially, the Regge limit was used as an input in this procedure, but when the ansatz can be constrained by other means, it yields a prediction. See also Ref.~\cite{DelDuca:2019tur} for recent work on the Regge limit of multi-particle amplitudes.
While most studies in this theory are restricted to massless scattering, in Ref.~\cite{Bruser:2018jnc} an interesting pattern of exponentiation was observed for certain massive scattering amplitudes. There, not only the leading, but also the subleading power terms were found to be governed by anomalous dimensions of a certain Wilson line operator. Power corrections
to energy-energy correlators have relatedly revealed a surprisingly simple pattern \cite{Moult:2019vou}.

Despite the huge progress and new insights obtained, most of the studies in $\mathcal{N}=4$ super Yang-Mills 
remain restricted to the planar limit, i.e. the limit of large 't~Hooft coupling.
In this case, amplitudes in ${\mathcal N=4}$ sYM have a dual conformal symmetry~\cite{Drummond:2006rz},
which heavily restricts their variable dependence, and similarly restricts the transcendental functions appearing.
While this is extremely interesting and helpful, it raises the question as to
how general and universal the structures that are found in this limit are. 

On the other hand, much less is known on non-planar amplitudes.
There are various motivations for studying the latter.
One reason is that while the 't~Hooft expansion is conceptually important, it is in general unclear whether
the non-planar terms are numerically subleading, especially in QCD, where $N_{c}=3$.
Another reason is that non-planar results are important in trying to understand how to make use of integrability, if possible, beyond the planar limit in ${\mathcal N=4}$ super Yang-Mills~\cite{Bern:2017gdk,Bern:2018oao,Chicherin:2018wes,Ben-Israel:2018ckc}.
Furthermore, it is interesting in itself to understand how the Regge limit interplays
with the much richer colour structures at the non-planar level. Recent conceptual
advances make it possible to predict some of these terms. We find it interesting
to work out some of these predictions, and to test them against explicit perturbative results.
Finally, in the context of (super)gravity theories, there is no notion of a planar limit,
and therefore any attempt at understanding scattering amplitudes in these theories
necessarily includes both planar and non-planar terms. 

The conceptual progress in understanding the Regge limit in quantum field theory~\cite{Caron-Huot:2013fea,DelDuca:2014cya,Caron-Huot:2017fxr} lead to predictions that were successfully compared against explicit three-loop results for the full-colour four-gluon amplitudes in $\mathcal{N}=4$ super Yang-Mills~\cite{Henn:2016jdu}.
Moreover, there is recent work
on understanding certain terms in the Regge limit in supergravity theories~\cite{Bartels:2012ra,SabioVera:2019edr,SabioVera:2019jqe,DiVecchia:2019myk,DiVecchia:2019kta},
and perturbative data for the four-graviton amplitude is available up to three loops~\cite{Naculich:2008ew,Brandhuber:2008tf,BoucherVeronneau:2011qv,Henn:2019rgj}.

Until recently, non-planar studies at two loops were limited to four external particles, due to the enormous technical difficulties of
dealing with higher-point scattering amplitudes at two loops. One major bottleneck had to do with dealing with the Feynman integrals,
that are transcendental functions of four dimensionless parameters, and (especially at the non-planar level) have an intricate analytic structure. Recently, this bottleneck was overcome~\cite{Chicherin:2017dob,Abreu:2018rcw,Chicherin:2018mue,Abreu:2018aqd,Chicherin:2018old} and various full-colour amplitudes are now available~\cite{Abreu:2018aqd,Chicherin:2018yne,Chicherin:2019xeg,Abreu:2019rpt,Badger:2019djh}.

The initial computations of both the five-particle amplitudes in $\mathcal{N}=4$ super Yang-Mills and $\mathcal{N}=8$ supergravity were done at the symbol level.
This means, roughly speaking, that the analytic structure of the result was found, but certain `integration constants' were dropped.  
Still, the symbol result allowed to study the Regge limit, and an interesting observation was made: in $\mathcal{N}=4$ super Yang-Mills,
the symbol of the five-particle amplitude vanishes at leading power in the multi-Regge limit~\cite{Abreu:2018aqd,Chicherin:2018yne}!
This observation certainly warrants further investigation, and it is interesting to ask whether the vanishing is exact, or whether 
the answer is rather of the type `transcendental constant' (that would be dropped in the symbol) $\times$ `lower-weight transcendental function'.
Such terms are usually referred to as `beyond-the-symbol' terms in the literature.
In this paper, we perform this analysis, and provide the Regge limit (at function level) for both maximally supersymmetric theories.

We also extend the ideas of~\cite{Caron-Huot:2017fxr} to the five-particle case, and work out predictions for the Regge limit in certain colour channels. We successfully compare the predictions against the result of the explicit perturbative computation.

The paper is structured as follows. In Section~\ref{sec:PentagonFunctions} we describe the kinematics and the pentagon functions, namely the function space relevant for the scattering of five massless particles up to two-loop order. In Section~\ref{sec:Assembly} we introduce the two-loop scattering amplitudes in \sym theory and $\mathcal{N}=8$ supergravity. We discuss how, starting from the integrands present in the literature, we obtain integrated expressions of manifestly uniform transcendental weight. Moreover, we briefly review the factorisation of infrared singularities, and define finite hard functions in both theories. Section~\ref{sec:MultiReggeLimit} is devoted to the multi-Regge limit. In particular, we show how we parametrise it and how we compute the asymptotics of the pentagon functions. We present our results first for the \sym hard function at one and two loops, in Section~\ref{sec:sYM}. For certain colour structures, the computation of the multi-Regge limit is vastly simplified by using the Balitsky-Fadin-Kuraev-Lipatov (BFKL) effective theory, as we discuss in Section~\ref{sec:BFKL}. Finally, the multi-Regge asymptotics of the $\mathcal{N}=8$ supergravity hard function at one and two loops is presented in Section~\ref{sec:SUGRA}. We draw our conclusions in Section~\ref{sec:Conclusions}.

\section{Kinematics and pentagon functions}
\label{sec:PentagonFunctions}

We study the scattering of five massless particles and follow the same notation used in \cite{Chicherin:2017dob, Chicherin:2018mue,Chicherin:2018old}. The momenta, which we label by $p_i^{\mu}$, are subject to on-shellness, $p_i^2=0 \ \forall i=1,\ldots,5$, and momentum conservation, $\sum_{i=1}^5 p_i^{\mu} = 0$. The kinematics can be described in terms of five independent Mandelstam invariants,  
\begin{align}
X = \{s_{12}, s_{23}, s_{34}, s_{45}, s_{51}\}\,,
\end{align}
with $s_{ij}=2 p_i \cdot p_j$. It is also convenient to introduce the pseudo-scalar invariant 
\begin{align}
\eps_5=\text{tr}_5(\gamma_5 \slashed{p}_4  \slashed{p}_5  \slashed{p}_1  \slashed{p}_2) = [12]\vev{23}[34]\vev{41}-\vev{12}[23]\vev{34}[41] \, .
\end{align}
The latter can be related to the $s_{ij}$ through $\Delta = (\eps_5)^2$, where $\Delta$ is the Gram determinant of the external momenta $\Delta = \text{det}(s_{ij}|_{i,j=1}^4)$. Note that we take the external states and momenta to live in four-dimensional Minkowski space, and perform the loop integrations in $D=4-2 \eps$ dimensions to regularise the divergences.

Scattering amplitudes depend on the kinematics through rational and special functions. For massless five-particle scattering up to two loops, the latter were conjectured~\cite{Chicherin:2017dob} and later shown~\cite{Abreu:2018rcw,Chicherin:2018mue,Abreu:2018aqd,Chicherin:2018old} to belong to a class of polylogarithmic functions called pentagon functions~\cite{Chicherin:2017dob}.

The pentagon functions can be conveniently written as $\mathbb{Q}$-linear combinations of Chen iterated integrals~\cite{chen1977},
\begin{align}
\label{eq:IteratedIntegralSingle}
\left[W_{i_1},\ldots, W_{i_n} \right]_{X_0}(X) = \int_{\gamma} d\log W_{i_n}(X') \, \left[W_{i_1},\ldots,W_{i_{n-1}}\right]_{X_0}(X')\, ,
\end{align}
where the integration contour $\gamma$ connects the boundary point $X_0$ to $X$. The iteration starts with $[]_{X_0}=1$. The number of integrations is called transcendental weight. The $W_i$ are algebraic functions of the kinematics called letters. 
Not any integral of the form~\eqref{eq:IteratedIntegralSingle} actually corresponds to a function. In general, one has to consider $\mathbb{Q}$-linear combinations of such objects, such that the sum satisfies certain integrability conditions. The latter essentially state that the partial derivatives commute. As a result, the Chen iterated integrals are homotopy functionals: once the endpoints $X_0$ and $X$ are fixed, their value does not change under smooth deformations of the contour $\gamma$. On the other hand, if during the deformation the contour $\gamma$ crosses a pole, then the iterated integral picks up a residue. The Chen iterated integrals are thus multi-valued functions. Within a region of analyticity, the Chen iterated integrals depend only on the
letters and boundary points $X_0$ and $X$, which motivates our notation in Eq.~\eqref{eq:IteratedIntegralSingle}.

In the massless five-particle scattering amplitudes up to two loops there are 31 independent letters, defined in Eqs.~(2.5) and~(2.6) of Ref.~\cite{Chicherin:2017dob}. They form the so-called pentagon alphabet. 
All of them have a definite behaviour under parity conjugation: 26 have even parity,
\begin{align}
d \log (W_i)^* = + d \log W_i \, , \qquad i=1,\ldots,25 , 31 \, ,
\end{align}
and 5 have odd parity,
\begin{align}
\label{eq:OddLetters}
d \log (W_i)^* = - d \log W_i \, ,  \qquad i=26,\ldots,30 \, . 
\end{align}

The first entries of the iterated integrals encode their discontinuities. They are therefore subject to physical constraints: a scattering amplitude may have discontinuities only where two-particle Mandelstam invariants $s_{ij}$ vanish. As a result, only ten letters are allowed as first entries in the pentagon functions: \mbox{$\{ s_{ij} \}_{i<j=1}^5 = \{ W_i\}_{i=1}^5 \cup \{ W_i \}_{i=16}^{20}$}. 

Of the other parity-even letters, $\{W_i\}_{i=6}^{15}$ and $\{W_i\}_{i=21}^{25}$ are given by simple combinations of $s_{ij}$, obtained from $s_{34}+s_{45}$ by permutation of the external legs. The last 6 letters are genuine to the five-particle kinematics, as they depend on the psuedo-scalar invariant $\epsilon_5$. The five odd letters, $\{W_i\}_{i=26}^{30}$, can be written as cyclic permutations of
\begin{align}
\label{eq:W26}
W_{26} = \frac{\text{tr}\left[\left(1-\gamma_5\right)\slashed{p}_4 \slashed{p}_5 \slashed{p}_1 \slashed{p}_2\right]}{\text{tr}\left[\left(1+\gamma_5\right)\slashed{p}_4 \slashed{p}_5 \slashed{p}_1 \slashed{p}_2 \right]}\,,
\end{align}
and are therefore pure phases. Finally, $W_{31} = \epsilon_5$. Note that the presence of $\epsilon_5$ makes the dependence on the kinematics algebraic, rather than rational, since its absolute value can be written in terms of the $s_{ij}$ as the square root of the Gram determinant $\Delta$.

Deeper entries of the iterated integrals are related to iterated discontinuities. In this regard, it is interesting to note that certain pairs of letters never appear as the first two entries~\cite{Chicherin:2017dob,Abreu:2018aqd,Chicherin:2018old}. It would be of great interest to find the physical principle underlying this second entry condition, which at the moment is a mere observation.

We consider the kinematics to lie in the $s_{12}$ channel. This physical scattering region is defined by
\begin{align}
& \{s_{12}, s_{34}, s_{35}, s_{45}\} \ge 0\,, \\
& \{s_{13}, s_{14}, s_{15}, s_{23}, s_{24}, s_{25}\} \le 0\,, \\
& \Delta \le 0\,,
\end{align}
which correspond to positive $s$-channel energies, negative $t$-channel energies, and reality of the momenta, respectively.
Within this region, all the Feynman integrals are analytic. The homotopy invariance thus allows us to choose the most convenient contour for the integration, as long as it never leaves the $s_{12}$ channel. As boundary point we choose
\begin{align}
\label{eq:SymmetryBasePoint}
X_0 = \{3, -1, 1, 1, -1\}\,, \qquad \epsilon_5\bigl|_{X_0}=i \sqrt{3}\,.
\end{align}
If the contour leaves the scattering region, care needs to be taken that the multi-valued functions are analytically continued to their sheet corresponding to the Feynman prescription (see e.g. Refs.~\cite{Gehrmann:2018yef,Henn:2019rgj}).

The pentagon functions can also be written in terms of polylogarithmic functions. Up to weight 2, logarithms and dilogarithms are sufficient. For instance, 
\begin{align}
[ W_{1} ]_{X_{0}}(X) &=  \log \left( s_{12}/3 \right) \,,\\
[ W_5 / W_2 , W_{12}/W_{2} ]_{X_{0}}(X) &= -{\rm Li}_{2}\left( 1- {s_{15}}/{s_{23}} \right) \,.
\end{align}
It is important to note that the pentagon alphabet can be rationalised. This is possible by using e.g. the momentum-twistor parametrisation of Ref.~\cite{Badger:2013gxa} or the $\beta_i$ variables of Ref.~\cite{Bern:1993mq}. As a result, the pentagon functions can be written in terms of Goncharov polylogarithms~\cite{2001math3059G} at any weight.

\section{The two-loop five-point $\mathcal N=4$ sYM and $\mathcal N=8$ supergravity amplitudes}
\label{sec:Assembly}

In this section we discuss the two-loop five-particle amplitudes in $\mathcal{N}=4$ super Yang-Mills theory and $\mathcal{N}=8$ supergravity. They have been computed at symbol level in Refs.~\cite{Abreu:2018aqd,Chicherin:2018yne,Chicherin:2019xeg,Abreu:2019rpt}. We obtain expressions for both the amplitudes in terms of rational functions of the kinematics and pure Feynman integrals. They thus exhibit manifestly uniform transcendental weight at all orders in $\eps$. After defining our notation, we discuss the structure of these expressions and show how we obtain them starting from the integrands presented in Ref.~\cite{Carrasco:2011mn}. Finally, we review the factorisation of infrared singularities in both theories, and define hard functions where the dimensional regulator $\eps$ can be removed.

We expand the five-point amplitude in $\mathcal{N}=4$ super Yang-Mills in the coupling constant $a = e^{-\epsilon \gamma_{\text{E}}} g^2/(4 \pi)^{2-\epsilon}$ as
\begin{align} \label{eq: def A5}
\mathcal{A}_5 = \delta^{(4)}\left(p_1+p_2+p_3+p_4+p_5 \right) \delta^{(8)}(Q) \, g^3 \sum_{\ell \ge 0} a^{\ell}  A_5^{(\ell)}  \, ,
\end{align}
where we extracted the overall momentum and super-momentum $Q$ conservation delta functions. In order to make the $SU(N_c)$ colour dependence explicit, we further decompose the five-point amplitudes $A_5^{(\ell)}$ up to two loops as
\begin{align}
\label{eq:A05color}
A_5^{(0)} & = \sum_{\lambda=1}^{12} A^{(0)}_{\lambda} \mathcal{T}_{\lambda}\, , \\
\label{eq:A15color}
A_5^{(1)} & = \sum_{\lambda=1}^{12} N_c A^{(1,0)}_{\lambda} \mathcal{T}_{\lambda} + \sum_{\lambda=13}^{22} A^{(1,1)}_{\lambda} \mathcal{T}_{\lambda} \, , \\
\label{eq:A25color}
A^{(2)}_5 & =\sum_{\lambda=1}^{12}\bigg(N_c^2 A_\lambda^{(2,0)}+A_\lambda^{(2,2)}\bigg) \mathcal T_\lambda
  +\sum_{\lambda=13}^{22}\bigg(N_c A_\lambda^{(2,1)}\bigg)\mathcal{T}_\lambda \, ,
\end{align}
where $\{\mathcal{T}_{\lambda}\}$, $\lambda=1,\ldots,22$, is the colour basis of Ref.~\cite{Edison:2011ta}. It contains $12$ single traces,
\begin{equation}
\label{eq:single_traces}
  \begin{array}{ccc}
   \mathcal T_1 = \Tr(12345) - \Tr(15432) \, , &\ \ & \mathcal T_2=\Tr(14325) -\Tr(15234)\, ,  \\
  \mathcal T_3 =\Tr(13425) - \Tr(15243)\, , & \ \ & \mathcal T_4=\Tr(12435) - \Tr(15342)\, ,\\
\mathcal T_5=\Tr(14235) -\Tr(15324)\, , &\ \ &  \mathcal T_6=\Tr(13245) - \Tr(15423) \, , \\
\mathcal T_7=\Tr(12543) - \Tr(13452)\, , & \ \ & \mathcal T_8=\Tr(14523) - \Tr(13254) \, ,\\
\mathcal T_9=\Tr(13524) - \Tr(14253)\, , & \ \ &  \mathcal T_{10}=\Tr(12534) -\Tr(14352) \, ,\\
\mathcal T_{11}=\Tr(14532) - \Tr(12354) \, , &  \ \ & \mathcal T_{12}=\Tr(13542) - \Tr(12453) \, ,\\
  \end{array}
\end{equation}
as well as $10$ double traces,
\begin{equation}
\label{eq:double_traces}
\begin{array}{ccc}
\mathcal T_{13}=\Tr(12) [\Tr(345) - \Tr(543)] \, , &  \ \ & \mathcal T_{14}=\Tr(23)[\Tr(451) - \Tr(154)] \, , \\
\mathcal T_{15}=\Tr(34) [\Tr(512) - \Tr(215)] \, , &  \ \ & \mathcal T_{16}=\Tr(45)[\Tr(123) - \Tr(321)] \, , \\
\mathcal T_{17}=\Tr(51) [\Tr(234) - \Tr(432)] \, , &  \ \ & \mathcal T_{18}=\Tr(13)[\Tr(245) - \Tr(542)] \, , \\
\mathcal T_{19}=\Tr(24) [\Tr(351) - \Tr(153)] \, , &  \ \ & \mathcal T_{20}=\Tr(35)[\Tr(412) - \Tr(214)] \, ,\\
\mathcal T_{21}=\Tr(41) [\Tr(523) - \Tr(325)] \, , &  \ \ & \mathcal T_{22}=\Tr(52)[\Tr(134) - \Tr(431)] \, ,
\end{array}
\end{equation}
with $\Tr(ij \cdots) = \Tr(\tilde{T}^{a_i} \tilde{T}^{a_j} \cdots)$, where $\tilde{T}^a = \sqrt{2} T^a$ and
$T^{a}$ are generators of $SU(N_c)$ in the fundamental representation normalised such that $\Tr(T^a T^b) = \frac{1}{2} \delta^{ab}$. 

The leading-colour components $A^{(\ell,0)}_n$ in this decomposition correspond to the planar part of the amplitude, since they receive contributions only from planar diagrams where the ordering of the external legs matches that of the generators in the corresponding trace. The complete amplitude, and therefore all the partial amplitudes $A_{\lambda}^{(\ell,k)}$, are symmetric under permutations of the external legs $S_5$.
Moreover, the partial amplitudes satisfy group-theoretic relations~\cite{Bern:1990ux,Edison:2011ta}, which follow from the decomposition in the basis $\{\mathcal{T}_{\lambda}\}$. At one loop, the double-trace components $A_{\lambda}^{(1,1)}$ can be written as linear combinations of the planar ones $A_{\lambda}^{(1,0)}$. At two loops, the colour-subleading single-trace components $A_{\lambda}^{(2,2)}$ can be expressed in terms of the planar $A_{\lambda}^{(2,0)}$ and of the double-trace $A_{\lambda}^{(2,1)}$ components.

The tree-level amplitude is given by the Parke-Taylor formula~\cite{Parke:1986,Nair:1988}
\begin{align}
\label{eq:A0sYM}
A^{(0)}_1 = \frac{1}{\vev{12} \vev{23} \vev{34} \vev{45} \vev{51}} \, ,
\end{align}
where we recall that the subscript refers to the colour decomposition~\eqref{eq:A05color}. The other single-trace components in Eq.~\eqref{eq:A05color} are simply obtained by permuting the external momenta in Eq.~\eqref{eq:A0sYM}.
Since the Parke-Taylor factors will appear many times in the rest of the paper, we introduce the short-hand notation 
\begin{align}
\label{eq:PTdefinition}
\text{PT}(i_1i_2i_3i_4i_5) = \frac{1}{\langle i_1 i_2 \rangle \langle i_2 i_3 \rangle \langle i_3 i_4 \rangle \langle i_4 i_5 \rangle \langle i_5 i_1 \rangle  }\,.
\end{align}
The integrand of the one-loop five-particle amplitude is given e.g. in Refs.~\cite{1997PhLB,Carrasco:2011mn}.

We expand the five-graviton amplitude in $\mathcal{N}=8$ supergravity in the gravitational coupling constant $\kappa$, with $\kappa^2 = 32 \pi G$, as
\begin{align}
\mathcal{M}_{5} = \delta^{(4)}\left(p_1+p_2+p_3+p_4+p_5 \right)  \delta^{(16)}(Q) \sum_{\ell \ge0 }  \bigg(\frac{\kappa}{2}\bigg)^{2 \ell + 3} \, \left( \frac{e^{- \eps \gamma_{\rm E} }}{(4 \pi)^{2-\eps}} \right)^{\ell} \, M^{(\ell)}_{5} \,,
\end{align}
where we have extracted the overall momentum and super-momentum $Q$ conservation delta functions. 
Note that $\kappa$ has dimension of $1/p$.
Note that there is no concept of colour in supergravity, and the partial amplitudes $M_5^{(\ell)}$ are therefore intrinsically non-planar. Just like in the $\mathcal{N}=4$ super Yang-Mills case they are invariant under permutations of the external legs. This symmetry can however be hidden in the explicit representation of the amplitude, as can be observed in the following expression for the tree-level amplitude~\cite{BERENDS198891},
\begin{align}
\label{eq:M0sugra}
M_5^{(0)} = s_{12} s_{34} \PT(12345) \PT(21435) + s_{13} s_{24} \PT(13245) \PT(31425) \, .
\end{align}
The integrand of the one-loop five-graviton amplitude can be found for instance in Refs.~\cite{Bern:1998sv,Carrasco:2011mn}.

\subsection{Expected structure of the two-loop amplitudes}
\label{sec:ExpectedStructure}

Having an insight into the final structure of the integrated amplitudes can simplify the computation dramatically. The two-loop five-point amplitudes in $\mathcal{N}=4$ super Yang-Mills and $\mathcal{N}=8$ supergravity, in particular, have been shown at symbol level~\cite{Abreu:2018aqd,Chicherin:2018yne,Chicherin:2019xeg,Abreu:2019rpt} to have \textit{uniform transcendental weight}~\cite{Henn:2013pwa}. The discussion below follows that of Refs.~\cite{Abreu:2018aqd,Chicherin:2018yne,Chicherin:2019xeg,Abreu:2019rpt}, where all of this was studied at symbol level.

In order to define this property of the amplitudes, it is convenient to start from the Feynman integrals which contribute to them. It was conjectured~\cite{Chicherin:2017dob} and then shown by explicit computation~\cite{Abreu:2018rcw,Chicherin:2018mue,Abreu:2018aqd,Chicherin:2018old} that all the massless five-point Feynman integrals up to two loops can be rewritten as linear combinations of \textit{pure} integrals~\cite{Henn:2013pwa}. 
An $\ell$-loop integral $\mathcal{I}^{(\ell)}_{\text{pure}}$ is pure if it has the very simple structure 
\begin{align}
\label{eq:PureForm}
\mathcal{I}^{(\ell)}_{\text{pure}}(X,\epsilon) = \frac{n}{\epsilon^{2 \ell}} \sum_{w=0}^{\infty} \epsilon^w h^{(w)}(X) \, ,
\end{align}
where $n$ is an arbitrary constant normalization factor, and $h^{(w)}$ is a weight-$w$ function (see Section~\ref{sec:PentagonFunctions}). At two loops, for instance, the Laurent expansion in $\epsilon$ of a pure integral starts with a constant leading pole, followed by logarithms at order $1/\epsilon^3$, and in general weight-$w$ functions at order $\epsilon^{w-4}$. 

As a result, a generic massless two-loop five-point (partial) amplitude $\mathcal{F}_5^{(2)}$ can be written as
\begin{align}
\mathcal{F}_5^{(2)} = \sum_i R_i(\lambda,\tilde{\lambda}, \epsilon) \, \mathcal{I}^{(2)\text{pure}}_i \, ,
\end{align}
where $\mathcal{I}^{(2)\text{pure}}_i$ are pure two-loop integrals, and the prefactors $R_i$ depend rationally on both the external spinors and $\epsilon$. If the latter do not depend on $\epsilon$, the (partial) amplitude $\mathcal{F}_5^{(2)}$ is said to have uniform transcendental weight. 

This property of the integrated amplitude is related to its integrand by a conjecture~\cite{ArkaniHamed:2010gh}. If the four-dimensional integrand\footnote{The analysis of the four-dimensional integrand is sometimes insufficient to determine the pureness of an integral. We refer the interested reader to Ref.~\cite{Chicherin:2018old} for progress towards a $D$-dimensional integrand analysis.} can be written in a so-called $d\log$-form~\cite{Lipstein:2012vs,Lipstein:2013xra,Arkani-Hamed:2014via,Bern:2015ple,Herrmann:2019upk} with constant prefactors and all square roots can be rationalised, then the integrated expression has uniform transcendental weight. In order to determine whether such a rewriting is possible, it is useful to study the poles~\cite{ArkaniHamed:2010gh,Henn:2013pwa,WasserMSc,Henn:2020lye}. A $d\log$ integrand, in fact, cannot have double poles. 

The absence of double poles has been shown for several integrands in $\mathcal{N}=4$ super Yang-Mills~\cite{ArkaniHamed:2012nw,Arkani-Hamed:2014via}, and all MHV amplitudes have been conjectured to have uniform transcendental weight~\cite{Bern:2005iz,Dixon:2011pw,ArkaniHamed:2012nw,KOTIKOV2007217}.
Their leading singularities~\cite{Cachazo:2008vp}, namely the rational prefactors of the pure integrals, are known~\cite{Arkani-Hamed:2014bca} to be given by Parke-Taylor tree-level amplitudes~\eqref{eq:PTdefinition} only\footnote{These properties are manifest in the representation of the four-dimensional integrand of the two-loop five-particle amplitude given by Ref.~\cite{Bern:2015ple}.}. In the five-point case, only six of them are linearly independent. Following Ref.~\cite{Bern:2015ple}, we choose the basis
\begin{equation}
\begin{aligned} \label{eq:PTbasis}
& {\rm PT}_1 = {\rm PT}(12345) \,, \qquad \quad {\rm PT}_2 = {\rm PT}(12354) \,, \\ 
& {\rm PT}_3 = {\rm PT}(12453) \,, \qquad \quad {\rm PT}_4 = {\rm PT}(12534) \,, \\ 
& {\rm PT}_5 = {\rm PT}(13425) \,, \qquad \quad {\rm PT}_6 = {\rm PT}(15423)\,. 
\end{aligned}
\end{equation}
The partial amplitudes of Eq.~\eqref{eq:A25color} are thus expected to have the structure
\begin{align}
\label{eq:sYM_expected}
A^{(2,k)}_{\lambda} = \sum_{i=1}^6 \sum_{j} a_{\lambda,ij}^{(2,k)} \, \text{PT}_i \, \mathcal{I}_j^{(2)\text{pure}} \, ,
\end{align}
where $a_{\lambda,ij}^{(k)} \in \mathbb{Q}$, and $\mathcal{I}_j^{(2)\text{pure}}$ are pure two-loop integrals.

Double or higher poles may appear in the integrands of the amplitudes in $\mathcal{N}=8$ supergravity in general. The two-loop five-graviton amplitude in particular, however, has been shown to be free of double poles at least at infinity~\cite{Herrmann:2018dja,Bourjaily:2018omh}. This hint of uniform transcendentality was indeed confirmed by the explicit computation of the symbol of the amplitude~\cite{Chicherin:2019xeg,Abreu:2019rpt}. The leading singularities $r^{(2)}_i$ form a 45-dimensional space, spanned by 40 permutations of
\begin{align}
\label{eq:r2_1}
r^{(2)}_1 = s_{13} s_{24} s_{35} s_{45} \text{PT}(13542) \text{PT}(31452)  \,,
\end{align}
and by
\begin{align}
\label{eq:r2_41}
r^{(2)}_{40+k} = \frac{s_{k \, k+1}}{\epsilon_5}\frac{[12][23][34][45][51]}{\vev{1 2}\vev{2 3}\vev{3 4}\vev{4 5}\vev{5 1}}\,, \qquad \text{for} \ k = 1,\ldots, 5\, ,
\end{align}
where the indices of the Mandelstam invariant $s_{k\, k+1}$ are defined modulo 5.
Our explicit choice of basis is provided in the ancillary files of Ref.~\cite{Chicherin:2019xeg}. The ensuing expected structure of the $\mathcal{N}=8$ supergravity amplitude then is
\begin{align}
\label{eq:sugra_expected}
M_5^{(2)} = \sum_{i=1}^{45} \sum_{j} m^{(2)}_{ij}  r^{(2)}_i  \mathcal{I}_j^{(2)\text{pure}} \, , 
\end{align}
where $m^{(2)}_{ij} \in \mathbb{Q}$.

\subsection{Two-loop integrands}
The integrands of two-loop five-point $\mathcal N=4$ super-Yang-Mills and $\mathcal
N=8$ supergravity amplitudes were obtained in Ref.~\cite{Carrasco:2011mn} using $D$-dimensional unitarity and colour-kinematics duality.
The external states and momenta live in the four-dimensional Minkowski space. The loop momenta live in $D=4-2\eps$ dimensions, and the internal states are treated in the Four-Dimensional-Helicity scheme.

 \begin{figure}[t]
  \begin{center}
    \includegraphics[width=0.3\columnwidth]{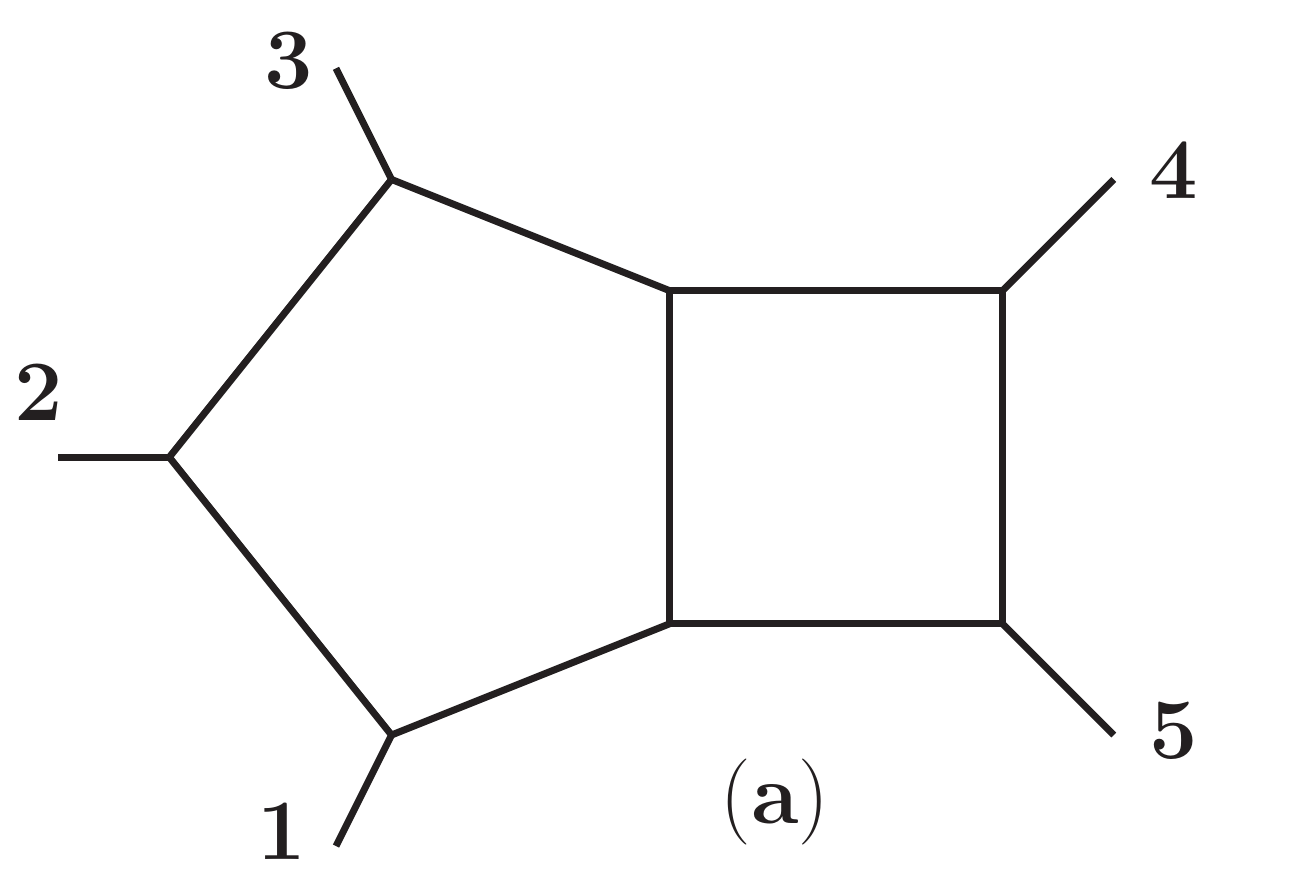}
    \includegraphics[width=0.3\columnwidth]{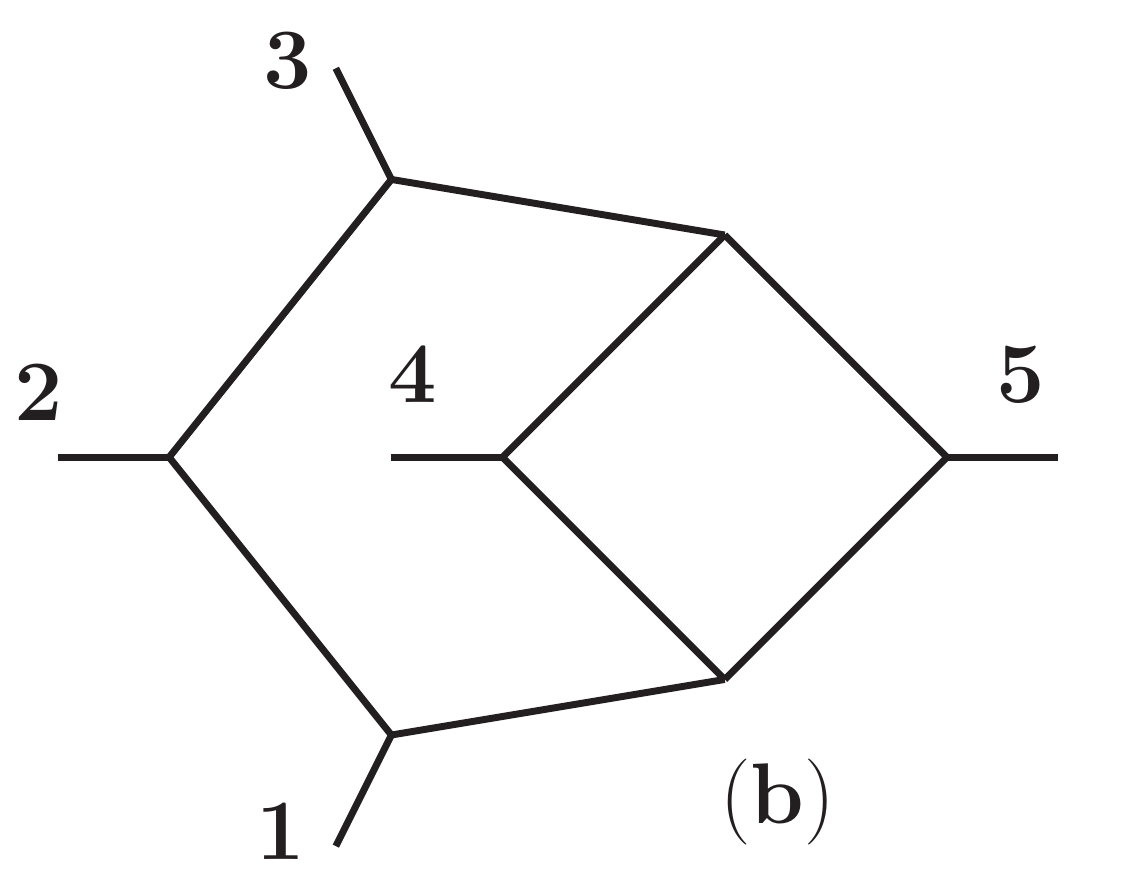}
\includegraphics[width=0.33\columnwidth]{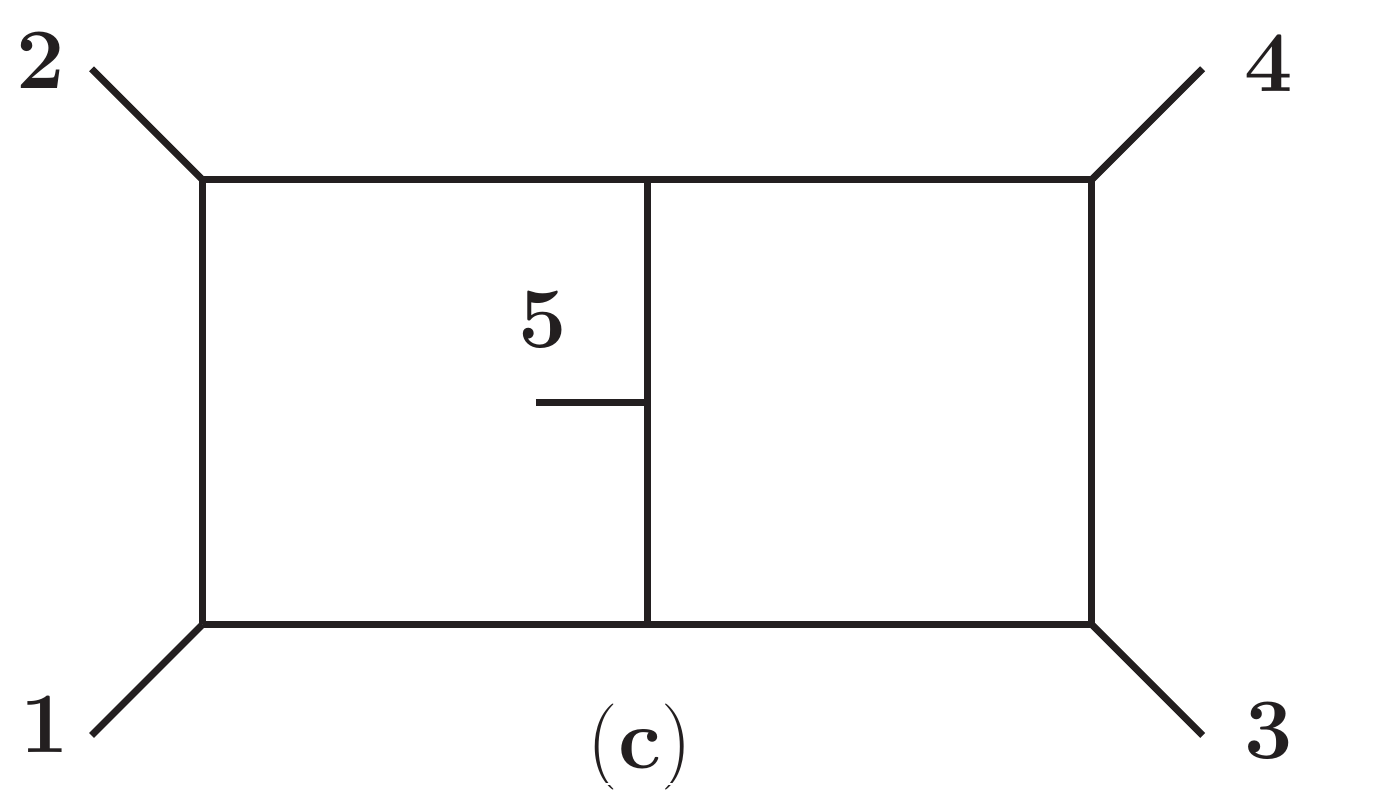}
\includegraphics[width=0.3\columnwidth]{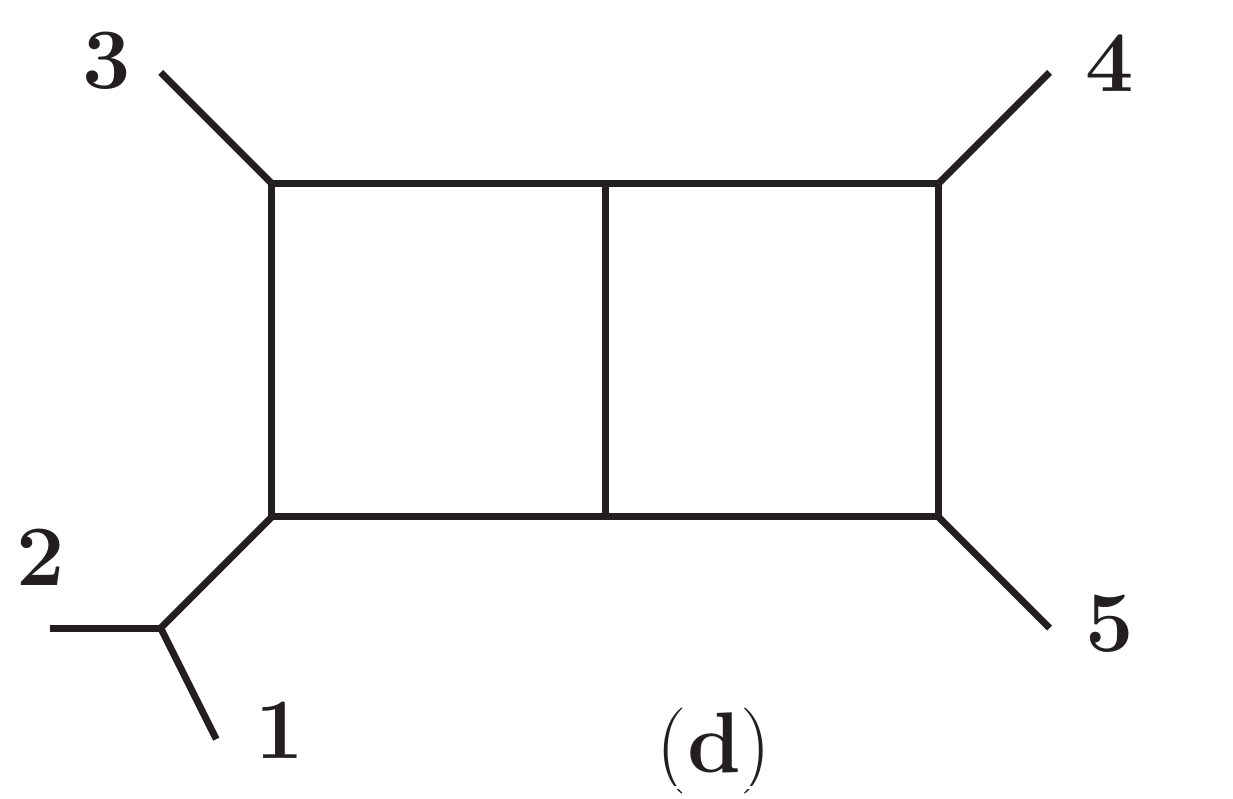}
\includegraphics[width=0.3\columnwidth]{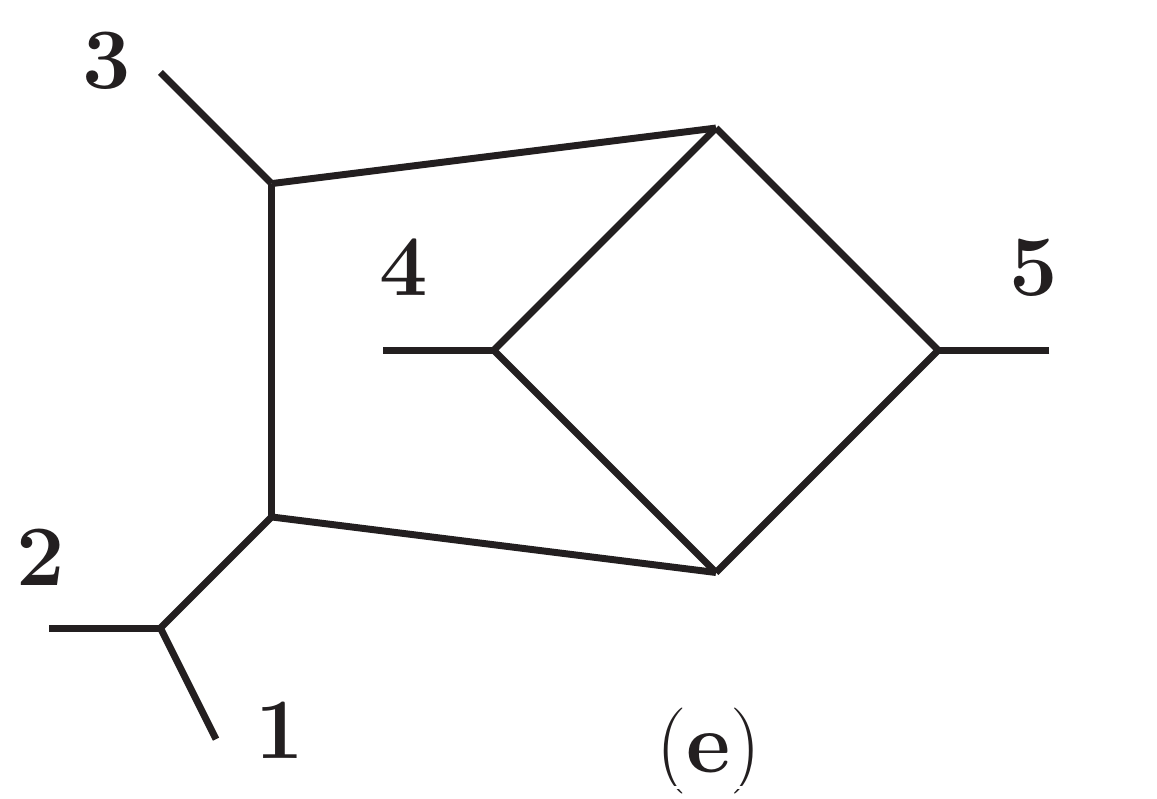}
\includegraphics[width=0.3\columnwidth]{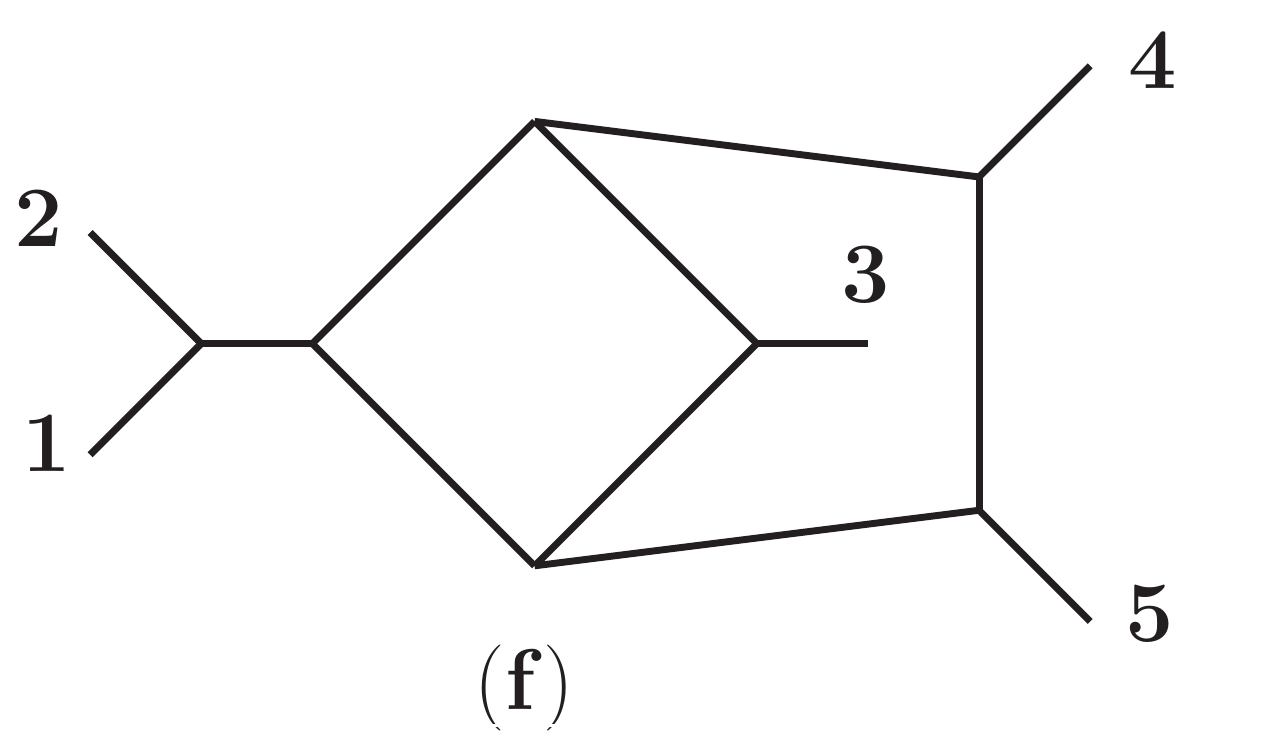}
    \caption{Diagrams in the representation of the integrands of the two-loop five-point amplitudes given by Ref.~\cite{Carrasco:2011mn}.}
    \label{fig:Diagrams}
  \end{center}
\end{figure}

The integrand of the two-loop five-point amplitude in $\mathcal N=4$ super-Yang-Mills can be written as
\begin{equation}
  \label{SYM_integrand}
  A_5^{(2)} = \sum_{S_5} \bigg(\frac{1}{2} \mathcal I^{(a)}_{\mathcal N=4} + \frac{1}{4}
  \mathcal I^{(b)}_{\mathcal N=4} + \frac{1}{4} \mathcal I^{(c)}_{\mathcal N=4} + \frac{1}{2}
  \mathcal I^{(d)}_{\mathcal N=4} + \frac{1}{4}
  \mathcal I^{(e)}_{\mathcal N=4} + \frac{1}{4} \mathcal I^{(f)}_{\mathcal N=4}\bigg) \, ,
\end{equation}
where the sum runs over the permutations of the external legs, and the integral upper indices correspond to the diagrams in Fig.~\ref{fig:Diagrams}. Schematically, each of the six integrals in Eq.~\eqref{SYM_integrand} has the form
\begin{equation}
  \label{eq:Integrals_SYM}
  \mathcal I^{(x)}_{\mathcal N=4}=\int \frac{d^Dl_1 }{i \pi^{D/2}} \frac{d^Dl_2 }{i \pi^{D/2}}
  \frac{c^{(x)} N^{(x)}}{D^{(x)}_1 \ldots D^{(x)}_8},\quad \text{for} \ \, x=a,b,\ldots, f\,.
\end{equation}
Here, the $D_i^{(x)}$ are the propagators associated with the graph $(x)$ in Fig.~\ref{fig:Diagrams} (for the graphs (d), (e) and (f) one of the propagators is given by $1/s_{12}$). The colour factor $c^{(x)}$ is a product of Lie-algebra structure constants, which we write as a vector in the colour basis given by Eqs.~\eqref{eq:single_traces} and~\eqref{eq:double_traces}. $N^{(x)}$ is a numerator in the Bern-Carrasco-Johansson form~\cite{Bern:2010ue}, which depends at most linearly in the loop momenta. For the explicit expressions of colour factors and numerators we refer to Eqs.~(4.15) and Table I of the original work~\cite{Carrasco:2011mn}, respectively.

As for $\mathcal{N}=8$ supergravity, the integrand of the two-loop five-graviton amplitude takes the same form as Eq.~\eqref{SYM_integrand},
\begin{equation}
  \label{SUGRA_integrand}
  M_5^{(2)} = \sum_{S_5} \bigg(\frac{1}{2} \mathcal I^{(a)}_{\mathcal N=8} + \frac{1}{4}
  \mathcal I^{(b)}_{\mathcal N=8} + \frac{1}{4} \mathcal I^{(c)}_{\mathcal N=8} + \frac{1}{2}
  \mathcal I^{(d)}_{\mathcal N=8} + \frac{1}{4}
  \mathcal I^{(e)}_{\mathcal N=8} + \frac{1}{4} \mathcal I^{(f)}_{\mathcal N=8}\bigg) \, ,
\end{equation}
but the integrals have different numerators and no colour factor,
\begin{equation}
  \label{eq:Integrals_sugra}
  \mathcal I^{(x)}_{\mathcal N=8}=\int \frac{d^Dl_1 }{i \pi^{D/2}} \frac{d^Dl_2 }{i \pi^{D/2}}
  \frac{\tilde{N}^{(x)}}{D^{(x)}_1 \ldots D^{(x)}_8},\quad \text{for} \ \, x=a,b,\ldots, f\,.
\end{equation}
The numerators $\tilde{N}^{(x)}$ are obtained by ``squaring" the $\mathcal{N}=4$ super Yang-Mills ones as shown in Eq.~(4.17) of Ref.~\cite{Carrasco:2011mn}, and thus depend at most quadratically on the loop momenta.

Note that the integrals~\eqref{eq:Integrals_SYM} and~\eqref{eq:Integrals_sugra}, and therefore all the results presented in this paper, are not dimensionless. They can be made dimensionless by multiplying them by a factor of $(\mu^2)^{2\epsilon}$, for some scale $\mu$.

\subsection{Pure integral bases}
The massless two-loop five-point integrals are organized into three integral families, whose propagator structures are given by the graphs (a), (b) and (c) in Fig.~\ref{fig:Diagrams}.
The integrals $a$ and $d$ belong to the planar pentagon-box family, spanned by $61$ master integrals, 
$\{I_i^{(a)}\}_{i=1}^{61}$, computed in Refs.~\cite{Gehrmann:2015bfy, Papadopoulos:2015jft, Gehrmann:2018yef}. The non-planar hexagon-box family, which includes the integrals $b$ and $e$, is spanned by 73 master integrals, $\{I_i^{(b)}\}_{i=1}^{73}$, calculated in Ref.~\cite{Chicherin:2018mue} (see also~\cite{Chicherin:2017dob,Chicherin:2018ubl,Chicherin:2018wes,Abreu:2018rcw}). Finally, the integrals $c$ and $f$ belong to the non-planar double-pentagon family, which has $108$ master integrals, $\{I_i^{(c)}\}_{i=1}^{108}$, computed in Refs.~\cite{Chicherin:2018old,Badger:2019djh} (see also~\cite{Abreu:2018aqd}). Note that the integrals $d$, $e$ and $f$ effectively have four-point kinematics, and are also known from Refs.~\cite{Gehrmann:2000zt,Gehrmann:2001ck}.

Our strategy for integrating the integrands~\eqref{SYM_integrand} and~\eqref{SUGRA_integrand} is the following. First, we use integration-by-parts (IBP) identities~\cite{Chetyrkin:1981qh} to rewrite the summands of Eqs.~\eqref{SYM_integrand} and~\eqref{SUGRA_integrand} in terms of master integrals. Note that at this stage we do not perform any permutation of the external legs. Since the numerator degree is low, we can use either the public available IBP packages like FIRE6~\cite{Smirnov:2019qkx}, Kira~\cite{Maierhoefer:2017hyi} and Reduze2~\cite{vonManteuffel:2012np}, or private IBP solvers with novel approaches~\cite{Ita:2015tya, Abreu:2018jgq, Boehm:2018fpv, Bendle:2019csk,Guan:2019bcx}.
The resulting form of the super Yang-Mills amplitude is
\begin{equation}
  A_5^{(2)} = \sum_{S_5} \left[ \sum_{\lambda=1}^{22} \left(
  \sum_{j=1}^{61} c_{\lambda,j}^{(a)} \, I^{(a)}_j+
  \sum_{j=1}^{73} c_{\lambda,j}^{(b)} \, I^{(b)}_j+ 
  \sum_{j=1}^{108} c_{\lambda,j}^{(c)} \, I^{(c)}_j
\right) \mathcal{T}_{\lambda} \right] \, ,
\end{equation}
and similarly for the supergravity amplitude. The prefactors $c_{\lambda,j}^{(x)} = c_{\lambda,j}^{(x)}(\lambda, \tilde{\lambda}, N_c, \epsilon)$ depend on $N_c$, on the spinor products of the external momenta, and on $\epsilon$. 

The choice of master integrals at this stage is somewhat arbitrary, and follows from the algorithm used in the solution of the IBP identities. It is extremely convenient to make a specific choice, namely to transform to a basis of pure master integrals.
The pure master integrals $\tilde{I}^{(x)} = \{\tilde{I}^{(x)}_i\}_{i=1}^{n_x}$ of each family $x=a,b,c$ satisfy a differential equation in the canonical form~\cite{Henn:2013pwa}
\begin{align}
\label{eq:CanonicalDE}
d \tilde{I}^{(x)}\left(X, \epsilon\right) = \epsilon \sum_{i=1}^{31} A^{(x)}_i d\log W_i(X) \cdot \tilde{I}^{(x)}\left(X, \epsilon \right) \, ,
\end{align}
where the $A_i^{(x)}$ are constant rational matrices and the $W_i$ are letters of the pentagon alphabet~\cite{Chicherin:2017dob,Chicherin:2018old} reviewed in Section~\ref{sec:PentagonFunctions}. Once the boundary values are known, the solution of Eq.~\eqref{eq:CanonicalDE} in terms of iterated integrals or Goncharov polylogarithms~\cite{Goncharov:2010jf,Duhr:2011zq} is straightforward. Different choices of pure master integral bases can be found in the references mentioned above together with the differential equations they satisfy. We use the ones of Ref.~\cite{Badger:2019djh}, where also the boundary values are computed for all permutations of the external legs at the kinematic point given by Eq.~\eqref{eq:SymmetryBasePoint}.
 
In order to perform the change of master integrals, we first reduce the pure bases to the bases chosen by the IBP solver. This way we determine the transformation matrices $T^{(x)}$,
\begin{align}
\tilde{I}^{(x)} = T^{(x)} \cdot I^{(x)}\,.
\end{align}
Then, we compute the inverse transformation matrices $\left(T^{(x)}\right)^{-1}$ using the sparse linear algebra method of Ref.~\cite{Boehm:2018fpv}, and use them to rewrite the amplitudes in terms of pure integrals,
\begin{equation}
\label{eq:A52pure}
  A_5^{(2)} = \sum_{S_5} \left[ \sum_{\lambda=1}^{22} \left(
  \sum_{j=1}^{61} \tilde{c}_{\lambda,j}^{(a)} \, \tilde{I}^{(a)}_j+
  \sum_{j=1}^{73} \tilde{c}_{\lambda,j}^{(b)} \, \tilde{I}^{(b)}_j+ 
  \sum_{j=1}^{108} \tilde{c}_{\lambda,j}^{(c)} \, \tilde{I}^{(c)}_j
\right) \mathcal{T}_{\lambda} \right] \, ,
\end{equation}
and similarly for the $\mathcal{N}=8$ supergravity amplitude. Note
that, at this stage, the prefactors of the integrals
$\tilde{c}_{\lambda,j}^{(x)}=\tilde{c}_{\lambda,j}^{(x)}(\lambda,
\tilde{\lambda}, N_c, \epsilon)$ still depend not only on the
kinematics and on $N_c$, but also on $\epsilon$. Similarly, the pure
integral prefactors of the $\mathcal{N}=8$ amplitude also depend on
$\epsilon$. Therefore, before summing up the different permutations, these amplitudes do not yet exhibit uniform transcendentality in a manifest way.

\subsection{Permutation of the external legs and integrated expressions}

As a final step, we need to sum over the permutations of the external legs. This step involves two issues, related to the rational functions and to the integrals, respectively. 

The integrals enter the amplitudes in all permutations of the external legs. As a result, we need to know them in all the kinematic regions. This can in principle by achieved via analytic continuation, see e.g. Refs.~\cite{Gehrmann:2018yef,Henn:2019rgj}, but this approach is cumbersome and error-prone. A different strategy was followed in Ref.~\cite{Badger:2019djh}. Each permutation of the required master integrals was considered separately, and computed directly in the $s_{12}$-channel. Permuting the differential equations is in fact straightforward, as they are rational in the kinematic variables. The boundary values were computed for all permutations of the external legs. In addition, relations between integrals of different families and with permuted external legs were found, in order to rewrite the amplitudes in terms of fewer pure integrals. We make use of these results of Ref.~\cite{Badger:2019djh}. No analytic continuation is needed.

The rational prefactors are trivial from the analytic point of view, but their proliferation in the sum over the permutations leads to a rapid growth in size of the expression. In order to tame this, we substitute the kinematic variables with random numbers in the rational prefactors. Note that we use rational numbers rather than floating-point numbers, so that there is no loss in precision. Then we make an ansatz of a $\mathbb{Q}$-linear combination of the known leading singularities --~see Section~\ref{sec:ExpectedStructure}~-- for the prefactor of each pure integral, and fix the unknown coefficients with just 6 (45) independent random evaluations of the $\mathcal{N}=4$ super Yang-Mills ($\mathcal{N}=8$ supergravity) amplitude. Additional evaluations are used to validate the result.

Note that the dependence on $\epsilon$ in the rational prefactors of the pure integrals drops out only after summing up all permutations and removing the redundancy due to the relations between the pure integrals of different families and in different orientations. Only then do the amplitudes become uniformly transcendental in a manifest way.

Finally, we obtain expressions for the two-loop five-point amplitudes in $\mathcal{N}=4$ super Yang-Mills and $\mathcal{N}=8$ supergravity of the form given by Eq.~\eqref{eq:sYM_expected} and~\eqref{eq:sugra_expected}, respectively. The leading singularities are those given in Section~\ref{sec:ExpectedStructure}, and the pure integrals are a set of the pure master integrals spanning the three relevant integral families in all orientations of the external legs.

We therefore have full analytical and numerical control over the amplitudes. Using the differential equations and the boundary constants of Ref.~\cite{Badger:2019djh}, we can straightforwardly rewrite them in terms of iterated integrals or Goncharov polylogarithms, and evaluate them anywhere in the physical scattering regions.

\subsection{Infrared factorisation and hard functions}
\label{sec:IRfactorization}

The infrared --~soft and collinear~-- divergences of scattering amplitudes factorise in well known ways in both gauge and gravity theories. As a result, the infrared-singular part of an amplitude is entirely determined by lower-loop information. This not only constitutes a useful check on amplitude calculations, but also allows to define an infrared-safe hard or remainder function, where the infrared singularities are removed. Experience shows that the hard functions exhibit a much simpler structure than the amplitudes. In Section~\ref{sec:IRsYM} we review the infrared factorisation of massless scattering amplitudes in gauge theories, whereas Section~\ref{sec:IRsugra} is devoted to the infrared structure of graviton amplitudes.

\subsubsection{$\mathcal{N}=4$ super Yang-Mills}
\label{sec:IRsYM}

The infrared singularities of massless scattering amplitudes in gauge theories factorise to all orders in the coupling according to the formula~\cite{Catani:1998bh, Sterman:2002qn, Dixon:2008gr, Becher:2009cu, Almelid:2015jia}
\begin{align}
\label{eq:DipoleFormula}
\mathcal{A}_5\left(\frac{s_{ij}}{\mu^2}, a(\mu^2), \epsilon \right) = \textbf{Z}_5\left(\frac{s_{ij}}{\mu_F^2}, a(\mu_F^2), \epsilon \right) \mathcal{A}_5^f\left(\frac{s_{ij}}{\mu^2}, \frac{\mu^2}{\mu_F^2}, a(\mu^2), \epsilon \right) \, ,
\end{align}
where the operator $\textbf{Z}_5$ captures all the poles in $\epsilon$, and the remaining amplitude $\mathcal{A}^f_5$ is therefore finite. Here, $\mu$ and $\mu_F$ are the renormalisation and factorisation scale, respectively, which for simplicity we choose to be equal. We will often choose $\mu = 1$, as the explicit dependence can be recovered from dimensional analysis.
Note that, since we treat the amplitudes as vectors in colour space, the pole operator $\textbf{Z}_5$ is a matrix, which we denote in bold face.

Letting $\epsilon \to 0$ in the finite amplitude defines the hard or remainder function,
\begin{align}
\mathcal{H}_5 = \lim_{\epsilon \to 0} \mathcal{A}^f_5 \, .
\end{align}
We adopt the $\overline{\text{MS}}$ scheme, namely we keep only the pure pole part in the operator $\mathbf{Z}_5$ and neglect the finite terms.
The operator $\mathbf{Z}_5$ may then be written as the path ordered exponential of an anomalous dimension,
which up to two-loop order is given by a concise ``dipole" form: 
\begin{align} \label{eq:dipole}
\mathbf{\Gamma}_5 \equiv
-\gamma_{\rm cusp}\sum_{i< j}^5 \left(\textbf{T}_i  \cdot \textbf{T}_j \right) \log \left(\frac{-s_{ij}}{\mu^2} \right)
+\sum_{i=1}^5 \gamma_{\rm c} \, ,
\end{align}
where the operators $\textbf{T}^a_i$ insert a colour generator in the adjoint representation on the $i^{\text{th}}$ leg.
Explicitly, their action on the generators of $SU(N_c)$ $T^{a_i}$ is
\begin{align}
\label{eq:ColourInsertion}
\textbf{T}_i^b \circ T^{a_i} = \begin{cases} 0\, , & j\neq i\, , \\ - i f^{b a_i c_i} T^{c_i}\, , & j=i \, . \\ \end{cases}
\end{align}
The renormalisation operator is then given by (using the fact that the $\beta$-function of $\mathcal{N}=4$ super Yang-Mills vanishes)~\cite{Becher:2009cu}:
\begin{align}
\log \textbf{Z}_5 =
a \left( \frac{\Gamma_5'^{(1)}}{4\epsilon^2}+\frac{\mathbf{\Gamma}_5^{(1)}}{2\epsilon}\right)+
a^2 \left( \frac{\Gamma_5'^{(2)}}{16\epsilon^2}+\frac{\mathbf{\Gamma}_5^{(2)}}{4\epsilon}\right) +O(a^3) \,,
\end{align}
where $\mathbf{\Gamma}_5^{(\ell)}$ is the coefficient of $a^\ell$ in $\mathbf{\Gamma}_5$ and
\begin{align}
\Gamma_5' = \mu\frac{\partial}{\partial \mu} \mathbf{\Gamma}_5 = 
2\gamma_{\rm cusp}\sum_{i< j}^5 \left(\textbf{T}_i  \cdot \textbf{T}_j \right) = -5C_A\gamma_{\rm cusp} \,,
\end{align}
with $C_A=N_c$. 
Finally, $\gamma_{\rm cusp}$ is the cusp anomalous dimension normalised
by the quadratic Casimir in the adjoint representation $C_A$~\cite{Korchemsky:1985xj,Korchemskaya:1992je,Moch:2004pa,Beisert:2006ez,Bern:2006ew,Henn:2019swt,vonManteuffel:2020vjv}, 
\begin{equation} \gamma_{\rm cusp} =4a - \frac{4\pi^2}{3} C_A a^2 +O(a^3)\,,
\end{equation}
and $\gamma_{\rm c}= 2\zeta_3C_A^2a^2 +O(a^3)$ is the collinear anomalous dimension.

The analytic continuation of the logarithms in $\mathbf{\Gamma}_5$~\eqref{eq:dipole} to the desired region is achieved by adding a small imaginary part to each timelike $s_{ij}$,
\begin{align}
\label{eq:LogAnalyticContinuation}
\log\left(-s_{ij} - i0^+\right) = \begin{cases} \log s_{ij} - i \pi \, , & \text{if} \ s_{ij}>0\, , \\ \log\left(-s_{ij}\right)\, , & \text{if} \ s_{ij}<0 \,.\\\end{cases}
\end{align}
Denoting $\mathcal{A}^{(\ell)}_{5;k}$ the order-$\epsilon^k$ term of $\mathcal{A}^{(\ell)}_{5}$,
the one and two-loop hard functions in the $\overline{\text{MS}}$ scheme are thus explicitly given as 
\begin{align}
\label{eq:H1}
& \mathcal{H}^{(1)}_5 = \mathcal{A}^{(1)}_{5;0} \, , \\
\label{eq:H2}
& \mathcal{H}^{(2)}_5 = \mathcal{A}^{(2)}_{5;0} + 5 C_A \, \mathcal{A}^{(1)}_{5;2}
+ 2\sum_{i<j}^5  \left(\textbf{T}_i  \cdot \textbf{T}_j \right) \log \left(\frac{-s_{ij}}{\mu^2} \right)\, \mathcal{A}^{(1)}_{5;1} \, .
\end{align}
The subleading coefficients of the one-loop amplitudes are essential in this formula.

\subsubsection{$\mathcal{N}=8$ supergravity}
\label{sec:IRsugra}

Perturbative gravity has a much simpler infrared structure as compared to Yang-Mills theories. Graviton scattering amplitudes are in fact free of collinear divergences~\cite{Weinberg:1965nx}, and exhibit soft divergences only. This results in a single pole in the dimensional regulator $\epsilon = (4-D)/2$ per loop order, rather than a double pole as in Yang-Mills theories. Moreover, due to the absence of colour, the renormalization factor is a simple exponential (as opposed to a path-ordered exponential).
The divergences exponentiate in a remarkably simple way~\cite{Weinberg:1965nx, Dunbar:1995ed, Naculich:2008ew, Naculich:2011ry, White:2011yy, Akhoury:2011kq,Beneke:2012xa},
\begin{align}
\label{eq:IRsugra}
\mathcal{M}_5 = \mathcal{S}_5 \, \mathcal{M}^f_5\, . 
\end{align}
The gravitational soft function $\mathcal{S}_5$ captures all divergences due to soft graviton exchanges, and is obtained to all orders in the coupling by exponentiating the infrared divergence of the one-loop amplitude,
\begin{align}
\label{eq:SoftFactorSugra}
\mathcal{S}_5 = \exp{\left[\frac{\sigma_5}{\epsilon}\right]}\,, \qquad \qquad \sigma_5 = \left(\frac{\kappa}{2} \right)^2 \sum_{j=1}^5 \sum_{i<j} s_{ij} \log\left(\frac{-s_{ij}}{\mu^2} \right) \, ,
\end{align}
where $\mu$ is a factorization scale. We choose $\mu=1$ for simplicity. The soft divergences of graviton amplitudes are in this sense one-loop exact. The analytic continuation of the logarithms to the desired scattering region is given by Eq.~\eqref{eq:LogAnalyticContinuation}.

Just like in the Yang-Mills case, we can let $\epsilon \rightarrow 0$ in the finite amplitude $\mathcal{M}^f_5$, and in this way define an infrared-safe hard or remainder function,
\begin{align}
\mathcal{F}_5 \equiv \underset{\epsilon \to 0}{\lim} \, \mathcal{M}^f_5 \,.
\end{align}
The one and two-loop contributions are given by
\begin{align}
\label{eq:F1}
& \mathcal{F}^{(1)}_{5} = \mathcal{M}^{(1)}_{5;0} \, , \\
\label{eq:F2}
& \mathcal{F}^{(2)}_{5} = \mathcal{M}^{(2)}_{5;0}- \sigma_5 \, \mathcal{M}^{(1)}_{5;1} \,,
\end{align}
where $\mathcal{M}^{(\ell)}_{5;k}$ is the order-$\epsilon^k$ term of $\mathcal{M}^{(\ell)}_{5}$.

\subsection{The two-loop hard functions}
\label{sec:HardFunctions}

In this section we present our results for the two-loop five-particle hard functions in $\mathcal{N}=4$ super Yang-Mills and $\mathcal{N}=8$ supergravity. We first discuss their structure, and then provide numerical reference values for future cross-checks.

Let us begin with $\mathcal{N}=4$ super Yang-Mills.
In the definition of the two-loop hard function~\eqref{eq:H2}, the one-loop amplitude is needed up to order $\epsilon^2$. In order to obtain it, we started from the integrand of Refs.~\cite{Carrasco:2011mn}, and followed the same procedure described for the two-loop amplitude. The result has manifestly uniform transcendental weight, with the same set of rational functions as the two-loop amplitude,
\begin{align}
\label{eq:sYM_expected_1loop}
A_{\lambda}^{(1,k)} = \sum_{i=1}^6 \sum_j a_{\lambda,ij}^{(1,k)} \, \PT_i \, \mathcal{I}_j^{(1) \text{pure}} \,,
\end{align}
where $a_{\lambda,ij}^{(1,k)}\in \mathbb{Q}$ and $\mathcal{I}_j^{(1)\text{pure}}$ are pure one-loop integrals.
Putting together the one and two-loop amplitudes as given by Eqs.~\eqref{eq:sYM_expected_1loop} and~\eqref{eq:sYM_expected} gives expressions for the one and two-loop hard functions of the form
\begin{align}
\label{eq:sYM_expected_H1}
\mathcal{H}^{(1,k)}_{\lambda} = \sum_{i=1}^6 \sum_j b_{\lambda,ij}^{(1,k)} \, \PT_i \, P^{(2)}_j \,,\\
\label{eq:sYM_expected_H2}
\mathcal{H}^{(2,k)}_{\lambda} = \sum_{i=1}^6 \sum_j b_{\lambda,ij}^{(2,k)} \, \PT_i \, P^{(4)}_j \,,
\end{align}
where $b_{\lambda,ij}^{(\ell,k)}\in \mathbb{Q}$ and $P^{(w)}_j$ are weight-$w$ pentagon functions. We adopted for the hard functions the same colour decomposition as for the amplitudes, given by Eqs.~\eqref{eq:A15color} and~\eqref{eq:A25color}. We find that the letter $W_{31}$ of the pentagon alphabet, present in the amplitudes, completely drops out of the hard function, as was already observed at symbol level~\cite{Abreu:2018aqd,Chicherin:2018yne}.

We have full analytical control over the two-loop hard function in the form given by Eq.~\eqref{eq:sYM_expected_H2}. This allowed us to compute its asymptotic behaviour in the multi-Regge limit, as we discuss in Section~\ref{sec:sYM}.  We can also evaluate the hard function anywhere in the physical scattering region. In order to facilitate future cross-checks, we provide numerical values at the reference point
\begin{align}
\label{eq:kinpoint_s12}
X_R = \left\{ \frac{13}{4}\;, -\frac{9}{11} \;, \frac{3}{2} \;, \frac{3}{4} \;, -\frac{2}{3} \right\} \, , \quad \text{with} \ \ \eps_5 = i \frac{\sqrt{222767}}{264} \, ,
\end{align}
in Table~\ref{tab:H2}. The explicit expression of the one and two-loop hard function in terms of the pentagon functions introduced in Ref.~\cite{Chicherin:2020oor} can be obtained at
\begin{center}
\href{https://pentagonfunctions.hepforge.org/downloads/2l_5pt_hardfunctions_N=4_N=8.tar.gz}{https://pentagonfunctions.hepforge.org/downloads/2l\_5pt\_hardfunctions\_N=4\_N=8.tar.gz}.
\end{center}
Although we are not computing a cross section and thus we cannot assess the final impact of the non-planar corrections on a theory prediction, we still find it interesting to note that the non-planar colour components of the two-loop hard function are of the same order of magnitude as the planar ones, as can be seen in Table~\ref{tab:H2}.

\begin{table}[t!]
\begin{center}
\begin{tabular}{|c|c|c|c|}
\hline 
& $N_c^2$ & $N_c$ & $N_c^0$ \\
\hline 
${\cal T}_1$ & $74.92986-61.83635 i $ &$ 0$ & $-617.3565+294.7986 i$ \\
\hline 
${\cal T}_2$ & $92.3051+108.9834 i$ & $0$ & $-1024.0932+532.1760 i$ \\
\hline 
${\cal T}_3$ & $-49.51614+73.37582 i$ & $0$ & $258.3246+558.5523 i$ \\
\hline 
${\cal T}_4$ & $7.50918+52.48750 i$ & $0$ & $427.1264+340.3532 i$ \\
\hline 
${\cal T}_5$ & $-95.8105-124.8597 i$ & $0 $& $-73.4024-741.5020 i$ \\
\hline 
${\cal T}_6$ & $-134.93821+4.43862 i $& $0$ & $853.1018-590.6476 i$ \\
\hline 
${\cal T}_7$ & $-12.39259+33.13533 i$ & $0$ & $494.0699+262.7033 i$ \\
\hline 
${\cal T}_8$ & $37.35506+120.68054 i$ & $0$ & $87.3332+500.0807 i$ \\
\hline 
${\cal T}_9$ & $80.04433+33.19817 i$ & $0$ & $-839.1711+349.2263 i$ \\
\hline 
${\cal T}_{10}$ & $50.71731-21.09889 i$ & $0 $& $-670.3692+131.0271 i$ \\
\hline 
${\cal T}_{11}$ & $-39.34196-85.68420 i$ &$ 0$ & $-263.6325-106.3503 i$ \\
\hline 
${\cal T}_{12}$ & $-27.72786+22.45736 i$ & $0$ & $662.8718+44.5041 i$ \\
\hline 
${\cal T}_{13}$ & $0$ & $-125.2669+216.9434 i$ & $0$ \\
\hline 
${\cal T}_{14}$ & $0$ &$ -696.3813-209.4301 i$ &$ 0$ \\
\hline 
${\cal T}_{15}$ & $0$ &$ -344.4732+447.8376 i$ & $0$ \\
\hline 
${\cal T}_{16}$ & $0$ &$ -127.9880+116.6798 i$ & $0$ \\
\hline 
${\cal T}_{17}$ & $0$ &$ -444.5692-325.7655 i $&$ 0$ \\
\hline 
${\cal T}_{18}$ & $0$ &$ -510.7351-321.1812 i$ & $0$ \\
\hline 
${\cal T}_{19}$ & $0$ &$ 459.3389+210.4025 i $& $0$ \\
\hline 
${\cal T}_{20}$ & $0$ & $-120.7437+267.2953 i$ &$ 0$ \\
\hline 
${\cal T}_{21}$ & $0$ &$ 711.4669+60.1616 i$ & $0$ \\
\hline 
${\cal T}_{22}$ & $0$ & $-460.7431-329.6070 i$ & $0$ \\
\hline 
\end{tabular}
\end{center}
\caption{Numerical values in the kinematic point~\p{eq:kinpoint_s12} of the two-loop five-particle hard function in $\mathcal{N}=4$ super Yang-Mills normalised by the Parke-Taylor factor $\PT_1$~\p{eq:PTbasis}, ${\cal H}_5^{(2)}/{\rm PT}_1$. The rows correspond to the decomposition in the colour basis given by Eqs.~\p{eq:single_traces} and~\p{eq:double_traces}, and the columns to the power of $N_c$.}
\label{tab:H2}
\end{table}

The two-loop hard function $\mathcal{F}_5^{(2)}$ in $\mathcal{N}=8$ supergravity is defined by Eq.~\eqref{eq:F2} in terms of the finite part of the two-loop amplitude $M^{(2)}_5$~\eqref{eq:sugra_expected}, of the order-$\eps$ part of the one-loop amplitude $M^{(1)}_5$~\eqref{eq:sugraM1}, and of the soft factor $\sigma_5$~\eqref{eq:SoftFactorSugra}. It is worth stressing that the logarithms in the latter have to be analytically continued to the $s_{12}$ channel as prescribed by Eq.~\eqref{eq:LogAnalyticContinuation}.
We obtained the one-loop amplitude by applying the same procedure described for the two-loop amplitude on the one-loop integrand presented in Ref.~\cite{Carrasco:2011mn}. The result is the one-loop analogue of Eq.~\eqref{eq:sugra_expected}, namely an expression with uniform transcendental weight,
\begin{align}
\label{eq:sugraM1}
\mathcal{M}^{(1)} = \sum_{i=1}^{16} m^{(1)}_{ij}  r^{(1)}_i  \mathcal{I}_j^{(1)\text{pure}},
\end{align}
where $\mathcal{I}_j^{(1)\text{pure}}$ are pure one-loop five-point integral, $r^{(1)}_i$ are rational functions of the kinematics, and $m^{(1)}_{ij} \in \mathbb{Q}$. Since the coupling constant $\kappa^2$ is dimensionful in gravity theories, the rational functions are different at each loop order. At one loop, they form a 16-dimensional linear space spanned by 15 $\mathbb{Q}$-linearly independent permutations of
\begin{align}
r^{(1)}_1 = s_{12} s_{23} s_{34}  {\rm PT}(34125){\rm PT}(43215) \, ,
\end{align}
and by
\begin{align}
r^{(1)}_{16} = \frac{1}{\epsilon_5}\frac{[12][23][34][45][51]}{\vev{12}\vev{23}\vev{34}\vev{45}\vev{51}}\,.
\end{align} 
Substituting the one and two-loop amplitudes as given by Eqs.~\eqref{eq:sugraM1} and~\eqref{eq:sugra_expected} into Eq.~\eqref{eq:F2} gives an expression for the two-loop hard function of the form
\begin{align}
\label{eq:F1ut}
\mathcal{F}^{(1)}_5 = \sum_{i=1}^{15} \sum_j c^{(1)}_{ij} r^{(1)}_i P_j^{(2)}\,, \\ 
\label{eq:F2ut}
\mathcal{F}^{(2)}_5 = \sum_{i=1}^{40} \sum_j c^{(2)}_{ij} r^{(2)}_i P_j^{(4)}\,,
\end{align}
where $c^{(\ell)}_{ij}\in \mathbb{Q}$ and $P_j^{(w)}$ are weight-$w$ pentagon functions. As was already noted in Refs.~\cite{Chicherin:2019xeg,Abreu:2019rpt}, the rational functions $r^{(1)}_{16}$ at one loop and $r^{(2)}_{40+k}$ with $k=1,...,5$ at two loops drop out of the hard function. Moreover, the letter $W_{31}$ does not appear neither in the amplitudes nor in the hard functions, as observed at symbol level~\cite{Chicherin:2019xeg,Abreu:2019rpt}.
We complete this section by providing the numerical value of the two-loop hard function ${\cal F}^{(2)}_5$ at the kinematic point $X_R$~\p{eq:kinpoint_s12}, 
\begin{align}
\frac{{\cal F}_5^{(2)}}{\PT_1^2} \Biggl|_{X_R} =  -1211.9365 -215.6087 i \,,
\end{align}
where the normalisation was chosen to cancel the helicity weight. The explicit expression of the one and two-loop hard function in terms of the pentagon functions introduced in Ref.~\cite{Chicherin:2020oor} can be obtained at
\begin{center}
\href{https://pentagonfunctions.hepforge.org/downloads/2l_5pt_hardfunctions_N=4_N=8.tar.gz}{https://pentagonfunctions.hepforge.org/downloads/2l\_5pt\_hardfunctions\_N=4\_N=8.tar.gz}.
\end{center}

\section{Multi-Regge limit and pentagon functions}
\label{sec:MultiReggeLimit}

\subsection{Multi-Regge kinematics}

The multi-Regge kinematics~\cite{Kuraev:1976ge,DelDuca:1995hf} is defined as a scattering process where the final-state particles are strongly ordered in rapidity and have comparable transverse momenta. We work in the $s_{12}$ channel, and assume that the two incoming particles travel along the $z$-axis. We introduce the light-cone coordinates $p_j = (p_j^+, p_j^-, {\bf p}_j )$, with $p_j^{\pm} = p_j^0 \pm p_j^3$ and the complexified transverse momenta ${\bf p}_j = p_j^1 + i p_j^2$.
Then, the strong ordering in rapidities with comparable transverse momenta,
\begin{align}
|{\bf p}_3| \simeq |{\bf p}_4| \simeq |{\bf p}_5|\,, \label{eq:reggekin1}
\end{align}
translates into strong orderings in light-cone components, 
\begin{align}
\label{eq:reggekin2}
&|p_3^+| \gg |p_4^+| \gg |p_5^+| \,, \qquad
|p_3^-| \ll |p_4^-| \ll |p_5^-| \,.
\end{align}

\begin{figure}
\begin{center}
\includegraphics[width=0.5\textwidth]{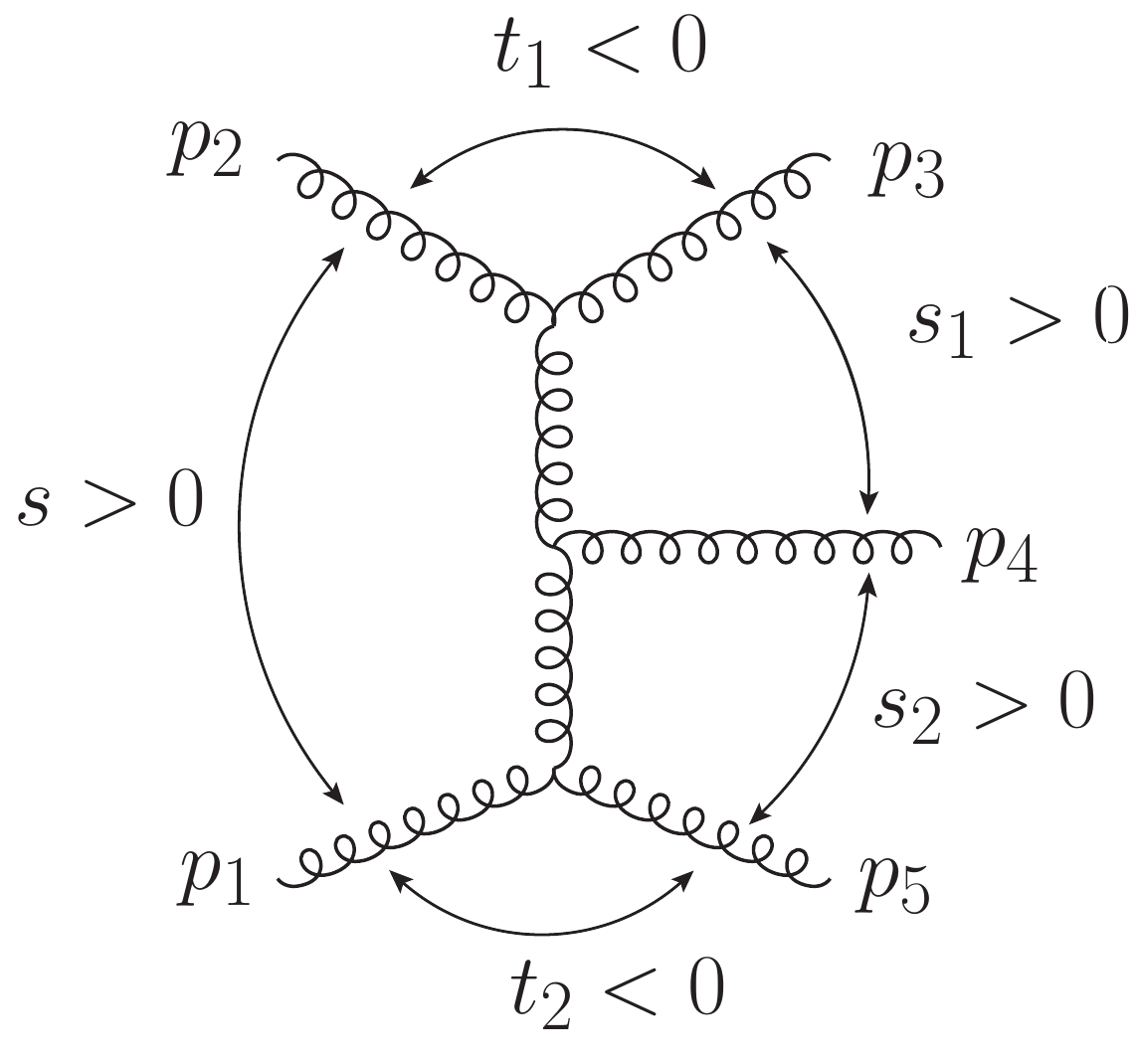}
\end{center}
\caption{Pictorial representation of the multi-Regge kinematics in the $s_{12}$ channel.} 
\label{fig:MRK}
\end{figure}

We implement the constraints~\eqref{eq:reggekin1} and~\eqref{eq:reggekin2} by introducing a parameter $x$, which regulates the size of the light-cone components as 
\begin{equation}
\label{eq:scaling}
\begin{aligned}
& |p_1^-| \sim |p_2^+| \sim |p_3^+| \sim |p_5^-| \sim \mathcal{O}\left(\frac{1}{x}\right)\, , \\
& |p_4^+| \sim |p_4^-| \sim |\textbf{p}_3| \sim |\textbf{p}_4| \sim |\textbf{p}_5| \sim \mathcal{O}\left(1\right)\, ,  \\
& |p_5^+| \sim |p_3^-| \sim \mathcal{O}\left(x\right)\,. 
\end{aligned}
\end{equation}
Multi-Regge kinematics is achieved in the limit $x\to 0$. 

The scalings given by Eq.~\eqref{eq:scaling} are equivalent to the following parametrisation of the Mandelstam invariants,
\begin{align}
s_{12} = \frac{s}{x^2}\,,\quad \
s_{23} = t_1 \,,\quad \
s_{34} = \frac{s_1}{x} \,,\quad \
s_{45} = \frac{s_2}{x}\,,\quad \
s_{15} = t_2\,,  \label{regge}
\end{align}
where $t_1,t_2 <0$ and $s,s_1,s_2 >0$ are fixed in the limit. This can be seen by rewriting them in terms of the light-cone components of the momenta in the multi-Regge kinematics,
\begin{equation}
\begin{aligned}
& s_{12} = p_1^- p_2^+= p_3^+ p_5^- \equiv \frac{s}{x^2} \, , \\ 
& s_{23} = - |{\bf p}_3|^2 \equiv t_1 \, , \\
& s_{15} = - |{\bf p}_5|^2 \equiv t_2 \, , \\
& s_{34} = p_3^+ p_4^- \equiv \frac{s_1}{x}\, , \\
& s_{45} = p_4^+ p_5^- \equiv \frac{s_2}{x} \, ,
\end{aligned}
\end{equation}
The transverse momenta can be chosen to be
\begin{align}
{\bf p}_3 =  z \sqrt{\frac{s_1 s_2}{s}} \;,\ \
\bar{\bf p}_3 =  \zbar \sqrt{\frac{s_1 s_2}{s}} \;,\ \
{\bf p}_5 =  (1-z) \sqrt{\frac{s_1 s_2}{s}} \;,\ \
\bar{\bf p}_5 =  (1-\zbar) \sqrt{\frac{s_1 s_2}{s}}\,,
\end{align}
where ${\bf p}_1={\bf p}_2=0$ and we have introduced the complex variables $z$ and $\zbar$, defined by 
\begin{align}
z \bar{z} = - \frac{t_1 s}{s_1 s_2}\;,\qquad 
(1-z)(1-\bar{z}) = - \frac{t_2 s}{s_1 s_2} \,.
\end{align}
In the physical scattering region $\bar{z}$ is the complex conjugate of $z$. A pictorial representation of the multi-Regge kinematics is shown in Fig.~\ref{fig:MRK}.

\subsection{Multi-Regge limit of the pure integrals}
\label{sec:PentagonMR}

As already observed in Refs.~\cite{Chicherin:2018yne,Chicherin:2019xeg}, the pentagon alphabet becomes very simple at the leading order in the multi-Regge limit.
First of all, it becomes rational. The Gram determinant, in fact, becomes a perfect square
\begin{align}
\label{eq:DeltaMR}
\Delta = \eps_5^2 \underset{x\to 0}{\sim} \frac{s^2_1 s^2_2 (z-\zbar)^2}{x^4} + \mathcal{O}\left(\frac{1}{x^3}\right) \, ,
\end{align}
which allows to write the pseudo-scalar invariant $\eps_5$ as a rational function. We choose the branch of the square root as
\begin{align}
\eps_5  \underset{x\to 0}{\sim}  \frac{s_1 s_2 (z - \zbar)}{x^2} + \mathcal{O}\left(\frac{1}{x}\right)  \,. \label{eq:e5param}
\end{align}
Note that Eq.~\eqref{eq:DeltaMR} explains why $z$ and $\zbar$ are complex conjugate, because
reality of the momenta in fact requires that $\Delta<0$. This implies that $z-\zbar$ is purely imaginary, and therefore that $z$ and $\zbar$ are complex conjugate of each other.

Moreover, the pentagon alphabet reduces to 12 letters only, factorised into four independent sub-alphabets:
\begin{align}
 & \{ x \}\,, \label{eq:alph1} \\
 & \left\{ \frac{s_1 s_2 }{s} \right\}\,,\label{eq:alph2} \\
 & \{ s_1 , s_2, s_1 - s_2 , s_1 + s_2 \}\,, \label{eq:alph3} \\
 & \{ z, \zbar , 1 - z , 1- \zbar , z - \zbar, 1-z - \zbar \}\,. \label{eq:alph4}
\end{align}
The functional structure of massless two-loop five-particle amplitudes is therefore extremely simple in the multi-Regge kinematics. The alphabets~\eqref{eq:alph1} and~\eqref{eq:alph2} simply correspond to logarithms, whereas the alphabets~\eqref{eq:alph3} and~\eqref{eq:alph4} encode the harmonic polylogarithms~\cite{Remiddi:1999ew} and the two-dimensional harmonic polylogarithms~\cite{Gehrmann:2001jv}, respectively.

We now discuss how to compute the asymptotic expansion of the pure master integrals in the multi-Regge limit $x\to 0$. A thorough discussion of this topic can be found e.g. in Ref.~\cite{wasow1965asymptotic}.

Let $\vec{f}$ be a basis of pure master integrals. In order to simplify the notation, let us denote the set of variables $s,s_1,s_2,z,\zbar$ cumulatively by $y$. Then, $\vec{f}$ satisfies the system of differential equations
\begin{align}
\label{eq:DEinitial}
\begin{cases}
\vspace{0.1cm}
\dfrac{\pa}{\pa x} \vec{f}(x,y,\epsilon) = \epsilon A_x(x,y) \vec{f}(x,y,\epsilon)\, , \\
\dfrac{\pa}{\pa y} \vec{f}(x,y,\epsilon) = \epsilon A_y(x,y)\vec{f}(x,y,\epsilon) \, .
\end{cases}
\end{align}  
The matrix $A_x$ has a regular singular point at $x=0$, 
\begin{align}
A_x(x,y) = \frac{A_0}{x}+\sum_{k \geq 0} x^{k} A_{k+1}(y)\,,
\end{align}
and the residue at $x = 0$ is a matrix of rational numbers $A_0$. The matrix $A_y$ is regular at $x = 0$.

We perform a gauge transformation with a holomorphic matrix $T(x,y,\epsilon)$,
\begin{align}
\vec f(x,y,\epsilon) = T(x,y,\epsilon) \vec g(x,y,\epsilon)\, .
\end{align} 
We construct the latter so that the new basis $\vec{g}$ satisfies a simplified differential equation with respect to $x$,
\begin{align}
\dfrac{\pa}{\pa x} \vec{g}(x,y,\epsilon) = \epsilon \frac{A_0}{x} \vec{g}(x,y,\epsilon)\, .
\end{align}
This requires that the transformation matrix obeys the differential equation
\begin{align}
\label{eq:DEforT}
T^{-1}(\epsilon A_x T - \pa_x T) = \epsilon \frac{A_0}{x} \, .
\end{align}
In order to solve the latter, we series expand $T(x,y,\epsilon)$ around $x=0$,
\begin{align}
\label{eq:TxExp}
T(x,y,\epsilon) = \mathds{1}+\sum_{k \geq 1}x^k T_k(y,\epsilon)\,.
\end{align} 
We are free to choose the transformation matrix such that it becomes the identity at $x=0$.
Substituting Eq.~\eqref{eq:TxExp} into Eq.~\eqref{eq:DEforT} gives a system of contiguous relations,
\begin{equation}
\label{eq:ContiguousRelations}
\epsilon A_k(y) + \epsilon A_0 T_k(y,\epsilon) - \epsilon T_k(y,\epsilon) A_0 - k T_k(y,\epsilon) + \epsilon 
\sum_{j=1}^{k-1} A_{k-j}(y) T_j(y,\epsilon) = 0 \;,\;\; \forall k \geq 1\, .
\end{equation}
Let us note that these equations imply that $T_k(y,\epsilon) = {\cal O}(\epsilon)$. We find it convenient to further series expand $T_k(y,\epsilon)$ in $\epsilon$,
\begin{align}
T_k(y,\epsilon) =\sum_{m\geq 1} \epsilon^k \, T_{k,m}(y) \, .
\end{align} 
Then, the contiguous relations~\eqref{eq:ContiguousRelations} take the form
\begin{equation}
\begin{aligned}
& T_{k,1}(y) = \frac{1}{k} A_k(y) \, \\
& T_{k,m}(y) = \frac{1}{k} \left[ A_0 T_{k,m-1}(y) - T_{k,m-1}(y) A_0 +\sum_{j=1}^{k-1} A_{k-j}(y) T_{j,m-1}(y) \right]  \,,\;\;\; \forall m > 1\, ,
\end{aligned}
\end{equation}
which can be solved order by order in $x$ and $\epsilon$, giving an explicit expression for $T$ as a double series,
\begin{align}
T(x,y,\epsilon) = 1+ \sum_{k \geq 1}\sum_{m \geq 1} x^k \epsilon^m T_{k,m}(y)\,.
\end{align} 

After the gauge transformation the system of differential equations for $\vec{g}$ takes the form
\begin{align}
\begin{cases}
\vspace{0.1cm}
\dfrac{\pa}{\pa x} \vec{g}(x,y,\epsilon)  = \epsilon \dfrac{A_0}{x} \vec{g}(x,y,\epsilon) \\
\dfrac{\pa}{\pa y} \vec{g}(x,y,\epsilon)  = B(x,y,\epsilon)\vec{g}(x,y,\epsilon)
\end{cases}
\end{align} 
where $B(x,y,\epsilon) = T^{-1}(\epsilon A_y T-\pa_y T)$. We want to solve this system of differential equations using a boundary point in the multi-Regge kinematics, namely $x=0$. In order to do so, we integrate the system along the piecewise path
\begin{align}
(x=0,y=y_0) \longrightarrow (x=0,y) \longrightarrow (x,y) \, ,
\end{align}  
for some $y_0$ in the $s_{12}$ channel.
In other words, we restore the dependence first on $y$ and then on $x$. Since $B(x=0,y,\epsilon) = \epsilon A_y(0,y)$, we find
\begin{align}
\vec g(x,y,\epsilon) = x^{\epsilon A_0}\, \mathbb{P}\exp\left[ \epsilon \int_{y_0}^y A_y(0,y') dy' \right]\vec g_0 (\epsilon) \, .
\end{align}
Finally, after the gauge transformation, we obtain the solution of the initial system of differential equations~\eqref{eq:DEinitial},
\begin{align}
\label{eq:AsymptoticSolution}
\vec f(x,y,\epsilon) = T(x,y,\epsilon)\, x^{\epsilon A_0}\, \mathbb{P}\exp\left[ \epsilon \int_{y_0}^y A_y(0,y') dy' \right]\vec g_0(\epsilon) \, ,
\end{align}
where $\vec g_0(\epsilon)$ are the boundary constants $\vec g_0(\epsilon) = \vec g(y=y_0,\epsilon)$ for the equation 
\begin{align}
\frac{\pa}{\pa y} \vec g(y,\epsilon) = \epsilon A_y(0,y)\vec g(y,\epsilon)\,.
\end{align}

Our choice for the boundary point in the multi-Regge kinematics is
\begin{align}
\label{eq:BasePointMR}
y_0 = \biggl( s = 1\,,  s_1 = 1 \,,  s_2 = 1\,, z = e^{\frac{i \pi}{3}}\, , \zbar = e^{-\frac{i \pi}{3}} \biggr) \, ,
\end{align}
which corresponds to $t_1 = t_2 = -1$. We computed the boundary constants $\vec{g}_0(\epsilon)$ by solving the canonical differential equation~\eqref{eq:CanonicalDE} asymptotically in the limit $x\to 0$ starting from $X_0$~\eqref{eq:SymmetryBasePoint}, where the values of the integrals are known from Ref.~\cite{Badger:2019djh}. Note that the power corrections in $x$ are not needed for this purpose. The integrals develop logarithmic singularities, which match the matrix exponential $x^{\eps A_0}$ in Eq.~\eqref{eq:AsymptoticSolution}. What remains after the latter are removed are the boundary constants $\vec{g}_0(\eps)$ at $y_0$.
For the numerical evaluation of the Goncharov polylogarithms we used the algorithms of Ref.~\cite{Vollinga:2004sn}, implemented within the {\sc GiNaC} framework~\cite{Bauer:2000cp}.

In this way we obtain explicit analytic expressions for the boundary constants $\vec g_0(\epsilon)$ at $y_0$, for all the relevant integral families.
Thanks to the simple functional dependence implied by the alphabets~\eqref{eq:alph1}~-~\eqref{eq:alph4}, in fact, the values of the integrals at the point $y_0$~\eqref{eq:BasePointMR} can be anticipated to be harmonic polylogarithms of argument 1, and two-dimensional harmonic polylogarithms of arguments $z$ and $\zbar$ given by Eq.~\eqref{eq:BasePointMR}. This makes it rather easy to fit the numerical values to analytic transcendental numbers. We used {\sc Mathematica}'s built-in function {\tt FindIntegerNullVector}. A basis of $\mathbb{Q}$-linearly independent constants which spans the values of all integrals at $y_0$ up to weight 4 is shown in Table~\ref{tab:Constants}. Note that the constants could be simplified even further by going to $z=\zbar=0$ (or equivalently $t_1 = t_2 = 0$), but we prefer $y_0$~\eqref{eq:BasePointMR} because it is less singular.
\begin{table}
\begin{center}
\begin{tabular}{|c|c|c|}
\hline
Weight & Real & Imaginary \\ \hline
1 & 0 & $i \pi$ \\ \hline
2 & $\pi^2$ & $i \,{\rm Im \, Li}_2(e^{\frac{i \pi}{3}}) $ \\ \hline
3 & $\zeta_3,\, \pi \,{\rm Im \, Li}_2(e^{\frac{i \pi}{3}}),\, \pi^2 \log(3) $ & $i \pi^3,\,  i \pi \log^2(3) - 48 i\, {\rm Im\, Li}_{3}\left(\frac{i}{\sqrt{3}}\right)$ \\ \hline
\multirow{2}{*}{4} & $\pi^4,\, \left({\rm Im \, Li}_2(e^{\frac{i \pi}{3}}) \right)^2,$ & $i \pi \zeta_3, \, i \pi^2 \,{\rm Im \, Li}_2(e^{\frac{i \pi}{3}}) , \, i \,{\rm Im \, Li}_4(e^{\frac{i \pi}{3}}),\, i \pi^3 \log(3),$ \\
& $\, \pi^2 \log^2(3) ,\,  \pi \, {\rm Im\, Li}_{3}\left(\frac{i}{\sqrt{3}}\right) $ & $i\pi \log^3(3) + 288 i\, {\rm Im\, Li}_4\left(\frac{i}{\sqrt{3}}\right) $
\\ \hline
\end{tabular}
\end{center}
\caption{Basis of $\mathbb{Q}$-linearly independent constants appearing in the values of the pentagon integrals at the base point in the multi-Regge kinematics $y_0$~\eqref{eq:BasePointMR}.}
\label{tab:Constants}
\end{table}

This procedure allows to solve the differential equations in canonical form~\eqref{eq:CanonicalDE} asymptotically starting from a boundary point in the multi-Regge limit, or in general from any regular singular point. The result contains divergent logarithms of $x$, generated in Eq.~\eqref{eq:AsymptoticSolution} by the matrix exponential $x^{\eps A_0}$. The path-ordered exponential, on the other hand, produces iterated integrals which depend on the kinematic variables $s_1,s_2,s,z$ and $\zbar$. Finally, the transformation matrix $T(x,y,\epsilon)$ is responsible for power corrections in $x$.

Before moving on to the actual calculation of the multi-Regge limit of the amplitudes, it is important to mention a treacherous subtlety.
The base point $y_0$~\eqref{eq:BasePointMR} is in the upper half of the complex plane, namely $\text{Im}[z]>0$. The $s_{12}$ physical scattering region in the multi-Regge kinematics is defined by
\begin{align}
s > 0 \,, \qquad s_1> 0 \, , \qquad s_2 >0 \, , 
\end{align}
and by $\zbar$ being the complex conjugate of $z\in \mathbb{C}$. Although $z$ can span the whole complex plane, it is non-trivial to analytically continue from $\text{Im}[z]>0$, where the base point $y_0$ lies, to $\text{Im}[z]<0$. As we discuss in Section~\ref{sec:eps5surface}, certain non-planar Feynman integrals contributing to the amplitudes have discontinuities (and some even diverge) at $\eps_5 = 0$. This hypersurface in the multi-Regge kinematics corresponds to $z = \zbar$, namely $\text{Im}[z]=0$. Indeed, as we will describe in the next section, we find that the two-loop five-particle amplitudes in $\mathcal{N}=4$ super Yang-Mills and $\mathcal{N}=8$ supergravity are continuous but not real analytic across the real axis $\text{Im}[z]=0$. A real-analytic function is infinitely smooth and its Taylor series around any point has a finite radius of convergence.  The second derivatives of the two amplitudes in the multi-Regge limit are discontinuous. It is worth stressing that this feature can be observed only in $n$-particle scattering with $n\ge 5$, as no pseudo-scalar invariant like $\epsilon_5$ exists for $n <5$. 

Rather than attempting this perilous analytic continuation, we prefer to follow a different approach, which we find less error prone. We work in the two regions separately. We integrate the canonical differential equations~\eqref{eq:CanonicalDE} for the master integrals starting from $y_0$, and obtain an expression for the amplitudes valid in the upper half of the complex plane. Then we parity-conjugate the base point $y_0$ to obtain a base point in the lower half of the complex plane,
\begin{align}
\label{eq:BasePointMRlower}
\tilde{y}_0 = \biggl( s = 1\,,  s_1 = 1 \,,  s_2 = 1\,, z = e^{-\frac{i \pi}{3}}\, , \zbar = e^{\frac{i \pi}{3}} \biggr) \, .
\end{align}
Note that in both halves of the complex plane we choose the branch of the square root for $\epsilon_5$ as in Eq.~\eqref{eq:e5param}.
Since all the pure master integrals of the bases we use have definite parity, their values at $\tilde{y}_0$ can be obtained from those at $y_0$ simply by flipping the overall sign of the parity-odd integrals. Then, we integrate the differential equations starting from $\tilde{y}_0$, and obtain a representation of the amplitudes valid for $\text{Im}[z]<0$.

Another similar approach to construct a representation in the lower half of the complex plane consists in actually keeping $\text{Im}[z]>0$, but changing the branch of $\eps_5$ as
\begin{align}
\eps_5  \underset{x\to 0}{\sim} - \frac{s_1 s_2 (z - \zbar)}{x^2} + \mathcal{O}\left(\frac{1}{x}\right)  \,,
\end{align}
rather than Eq.~\eqref{eq:e5param}. This operation affects not only the values of the integrals at the base point --~by flipping the sign of the parity-odd integrals~-- but also the letters in the differential equation~\eqref{eq:CanonicalDE}, according to Eq.~\eqref{eq:OddLetters}. We find agreement between the two approaches.

\subsection{Feynman integrals with non-trivial analytic properties}
\label{sec:eps5surface}
\begin{figure}[t!]
    \centering
\begin{subfigure}[b]{0.4\textwidth}
    \includegraphics[width=\textwidth]{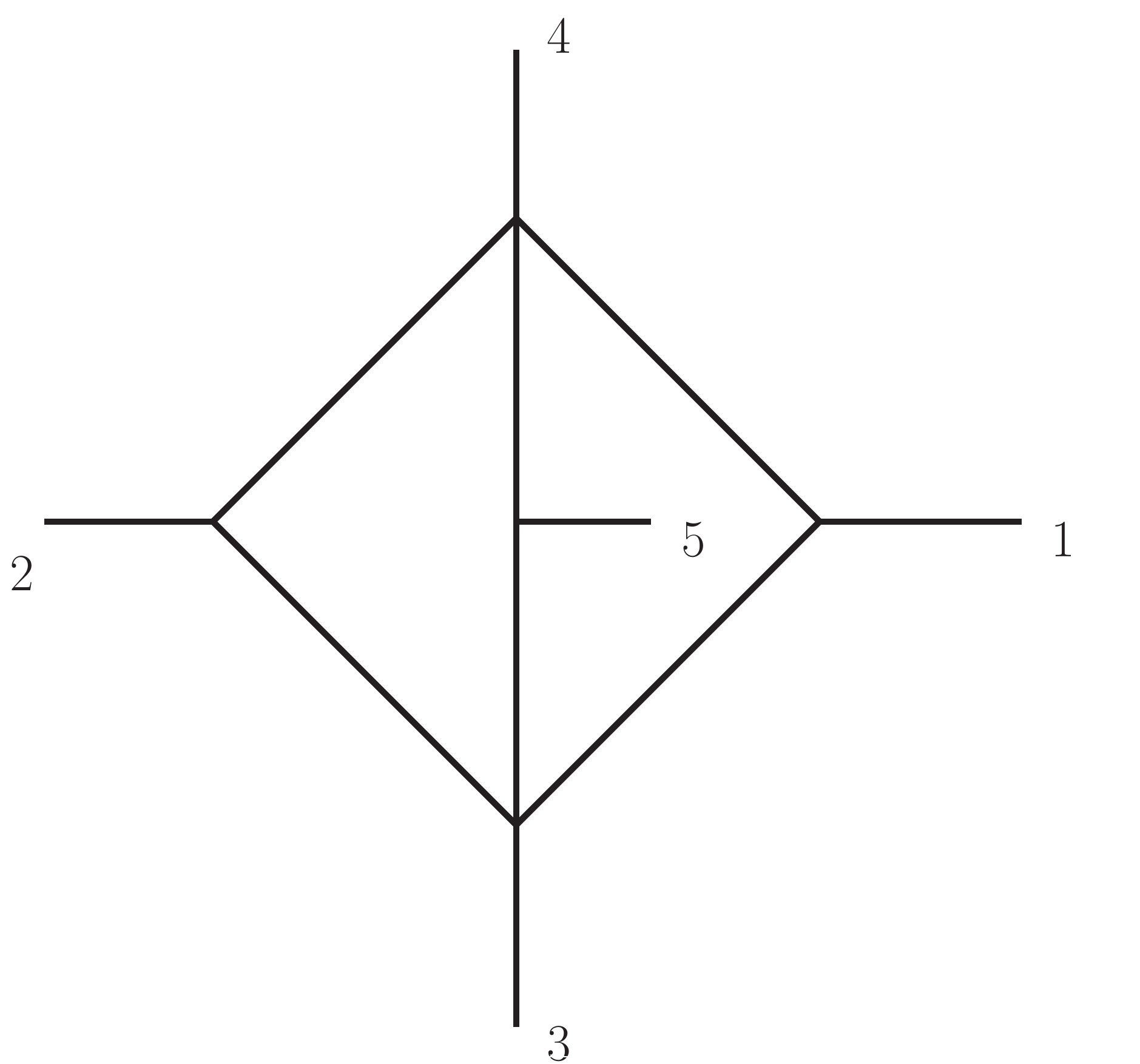}
    \caption{$\quad$}
	\label{I_odd_1}
\end{subfigure}
\qquad \qquad
\begin{subfigure}[b]{0.4\textwidth}
    \includegraphics[width=\textwidth]{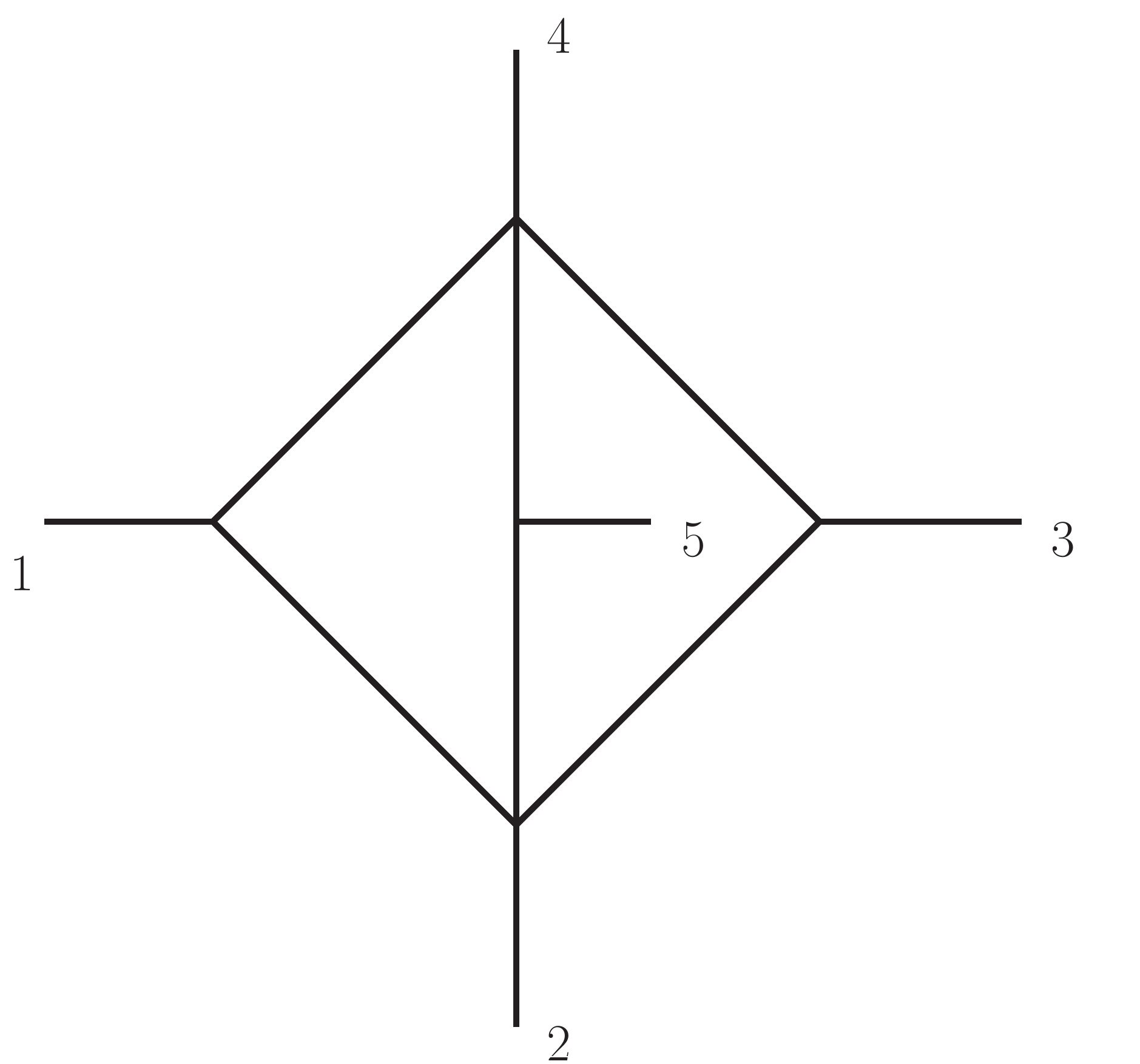}
    \caption{$\quad$}
	\label{I_odd_2}
\end{subfigure}
\caption{Examples of Feynman integrals to illustrate the analytic properties near the hypersurface $\eps_5=0$. The scalar integrals depicted in this figure are multiplied by a factor of $\eps_5$. As a consequence, they have uniform transcendental weight and odd parity. While the integral (a) does {\it not} vanish on the hypersurface $\epsilon_5=0$ approached from within the physical $s_{12}$ scattering region, the integral (b) does.}
\label{fig:XT246}
\end{figure}

In this subsection, we present examples of Feynman integrals to illustrate the analytic properties near the hypersurface $\eps_5 = 0$. We consider the non-planar two-loop integrals shown in Fig.~\ref{fig:XT246}. The scalar integrals depicted there are multiplied by a factor of $\eps_5$,
\begin{equation}
  \label{eq:IaDef}
  I_a=e^{2\gamma_E \epsilon}\int \frac{d^D l_1}{i \pi^{D/2}} \frac{d^D l_2}{i \pi^{D/2}}  \frac{\epsilon_5}{l_1^2 (l_1-p_2)^2 l_2^2 (l_2-p_1)^2 (l_1+l_2+p_3)^2 (l_1+l_2+p_3+p_5)^2}\,,
\end{equation}
and similarly for $I_b$.
Thanks to the factor of $\eps_5$ in the numerator, these integrals have odd parity and uniform transcendental weight. We consider them in the $s_{12}$ channel, and assume that $\Im[\eps_5] > 0$.

One might na\"ively expect that, given the factor of $\eps_5$ in the numerator, the integrals $I_a$ and $I_b$ vanish on the hypersurface $\eps_5=0$ which bounds the $s_{12}$ scattering region. $I_b$ does indeed vanish, but $I_a$ does not. Although they can be obtained one from another by permuting the external legs, the two integrals exhibit a completely different behaviour at $\eps_5=0$.
 
Let us start by considering the integral $I_b$ shown in Fig.~\ref{I_odd_2}. By solving the differential equations in the physical scattering region, we find the Laurent expansion
\begin{align}
  \label{eq:11}
  I_b=\frac{1}{\epsilon^2} f^{(2)}_b +\frac{1}{\epsilon} f^{(3)}_b + f^{(4)}_b +\mathcal O(\epsilon)\,.
\end{align}
For our purposes, it suffices to look at the weight-two part, $f^{(2)}_b$, which contains only dilogarithms,
\begin{equation}
\begin{aligned}
  \label{eq:12}
   f^{(2)}_b= \; & -3 \bigg[\Li_2\bigg(\frac{1}{W_{27} } \bigg)-\Li_2\bigg(W_{27} \bigg)+\Li_2\bigg(\frac{1}{W_{28}}\bigg) -\Li_2\bigg(\frac{1}{W_{27}W_{28} }\bigg)  \\
& -\Li_2\bigg(W_{28} \bigg) +\Li_2 \bigg (W_{27} W_{28}\bigg) \bigg]\,.
\end{aligned}
\end{equation}
Since all the odd letters become 1 if $\eps_5=0$, $f^{(2)}_b$ does indeed satisfy the na\"ive expectation of vanishing on the whole hypersurface $\epsilon_5=0$. Note that $f^{(2)}_b$ does not contribute in the multi-Regge limit,
\begin{equation}
  \lim_{x\to 0}f_b^{(2)} = 0\,.
\end{equation}

The integral $I_a$, shown in Fig.~\ref{I_odd_1}, exhibits a more interesting behaviour. It has the Laurent expansion
\begin{equation}
  \label{eq:3}
  I_a=\frac{1}{\epsilon^2} f^{(2)}_a +\frac{1}{\epsilon} f^{(3)}_a + f^{(4)}_a +\mathcal O(\epsilon)\,.
\end{equation}
In particular, the weight-two part is given by
\begin{equation}
  \label{eq:4}
  f^{(2)}_a= 3 \mathcal P_a +6 i\pi h_a\,, 
\end{equation}
where $\mathcal{P}_a$ is a combination of dilogarithms,
\begin{equation}
\begin{aligned}
  \label{fun_PA}
  \mathcal{P}_a= \; & \Li_2\bigg(\frac{W_{30}}{W_{27} W_{28}} \bigg)-\Li_2\bigg(\frac{W_{27}W_{28}}{W_{30} } \bigg)+\Li_2\bigg(\frac{W_{28}}{W_{26} W_{30}}\bigg) -\Li_2\bigg(\frac{W_{26}W_{30}}{W_{28} }\bigg)  \\
& +\Li_2\bigg(W_{26} W_{27}\bigg) -\Li_2 \bigg (\frac{1}{W_{26} W_{27}}\bigg) \, ,
\end{aligned}
\end{equation}
and $h_a$ has the expression
\begin{equation}
\begin{aligned}
  h_a  = \; & \log(W_{28})\Theta(a_{28}) +\big(\log(W_{28})-2i\pi \big) \Theta(-a_{28}) -i \pi \delta_{a_{28}}  \\
& -\log(W_{26})\Theta(a_{26}) -\big(\log(W_{26})-2i\pi\big) \Theta(-a_{26}) +i \pi \delta_{a_{26}}  \\
& -\log(W_{30})\Theta(-a_{30}) -\big(\log(W_{30})+2i\pi\big) \Theta(a_{30}) -i \pi \delta_{a_{30}} \,.
\label{fun_hA}
\end{aligned}
\end{equation}
In Eq.~\eqref{fun_hA}, the polynomial $a_{25+i}$, $i=1,\ldots 5$, is defined by rewriting the odd letters~\eqref{eq:W26} as
\begin{gather}
  \label{eq:odd_letters}
  W_{25+i}=\frac{a_{25+i}-\epsilon_5}{a_{25+i}+\epsilon_5},\quad i=1,\ldots, 5 \, ,
\end{gather}
$\Theta$ is the Heaviside theta function (with $\Theta(0)\equiv 0$), and $\delta_x$ is defined to be $1$ if $x=0$, and $0$ otherwise. 
Both $\mathcal P_a$ and $h_a$ are single-valued functions in the physical scattering region. It is easy to see from the explicit expressions~\eqref{fun_PA} and~\eqref{fun_hA} that, in a generic point on the hypersurface $\epsilon_5=0$, 
\begin{gather}
  \label{eq:8}
  \mathcal P_a\Bigl|_{\epsilon_5=0}=0\, , \qquad  h_a\Bigl|_{\epsilon_5=0}\not=0 \, .
\end{gather}
The integral $I_a$ therefore does not vanish if $\epsilon_5=0$ in general. 

We checked the analytic expressions \eqref{fun_PA} and \eqref{fun_hA} against numerical evaluations. The latter were carried out by using {\sc pySecDec}~\cite{Borowka:2017idc} to integrate numerically the concise integral representation given in Ref.~\cite{Chicherin:2017dob}, 
\begin{equation}
  \label{IA_rep}
  I_a=-\epsilon_5 e^{2 \epsilon \gamma_E} \bigg(-\frac{\Gamma^3(-\epsilon) \Gamma(2+2\epsilon)}{\Gamma(-3\epsilon)}\bigg) \int_0^1 d\alpha_1 \int_0^1 d\alpha_2 \int_0^1 d\alpha_3 \, F^{-2-2\epsilon}\,,
\end{equation}
where
\begin{equation}
  \label{eq:14}
  F = (-s_{23})\alpha_2 + (-s_{13})\alpha_3 + (-s_{35} )\alpha _1 + (-s_{25}) \alpha_1 \alpha _2 + (-s_{15})\alpha_1 \alpha_3 + (-s_{12}) \alpha_2 \alpha_3 \,.
\end{equation}
We found this representation to be more convenient than the usual Feynman parameters. We found agreement between the numerical evaluation and our analytic expression within the error estimates. 

The fact that $I_a$ does not vanish on the hypersurface $\eps_5 = 0$ has important consequences for its analytic structure. Since $I_a$ is an odd integral, it changes sign upon parity conjugation, which acts by flipping the sign of the pseudo-invariant $\eps_5$. As a result, approaching a point on the surface $\eps_5 = 0$ from within the scattering region but with a different sign of $\Im[\eps_5]$ may yield values with opposite signs,
\begin{align}
f^{(2)}_a\biggl|_{\Im[\eps_5] = 0^{\pm}} = \pm 12 \pi^2 \biggl(\Theta(-a_{28})-\Theta(-a_{26})+ \Theta(a_{30})\biggr)\,,
\end{align}
where the superscript $\pm$ denotes whether the hypersurface $\eps_5=0$ was reached along a path with $\Im[\eps_5]>0$ or $\Im[\eps_5]<0$, respectively.
We can thus conclude that the integral $I_a$ has a discontinuity across the hypersurface $\eps_5 = 0$, although we never leave the scattering region. The latter is made of two copies, one with $\Im[\eps_5]>0$ and one with $\Im[\eps_5]<0$, and the analytic continuation between the two appears to be non-trivial.

It is interesting to note that the integral $I_a$ in the multi-Regge limit gives rise to one of the functions which are needed to describe the asymptotics of the $\mathcal{N}=4$ super Yang-Mills and $\mathcal{N}=8$ supergravity amplitudes,
\begin{gather}
  \label{eq:9}
  \lim_{x\to 0} f_a^{(2)} = 6 i\pi g^{(1)}_6\,,
\end{gather}
where
\begin{gather}
  \label{eq:10}
  g^{(1)}_6 = \log (z)-\log (\bar z)\,.
\end{gather}
This function is well defined in the whole complex $z$ plane minus the negative real axis, where it is discontinuous.
Physically, the negative real axis represents the situation where the transverse momenta ${\bf p}_3$ and ${\bf p}_4$
are parallel. 

In this section we have seen that certain two-loop five-particle Feynman integrals can be discontinuous when crossing the hypersurface $\eps_5 = 0$ without leaving the physical scattering region, which is the kinematic region accessible in a hadron collider experiment. We stress however that this is a property of the individual Feynman integrals. The scattering amplitudes are expected to have no discontinuity throughout the entire physical scattering region. Indeed, we checked explicitly that the two-loop five-particle amplitudes in $\mathcal{N}=4$ super Yang-Mills and $\mathcal{N}=8$ supergravity are continuous at the point 
\begin{align}
X = \left\{3,-1+ \frac{\sqrt{3}}{2},1,1,-1\right\}\, , \qquad \quad \eps_5 = 0  \, ,
\end{align}
where some of the contributing Feynman integrals are discontinuous and even divergent. The sum over all the Feynman integrals in the amplitudes smoothens this singularity.

\subsection{Transcendental functions for the $\mathcal{N}=4$ super Yang-Mills amplitude in the multi-Regge limit}
\label{sec:sYMfunctions}

In this subsection we present a basis of the functions appearing in the one and two-loop $\mathcal{N}=4$ super Yang-Mills hard functions up to power corrections in the multi-Regge limit. They belong to the alphabet given by Eqs.~\eqref{eq:alph1}~-~\eqref{eq:alph4}. We expect physical singularities at $z =0$, $z=1$, $s_1=0$, $s_2=0$ and $s=0$, but some letters of the alphabet vanish at $z+\zbar = 1$, $z =\zbar$, and $s_1=s_2$. These spurious singularities manifest themselves in the functional representation, but drop out in the hard function. Moreover, some of the functions involved exhibit discontinuities across the real axis, although the hard function is continuous throughout the whole complex plane. In order to discuss the analytic structure of the functions, it is convenient to recall that we are working in the $s_{12}$-channel Regge kinematics, defined by 
\begin{align}
\label{eq:s12Regge}
s \ge 0\,, \qquad s_1 \ge 0\,, \qquad s_2 \ge 0\,, \qquad z = \zbar^* \in \mathbb{C}\,,
\end{align}
where the asterisk denotes complex conjugation.
We organise the discussion by transcendental weight.

\textbf{Weight 1.} At weight 1 we need five manifestly single-valued logarithms,
\begin{equation}
\label{eq:funW1sv}
\begin{aligned}
& g^{(1)}_1 =  \log(s_1)\,, \\
& g^{(1)}_2 =  \log(s_2)\,, \\
& g^{(1)}_3 = \log\left(\frac{s}{s_1 s_2}\right)\,, \\
& g^{(1)}_4 = \log(z \zbar)\,, \\ 
& g^{(1)}_5 =  \log((1-z)(1-\zbar)) \, , 
\end{aligned}
\end{equation} 
and two logarithms which require more attention,
\begin{equation}
\label{eq:funW1nsv}
\begin{aligned}
& g^{(1)}_6 = \log(z)-\log(\zbar)\,, \\
& g^{(1)}_7 =\log(1-z)- \log(1-\zbar) \,. 
\end{aligned}
\end{equation}
Let us show that $g^{(1)}_6$ and $g^{(1)}_7$ are well defined in both the upper and the lower half of the complex plane. 
For $g^{(1)}_6$, we parametrise $z$ using polar coordinates, $z = r e^{i \varphi}$ with $\varphi \in [0,\pi]$ for $\text{Im}[z]>0$, and $z = r e^{-i \varphi}$ with $\varphi \in [0,\pi]$ for $\text{Im}[z]<0$. Then, 
\begin{align}
g^{(1)}_6 = \begin{cases}
2 i \varphi \, , & \text{for} \ \text{Im}[z]>0\,, \\
-2 i \varphi \, , & \text{for} \ \text{Im}[z]<0\,. \\
\end{cases}
\end{align}
As for $g^{(1)}_7$, we parametrise $z$ as $z=1+r e^{i \varphi}$ for $\text{Im}[z]>0$, and $z=1+r e^{-i \varphi}$ for $\text{Im}[z]<0$. In both cases, $\varphi \in [0, \pi]$. Then,
\begin{align}
g^{(1)}_7 = \begin{cases}
-2 i \pi + 2 i \varphi \, , & \text{for} \ \text{Im}[z]>0\,, \\
2 i \pi + 2 i \varphi \, , & \text{for} \ \text{Im}[z]<0\,. \\
\end{cases}
\end{align}
We thus observe that both $g^{(1)}_6$ and $g^{(1)}_7$ have a discontinuity along the real axis, for $\text{Re}[z]<0$ and for $\text{Re}[z]>1$, respectively.
We stress however that these discontinuities cancel out in the hard function.

\textbf{Weight 2.}  At weight 2 only two irreducible functions are required. The single-valued dilogarithm,
\begin{align}
g^{(2)}_1 = D_2(z,\zbar)\equiv{\rm Li}_2(z) - {\rm Li}_2(\zbar) + \frac{1}{2} \left( \log(1-z) - \log(1-\zbar)\right) \log(z \zbar)\,, \label{eq:funW2sv}
\end{align}
and
\begin{align}
g^{(2)}_2 = {\rm Li}_2 (z) + {\rm Li}_2(\zbar) \,. \label{eq:funW2nsv}
\end{align}
The latter is manifestly real in the whole complex plane. It is continuous, but it is not real analytic along the real axis for $\text{Re}[z]>1$.

\textbf{Weight 3.} The most complicated functions appearing in the multi-Regge limit of the hard function have genuine transcendental weight 3. Two of them are manifestly single valued, 
\begin{equation}
\label{eq:funW3sv}
\begin{aligned}
& g^{(3)}_1 = D_3(z,\zbar)\equiv {\rm Li}_3(z) + {\rm Li}_3(\zbar) - \frac{1}{2} \log(z \zbar) \left( {\rm Li}_2(z) + {\rm Li}_2(\zbar)\right) \\
& \qquad \qquad \qquad \qquad  - \frac{1}{4} \log^2(z \zbar) \log((1-z)(1-\zbar)) \,, \\
& g^{(3)}_2 = D_3(1-z,1-\zbar) = {\rm Li}_3(1-z) + {\rm Li}_3(1-\zbar)  - \frac{1}{4} \log^2((1-z)(1- \zbar)) \log(z \zbar) \\
& \qquad \qquad \qquad \qquad \qquad \quad - \frac{1}{2} \log((1-z)(1- \zbar)) \left( {\rm Li}_2(1-z) + {\rm Li}_2(1-\zbar)\right)  \,, 
\end{aligned}
\end{equation}
and four require a more thorough analysis,
\def\dg#1{g'^{#1}}
\begin{equation}
\label{eq:funW3nsv}
\begin{aligned}
& g^{(3)}_3 = {\rm Li}_3(z) - {\rm Li}_3(\zbar)\,, \\
& g^{(3)}_4 = {\rm Li}_3(1-z) - {\rm Li}_3(1-\zbar)\,, \\
& g^{(3)}_5 = {\rm Li}_3\left(\frac{z \zbar}{(1-z)(1-\zbar)}\right) + \frac{1}{2}\log(1-z-\zbar) \log^2 \left( \frac{z \zbar}{(1-z)(1-\zbar)} \right) \,, \\
& g^{(3)}_6 = 2 \, {\rm Li}_3\left(\frac{z}{1-\zbar}\right) - 2 \, {\rm Li}_3\left(\frac{\zbar}{1-z}\right) - \log\left(\frac{z \zbar}{(1-z)(1-\zbar)}\right) D_2\left( \frac{z}{1-\zbar}, \frac{\zbar}{1-z} \right) + \dg{(3)}_6\,.
\end{aligned}
\end{equation}

The analysis of $g^{(3)}_3$ and $g^{(3)}_4$ is similar to that of $g^{(1)}_6$ and $g^{(1)}_7$: they are single valued in the lower and in the upper half of the complex plane separately, but have a discontinuity across the real axis for $\text{Re}[z]>1$ and for $\text{Re}[z]<0$, respectively.

The function $g^{(3)}_5$ is real-valued in the whole complex plane. The polylogarithm has a branch cut at $\text{Re}[z]>1/2$. Its imaginary part on the branch cut is however compensated by that of $\log(1-z-\zbar)$, so that the whole expression is real-valued. It is also worth noting that $g^{(3)}_5$ is analytic across $\text{Re}[z]=1/2$.

The analysis of $g^{(3)}_6$ is more involved, and will motivate the correction term $\dg{(3)}_6$.
The logarithm and $D_2$ are manifestly single valued in the complex plane.
However, the argument of the  weight-3 polylogarithms in $g^{(3)}_6$ is equal to unity over the whole line $\text{Re}[z]=1/2$:
\begin{align} \label{eq: ratio at 1/2}
\frac{z}{1-\zbar}\biggr|_{\text{Re}[z]=\frac{1}{2}} = 1\, .
\end{align}
As a result, it is impossible to cross the line $\text{Re}[z]=1/2$ while keeping $z=\zbar^*$ without encountering
a singularity of the polylogarithms.  If one performs a Taylor series expansion of the function around
the line $\text{Re}[z]=1/2$ from the left, one finds however that all logarithmic singularities cancel between the two polylogarithms,
and $g^{(3)}_6$ has a regular Taylor series. It is then natural to use this Taylor series to continue the function across the line $\text{Re}[z]=1/2$. We find that it is given there by the same polylogarithms as in Eq.~\eqref{eq:funW1sv} (evaluated in their standard branches), plus the following correction term:
\begin{align}
\label{eq:f36}
\dg{(3)}_6 = \frac{i\pi}{2} \left[ \left( g^{(1)}_4 - g^{(1)}_5 \right)^2 + \left( g^{(1)}_6 + g^{(1)}_7 \right)^2 \right] \sign\left(\Im[z] \right)  \Theta\left( \text{Re}[z] - \frac{1}{2}\right)
\end{align}
where $\Theta$ is the Heaviside step function and the sign function distinguishes the representation of the function in the upper and in the lower half of the complex plane. 
The complete function $g^{(3)}_6$, which includes the correction $\dg{(3)}_6$, is then real analytic around any point in the complex plane except for the real axis where $\text{Re}[z]>1$ and for $\text{Re}[z]<0$. We find that all functions appearing in the hard function of $\mathcal{N}=4$ super Yang-Mills have this property.

\subsection{Transcendental functions for the $\mathcal{N}=8$ supergravity amplitude in the multi-Regge limit}
\label{sec:SUGRAfunctions}

As compared to the $\mathcal{N}=4$ super Yang-Mills case, several new transcendental functions are needed to describe the multi-Regge asymptotics of $\mathcal{N}=8$ supergravity. We present them weight by weight in this section, and show that they are well defined and real-analytic in both the upper and the lower half of the complex $z$ plane.

\textbf{Weight 2.} At weight 2 we need the functions $g_1^{(2)}$ and $g_2^{(2)}$ defined by Eqs.~\eqref{eq:funW2sv} and~\eqref{eq:funW2nsv}, and
\begin{equation}
\begin{aligned}
\label{eq:funW2sugra}
& g^{(2)}_3 = {\rm Li}_2\left(\frac{z}{1-\zbar}\right)+{\rm Li}_2\left(\frac{\zbar}{1-z}\right) + \dg{(2)}_3\, , \\
& g^{(2)}_4 = D_2\left(\frac{z}{1-\zbar},\frac{\zbar}{1-z}\right) + \dg{(2)}_4\, ,
\end{aligned}
\end{equation}
where $D_2$ is the single-valued dilogarithm~\eqref{eq:Li2BlochWigner}.
Note that these ``new" functions are simply derivatives of the weight-3 functions introduced preceedingly.

The correction terms $\dg{(3)}_3$ and $\dg{(2)}_4$, which involve only logarithms and step functions,
make these functions real-analytic away from the real axis.
They are recorded in Eqs.~\eqref{eq:gprime_corrections}, alongside corrections terms for the other functions defined in this subsection.
$g^{(2)}_3$ is real-valued, but note that the arguments of the dilogarithms are complex and cross the real axis as $z$ varies in the lower or upper half of the complex plane. However, for $\text{Im}[z]\neq 0$ the arguments take real values only at $\text{Re}[z] = 1/2$, where it is unity as shown in Eq.~\eqref{eq: ratio at 1/2}.
As a result, for $\text{Im}[z]\neq 0$, the branch cut of the dilogarithm is never crossed and Eq.~\eqref{eq:funW2sugra} defines an unambiguous function.
$g_4^{(2)}$ is manifestly single valued in the whole complex plane, but the $D_2$ function contains singular logarithms
at $\text{Re}[z]=1/2$ which are canceled by $\dg{(2)}_4$.

\textbf{Weight 3.}
In addition to $g^{(3)}_1,\ldots, g^{(3)}_6$, defined by Eqs.~\eqref{eq:funW3sv} and~\eqref{eq:funW3nsv}, we need three new genuine weight-3 functions,
\begin{equation}
\label{eq:funW3sugra}
\begin{aligned}
& g^{(3)}_7  = D_3\left(\frac{z}{1-\zbar},\frac{\zbar}{1-z}\right) + \dg{(3)}_7\, , \\
& g^{(3)}_8  = D_3\left(\frac{z+\zbar-1}{z},\frac{z+\zbar-1}{\zbar}\right) + \dg{(3)}_8\,, \\
& g^{(3)}_9  = {\rm Li}_3 \left( \frac{1-z-\zbar}{(1-z)(1-\zbar)} \right) \,,\\
& g^{(3)}_{10}  = {\rm Li}_3 \left( \frac{z+\zbar-1}{z} \right) - {\rm Li}_3 \left( \frac{z+\zbar-1}{\zbar} \right) \, .
\end{aligned}
\end{equation}
$D_3$ is the single-valued trilogarithm defined in Eq.~\eqref{eq:funW3sv}, and thus $g^{(3)}_7$ and $g^{(3)}_8$ are single-valued in the whole complex plane.
The argument of $g^{(3)}_9$ is real-valued and never lies on the branch cut of the polylogarithm for $\text{Im}[z]\neq 0$,
\begin{align}
\label{eq:argument1}
\frac{1-z-\zbar}{(1-z)(1-\zbar)} < 1\, ,
\end{align}
Also the arguments of $g^{(3)}_{10}$, although complex, never cross the branch cut of the polylogarithm for ${\rm Im}(z) \neq 0$. In fact, they take real values only at $\text{Re}[z]=1/2$, where it vanishes,
\begin{align}
\label{eq:argument2}
\frac{z+ \zbar-1}{z} \biggr|_{\text{Re}[z]=1/2} = 0\, .
\end{align}
As a result, both $g^{(3)}_{9}$ and $g^{(3)}_{10}$ are well-defined in both the upper and the lower half of the complex plane.

\textbf{Weight 4.} In order to describe weight-4 part of the \sugra hard function we introduce genuine weight-4 functions, which were not needed in the \sym case. Some of them are manifestly single valued and involve only classical polylogarithms, including
purely logarithmic correction terms recorded below:
\begin{equation}
\label{eq:funW4sv}
\begin{aligned}
& g^{(4)}_1 = D_4(z,\zbar) \, , \\
& g^{(4)}_2 = D_4(1-z,1-\zbar)\, ,\\
& g^{(4)}_3 = D_4 \left(\frac{z}{z-1},\frac{\zbar}{\zbar-1}\right) \, , \\ 
& g^{(4)}_4 = D_4 \left(\frac{z}{1-\zbar},\frac{\zbar}{1-z}\right) + \dg{(4)}_4\,,\\
 & g^{(4)}_7 = D_4 \left(\frac{z+\zbar-1}{z}, \frac{z+\zbar-1}{\zbar}  \right) + \dg{(4)}_7\,, \\
 & g^{(4)}_8 = D_4 \left( \frac{1-z-\zbar}{1-z}, \frac{1-z-\zbar}{1-\zbar}  \right)+ \dg{(4)}_8\, ,
\end{aligned}
\end{equation}
 where
\begin{align}
\label{eq:D4}
D_4(z,\zbar) \equiv {\rm Li}_4(z)-{\rm Li}_4(\zbar) - \frac{1}{2} \log(z \zbar) ({\rm Li}_3(z)-{\rm Li}_3(\zbar))+ \frac{1}{12} \log^2(z \zbar) ({\rm Li}_2(z)-{\rm Li}_2(\zbar)) 
\end{align}
is single valued in the whole complex plane. Two functions, instead, contain multiple polylogarithms and require a more careful analysis of the branch cut structure,
\begin{equation}
\begin{aligned}
g^{(4)}_5 = & \ \text{Li}_{2,2} \left(\frac{z+\zbar-1}{\zbar}, \frac{\zbar}{\zbar-1} \right) - 2 \, \text{Li}_{2,2}\left(\frac{\zbar}{1-z}, 1-z \right) \\
& +2 \, {\rm Li}_{2,2}\left( \frac{1-z-\zbar}{(1-z)(1-\zbar)}, 1-z \right) + 6 \, {\rm Li}_{2,2}\left( \frac{1-z-\zbar}{(1-z)(1-\zbar)}, 1-\zbar \right) \\
 & -9 \, {\rm Li}_{2,2}\left( \frac{z+\zbar-1}{z \zbar}, z \right)  -11 \, {\rm Li}_{2,2}\left(\frac{z+\zbar-1}{z \zbar}, \zbar \right) \\
 &- 34 \, {\rm Li}_{4}(z) - 34 \, {\rm Li}_{4}(\zbar) + 20 \, {\rm Li}_{4}(1-z) + 20 \, {\rm Li}_{4}(1-\zbar)  \\
 & -11 \, {\rm Li}_{4}\left(\frac{z+\zbar-1}{z}\right) -11 \, {\rm Li}_{4}\left(\frac{z+\zbar-1}{\zbar}\right) \\
 & + \frac{7}{2} \, {\rm Li}_{4}\left(\frac{1-z-\zbar}{1-z}\right)  + \frac{7}{2} \, {\rm Li}_{4}\left(\frac{1-z-\zbar}{1-\zbar}\right) \\
& + 38 \left[ {\rm Li}_{4}\left(\frac{z \zbar}{(1-z)(1-\zbar)}\right)  - \frac{1}{6} \log(1-z-\zbar) \log^3\left( \frac{(1-z)(1-\zbar)}{z \zbar}\right) \right] \\
& -\frac{13}{2} \, {\rm Li}_{4}\left(\frac{z}{1-\zbar}\right) - \frac{13}{2} \, {\rm Li}_{4}\left(\frac{\zbar}{1-z}\right)
 -19 \, {\rm Li}_{4}\left(\frac{z}{z-1}\right) - 19 \, {\rm Li}_{4}\left(\frac{\zbar}{\zbar-1}\right) \\
& +13 \, {\rm Li}_4 \left(\frac{z+\zbar-1}{z \zbar} \right) + 13 \, {\rm Li}_4 \left( \frac{1-z - \zbar}{(1-z)(1-\zbar)} \right)+ \dg{(4)}_5 \,,
\end{aligned}
\end{equation}
\begin{equation}
\begin{aligned}
g^{(4)}_6 = \, & 2 \, {\rm Li}_{2,2}\left( \frac{1-z-\zbar}{(1-z)(1-\zbar)}, 1-z \right) + 2 \, {\rm Li}_{2,2}\left( \frac{1-z-\zbar}{(1-z)(1-\zbar)}, 1-\zbar \right) \\
 & -2 \, {\rm Li}_{2,2}\left(\frac{z+\zbar-1}{z \zbar}, z \right)  -2\, {\rm Li}_{2,2}\left(\frac{z+\zbar-1}{z \zbar}, \zbar \right) \\
 &- 8 \, {\rm Li}_{4}(z) - 8 \, {\rm Li}_{4}(\zbar) + 8 \, {\rm Li}_{4}(1-z) + 8 \, {\rm Li}_{4}(1-\zbar)  \\
 & -2 \, {\rm Li}_{4}\left(\frac{z+\zbar-1}{z}\right) -2 \, {\rm Li}_{4}\left(\frac{z+\zbar-1}{\zbar}\right) \\
 & + 2 \, {\rm Li}_{4}\left(\frac{1-z-\zbar}{1-z}\right)  + 2 \, {\rm Li}_{4}\left(\frac{1-z-\zbar}{1-\zbar}\right) \\
& + 11 \left[ {\rm Li}_{4}\left(\frac{z \zbar}{(1-z)(1-\zbar)}\right)  - \frac{1}{6} \log(1-z-\zbar) \log^3\left( \frac{(1-z)(1-\zbar)}{z \zbar}\right) \right] \\
& -2 \, {\rm Li}_{4}\left(\frac{z}{1-\zbar}\right) - 2 \, {\rm Li}_{4}\left(\frac{\zbar}{1-z}\right)
 -6 \, {\rm Li}_{4}\left(\frac{z}{z-1}\right) - 6 \, {\rm Li}_{4}\left(\frac{\zbar}{\zbar-1}\right) + \dg{(4)}_6\, .
\end{aligned}
\end{equation}

Let us take a closer look. 
The functions
\begin{equation}
\begin{aligned}
& {\rm Li}_4(z) + {\rm Li}_4(\zbar)\, \ {\rm Li}_4(1-z) + {\rm Li}_4(1-\zbar)\, , \\
& {\rm Li}_4 \left(\frac{z}{z-1}\right) + {\rm Li}_4\left(\frac{\zbar}{\zbar-1}\right) \,, \ 
{\rm Li}_4 \left(\frac{z}{1-\zbar}\right) + {\rm Li}_4 \left(\frac{\zbar}{1-z}\right) \,, \\
 & {\rm Li}_4 \left(\frac{z+\zbar-1}{z}\right)  + {\rm Li}_4\left(\frac{z+\zbar-1}{\zbar}  \right) \,, {\rm Li}_4\left( \frac{1-z-\zbar}{1-z}\right) + {\rm Li}_4 \left( \frac{1-z-\zbar}{1-\zbar}  \right)\,
\end{aligned}
\end{equation}
are real valued and well defined for ${\rm Im}(z) \neq 0$. Eq.~\eqref{eq:argument1} and a similar analysis show that 
\begin{align}
{\rm Li}_4 \left(\frac{z+ \zbar-1}{z \zbar} \right) \ , \quad
{\rm Li}_4 \left( \frac{1-z - \zbar}{(1-z)(1-\zbar)} \right) 
\end{align}
are also well defined in both the upper and the lower half of the complex plane.
The analysis of
\begin{align}
\label{eq:combination}
{\rm Li}_4 \left( \frac{ z \zbar}{(1-z )(1-\zbar)} \right) - \frac{1}{6} \log(1-z-\zbar) \log^3\left( \frac{(1-z)(1-\zbar)}{z \zbar}\right)
\end{align} 
is similar to that of $g^{(3)}_5$ discussed below Eq.~\eqref{eq:funW3nsv}. The $\text{Li}_4$ has a branch cut at $\Re[z]>1/2$. Its imaginary part is however compensated by that of $\log(1-z-\zbar)$, and the function~\eqref{eq:combination} is real-valued. Moreover, the spurious singularity of $\log(1-z-\zbar)$ at $\Re[z]=1/2$ is canceled by the zero of the other logarithm. Finally, the multiple polylogarithms $\text{Li}_{2,2}(a,b)$ have branch cuts at $\Re[b] \in [1, + \infty)$ and at $\Re[a b] \in [1, +\infty)$.
For the ${\rm Li}_{2,2}(a,b)$ functions appearing in $g^{(4)}_5$ and $g^{(4)}_6$,
\begin{equation}
\begin{aligned}
& {\rm Li}_{2,2} \left( \frac{1-z - \zbar}{(1-z)(1-\zbar)}, 1-z\right) \,,\
{\rm Li}_{2,2} \left( \frac{1-z - \zbar}{(1-z)(1-\zbar)}, 1-\zbar\right) \,, \
{\rm Li}_{2,2} \left(\frac{z+\zbar-1}{z \zbar}, z\right) \,,\\
& {\rm Li}_{2,2} \left(\frac{z+ \zbar-1}{z \zbar}, \zbar \right) \,, \
{\rm Li}_{2,2} \left(\frac{z + \zbar-1}{\zbar}, \frac{\zbar}{\zbar-1}\right) \,,\ 
{\rm Li}_{2,2} \left( \frac{\zbar}{1-z}, 1-z \right)\,,
\end{aligned}
\label{eq:Li22list}
\end{equation}
it is clear that the branch cut at $\Re[b]\in [1,+\infty)$ is never crossed for $\Im[z]\neq 0$. The same holds for the branch cut at $\Re[ab]\in [1,+\infty)$, as well. This follows from Eq.~\eqref{eq:argument2}  and from a similar analysis for the argument $(z+\zbar-1)/(\zbar-1)$. The latter is complex valued and crosses the real axis as $z$ varies in the upper or lower half of the complex plane. However, for $\Im[z]\neq 0$ it takes real values only at $\Re[z]=1/2$, where it vanishes,
\begin{align}
\frac{z+\zbar-1}{\zbar-1}\biggr|_{\Re[z]=\frac{1}{2}} = 0 \, .
\end{align}
Therefore, the branch cuts of the multiple polylogarithms~\eqref{eq:Li22list} are never crossed for $\Im[z]\neq 0$. 

\def\lzomz{\ell} 
Note that many of the functions introduced above are singular at $z+\zbar -1 = 0$. This is a spurious artifact of our representation. The \sugra hard function, just like the \sym one, is real-analytic across $\Re[z]=1/2$, and it is precisely the task
of the correction terms $g'$ to ensure that. Introducing the shorthand notations
\begin{align}
\label{eq:funW1sugra}
& \lzomz = \Theta(1- z - \zbar) \,\log(1-z-\zbar) + \Theta(z + \zbar - 1)\,\log(z+\zbar-1) \equiv \log(|1-z-\zbar|) \,, \\
& \theta = \sign\left(\Im[z] \right)\Theta\left( \Re[z] - \frac{1}{2}\right) \, ,
\end{align}
the corrections are given as
\begingroup
\allowdisplaybreaks
\begin{align} \label{eq:gprime_corrections}
& \dg{(2)}_3 = \left(g^{(1)}_4 - g^{(1)}_5\right) \lzomz + i \pi \left(g^{(1)}_6 + g^{(1)}_7\right) \theta  \,, \\
& \dg{(2)}_4 =\left(g^{(1)}_6 + g^{(1)}_7\right) \lzomz  \, ,\\
&  \dg{(3)}_7 =  \frac{1}{4} \lzomz \left( \left(g^{(1)}_4 - g^{(1)}_5\right)^2 + \left(g^{(1)}_6 + g^{(1)}_7\right)^2 \right) \,, \\
&  \dg{(3)}_8 = \lzomz \left( -\frac{\pi^2}{3} + g^{(2)}_3 + \frac{3}{4} (g^{(1)}_4)^2 -  g^{(1)}_4 g^{(1)}_5 + \frac{1}{4} (g^{(1)}_5)^2 -\frac{1}{4} (g^{(1)}_6)^2 + \frac{1}{4} (g^{(1)}_7)^2\right) \notag\\
&  \quad +   i\pi \lzomz \left(g^{(1)}_6 + g^{(1)}_7\right) \theta \,,  \\ 
& \dg{(4)}_4 = -\frac{1}{24} \lzomz \left(g^{(1)}_4 - g^{(1)}_5 - g^{(1)}_6 - g^{(1)}_7) (g^{(1)}_6 + g^{(1)}_7) (g^{(1)}_4 - g^{(1)}_5 + g^{(1)}_6 + g^{(1)}_7\right) \,, \\
& \dg{(4)}_5 = -\frac{13}{48} \lzomz \left(g^{(1)}_4 - g^{(1)}_5\right) \left( \left(g^{(1)}_4 - g^{(1)}_5\right)^2 + 3\left(g^{(1)}_6 + g^{(1)}_7\right)^2  \right) \notag\\
&\quad  - \frac{13}{48} i \pi \left(g^{(1)}_6 + g^{(1)}_7\right) \left( 3\left(g^{(1)}_4 - g^{(1)}_5\right)^2 + \left(g^{(1)}_6 + g^{(1)}_7\right)^2  \right) \theta   \,,\\
& \dg{(4)}_6 = -\frac{1}{12} \lzomz \left(g^{(1)}_4 - g^{(1)}_5\right) \left( \left(g^{(1)}_4 - g^{(1)}_5\right)^2 + 3\left(g^{(1)}_6 + g^{(1)}_7\right)^2  \right) \notag\\
&  \quad -  \frac{1}{12} i \pi \left(g^{(1)}_6 + g^{(1)}_7\right) \left( 3\left(g^{(1)}_4 - g^{(1)}_5\right)^2 + \left(g^{(1)}_6 + g^{(1)}_7\right)^2  \right) \theta \, , \\
& \dg{(4)}_7 = -\frac{1}{3} (\lzomz)^3 \left(g^{(1)}_6 + g^{(1)}_7\right) +  (\lzomz)^2 \left( \frac{1}{2} g^{(1)}_4 \left(g^{(1)}_6 + g^{(1)}_7\right) - \frac{1}{3} g^{(2)}_4 \right) \notag\\
&  \quad  + \lzomz\left(
g^{(3)}_{10} +\frac{1}{3} g^{(2)}_{4} g^{(1)}_{4} -\frac{1}{6} \left(g^{(1)}_{4}\right)^2 \left(g^{(1)}_{6} + g^{(1)}_{7}\right)  \right) \,, \\
& \dg{(4)}_8 = -\frac{1}{3} (\lzomz)^3 \left(g^{(1)}_6 + g^{(1)}_7\right) +  (\lzomz)^2 \left( \frac{1}{2} g^{(1)}_4 \left(g^{(1)}_6 + g^{(1)}_7\right) - \frac{1}{3} g^{(2)}_4 \right)\notag\\
&  \quad  + \lzomz\Biggl(
g^{(3)}_{10}   + \frac{1}{2} g^{(3)}_{6} - \frac{\pi^2}{6} \left(g^{(1)}_6 + g^{(1)}_7\right) +\frac{1}{2} g^{(2)}_{4} g^{(1)}_{4}  -\frac{1}{8} \left(g^{(1)}_4\right)^2 g^{(1)}_6  - \frac{1}{24} \left(g^{(1)}_5\right)^2 g^{(1)}_6 \notag\\
& \quad -\frac{1}{6} g^{(2)}_{4} g^{(1)}_{5} - \frac{1}{24} \left(g^{(1)}_6\right)^3 - \frac{1}{4} g^{(1)}_4 g^{(1)}_5 g^{(1)}_7 + \frac{1}{12} \left(g^{(1)}_5\right)^2 g^{(1)}_7 + \frac{1}{8}  g^{(1)}_6 \left(g^{(1)}_7\right)^2 + \frac{1}{12} \left(g^{(1)}_7\right)^3  \Biggr) \notag\\
& \quad  +  \frac{i\pi}{4} \lzomz \left( \left(g^{(1)}_4 - g^{(1)}_5\right)^2 + \left(g^{(1)}_6 + g^{(1)}_7\right)^2  \right) \theta \,.
\end{align}
\endgroup
All the $g$ functions, well-defined for $\Im[z]\neq 0$, are then real-analytic across $\Re[z]=1/2$.

\section{The multi-Regge limit of the $\mathcal{N}=4$ super Yang-Mills amplitude}
\label{sec:sYM}

In this section we discuss the multi-Regge limit of the full-colour two-loop five-particle amplitude in $\mathcal{N}=4$ super Yang-Mills theory. After discussing the behaviour of the rational functions in the limit, we introduce a colour decomposition based on the colour flowing in the $t$ channels. This highlights certain properties of the multi-Regge regime, and makes the expressions more compact. Then we present our results for the asymptotics of the hard function at one and two loops. Finally, we classify all the required transcendental functions, and discuss their analytic structure. We provide the explicit expressions in ancillary files.

The Regge limit of the planar part has already been investigated in Ref.~\cite{Bartels:2008ce}, where the simple form of the ABDK/BDS formula~\cite{Anastasiou:2003kj, Bern:2005iz,Drummond:2007au} was shown to be Regge-exact at five points. The multi-Regge limit of the non-planar part was then studied in Ref.~\cite{Abreu:2018aqd, Chicherin:2018yne}. The double-trace components in the colour basis given by Eqs.~\eqref{eq:single_traces} and~\eqref{eq:double_traces} were found to vanish at symbol level in the limit $x\to 0$. The analysis was then pushed to the subleading power terms, of which the leading-logarithmic contributions were provided analytically.

Here we present explicit analytic expressions for the divergent and finite contributions of the full-colour hard function. We neglect the terms which vanish as $x \log^k(x)$ in the limit $x\to 0$.

The starting point is the one and two-loop amplitudes in the form given by Eqs.~\eqref{eq:sYM_expected_1loop} and~\eqref{eq:sYM_expected}. The amplitudes consist of rational functions of the spinors and pure integrals.
The rational functions exhibit an extremely simple behaviour in the multi-Regge limit.
Since the latter was defined for helicity-free quantities only~\eqref{regge}, we normalise them by a Parke-Taylor factor. We choose $\PT_1$~\eqref{eq:PTbasis}.
Then, all but two of the six linearly independent Parke-Taylor prefactors~\eqref{eq:PTbasis} vanish, and the remaining ones become identical up to power corrections:
\begin{align}
\label{eq:PTregge}
\frac{\PT_i}{\PT_1} = r_i + \mathcal{O}(x) \, ,
\end{align}
with 
\begin{align}
\{r_i\}_{i=1}^6 = \{1,0,0,0,0,1\}.
\end{align}
The absence of poles in $x$ in the rational prefactors constitutes a major simplification, as it implies that the power corrections to the pure integrals can be neglected. In contrast, as we will see in Section~\ref{sec:SUGRA}, they need to be taken into account in the $\mathcal{N}=8$ supergravity case. We thus compute the asymptotics of the required pure integrals according to Eq.~\eqref{eq:AsymptoticSolution}. Finally, we assemble the two-loop hard function in the $\overline{\text{MS}}$-scheme according to Eq.~\eqref{eq:H2}.

\subsection{Colour flow in the multi-Regge limit}
\label{sec:Colour}

Scattering amplitudes in a gauge theory can be seen as vectors in colour space. In Section~\ref{sec:Assembly} we have introduced a basis of this vector space made of traces of generators in the fundamental representation. This basis makes the distinction between planar and non-planar components transparent, and highlights the permutation symmetries. However, when discussing the multi-Regge limit, it is more meaningful to consider a different colour decomposition. 
We expand the $\ell$-loop five-particle scattering amplitude $A_5^{(\ell)}$ into a colour basis $\{ \mathcal{S}_J\}$ where each element corresponds to a definite $t$-channel exchange,
\begin{align}
\label{eq:MexpansionS}
A_5^{(\ell)} = \sum_{J} A_J^{(\ell)} \mathcal{S}_J \, .
\end{align}
The sum in Eq.~\eqref{eq:MexpansionS} runs over all possible pairs $J = (r_1, r_2)$, where $r_1$ ($r_2$) labels the possible irreducible representations of the states propagating in the $t_1$ ($t_2$) channel. These are obtained by reducing the tensor products of the representations of particles $2$ and $3$, and $1$ and $5$, namely
\begin{align}
R_2 \otimes R_3 = \bigoplus_{r_1} r_1\, , \qquad \qquad \qquad R_1 \otimes R_5 = \bigoplus_{r_2} r_2 \, ,
\end{align}
where $R_i$ denotes the irreducible representation corresponding to particle $i$. Multiple occurrences of equivalent representations are counted as distinct. 
In the case of two adjoint representations the decomposition is
\begin{align}
\textbf{8}_a \otimes \textbf{8}_a = \textbf{1} \oplus \textbf{8}_s \oplus \textbf{8}_a \oplus \textbf{10} \oplus \overline{\textbf{10}} \oplus \textbf{\textbf{27}} \oplus \textbf{0}\, ,
\end{align}
where we use the subscripts $s$ and $a$ distinguish the antisymmetric adjoint representation $\textbf{8}_a$ from the 8-dimensional symmetric representation $\textbf{8}_s$.
Note that we label the representations with their $SU(3)$ dimensions, but we keep the expressions for generic $SU(N_c)$. As a result, we keep the ``null" representation $\textbf{0}$, although it does not contribute for $N_c=3$ since its dimensionality vanishes, 
\begin{align}
\text{dim}[\textbf{0}] = \frac{N_c^2 (N_c-3)(N_c+1)}{4}\,.
\end{align}

In order to better understand the colour decomposition~\eqref{eq:MexpansionS}, let us introduce colour operators associated with the colour flow in the $t$-channels~\cite{Dokshitzer:2005ig,DelDuca:2011ae},
\begin{align}
& \textbf{T}_{t_1} = \textbf{T}_2 + \textbf{T}_3\, , \\
& \textbf{T}_{t_2} = \textbf{T}_2 + \textbf{T}_3 + \textbf{T}_4 = - \textbf{T}_1 - \textbf{T}_5 \, ,
\end{align}
where we used colour conservation, $\sum_{i=1}^5 \textbf{T}_i = 0$. We recall that $\textbf{T}_i$ denotes the colour insertion operator, given for the adjoint representation by Eq.~\eqref{eq:ColourInsertion}.
It is apparent that the Casimir operators $\textbf{T}_{t_1}^2$ and $\textbf{T}_{t_2}^2$ commute with each other.
The colour factors $\mathcal{S}_J$, by definition, are their simultaneous eigenvectors:
\begin{align}
\textbf{T}^2_{t_k} \circ \mathcal{S}_J = C_{r_k} \mathcal{S}_J \, ,
\end{align}
for $k=1,2$. We recall that $J=(r_1,r_2)$, where $r_k$ is the representation in the $t_k$ channel.
The 22 allowed pairs are listed in Table~\ref{tab:Representations}.

\begin{table}
\begin{center}
\begin{tabular}{|c|c|c|c|}
\hline
$\mathcal{S}_a$ & $(r_1, r_2)$ & $C_{r_1}$ &  $C_{r_2}$ \\
\hline
1 & $(\textbf{1}, \textbf{8}_a)$ & $0$ & $2 N_c$ \\
2 & $(\textbf{8}_a, \textbf{1})$ & $N_c$ & $0$ \\
3 & $(\textbf{8}_a, \textbf{8}_a)$ & $N_c$ & $N_c$ \\
4 & $(\textbf{8}_a, \textbf{8}_s)$ & $ N_c$ & $N_c$ \\
5 & $(\textbf{8}_a, \textbf{0})$ & $N_c$ & $ 2 (N_c-1)$ \\
6 & $(\textbf{8}_a, \textbf{27})$ & $N_c$ & $2 (N_c+1)$ \\
7 & $(\textbf{8}_a, \textbf{10})$ & $N_c$ & $ 2 N_c$ \\
8 & $(\textbf{8}_s, \textbf{8}_a)$ & $N_c$ & $ N_c$ \\
9 & $(\textbf{8}_s, \textbf{8}_s)$ & $ N_c$ & $ N_c$ \\
10 & $(\textbf{8}_s, \textbf{10})$ & $ N_c$ & $2 N_c$ \\
11 & $(\textbf{0}, \textbf{8}_a)$ & $2(N_c-1)$ & $N_c$ \\
\hline
\end{tabular}
\hspace{0.5cm}
\begin{tabular}{|c|c|c|c|}
\hline
$\mathcal{S}_a$ & $(r_1, r_2)$ & $C_{r_1}$ &  $C_{r_2}$ \\
\hline
12 & $(\textbf{0}, \textbf{0})$ & $2 (N_c-1)$ & $2 (N_c-1)$ \\
13 & $(\textbf{0}, \textbf{10})$ & $2 (N_c-1)$ & $2 N_c$ \\
14 & $(\textbf{27}, \textbf{8}_a)$ & $2(N_c+1)$ & $N_c$ \\
15 & $(\textbf{27}, \textbf{27})$ & $2(N_c+1)$ & $2 (N_c+1)$ \\
16 & $(\textbf{27}, \textbf{10})$ & $2(N_c+1)$ & $2 N_c$ \\
17 & $(\textbf{10}, \textbf{8}_a)$ & $2 N_c$ & $N_c$ \\
18 & $(\textbf{10}, \textbf{8}_s)$ & $2 N_c$ & $N_c$ \\
19 & $(\textbf{10}, \textbf{0})$ & $2 N_c$ & $2 (N_c-1)$ \\
20 & $(\textbf{10}, \textbf{27})$ & $2 N_c$ & $2 (N_c+1)$ \\
21 & $(\textbf{10}, \textbf{10})_1$ & $2 N_c$ & $2 N_c$ \\
22 & $(\textbf{10}, \textbf{10})_2$ & $2 N_c$ & $2 N_c$ \\
\hline
\end{tabular}
\end{center}
\caption{Characterisation of the colour basis $\{ \mathcal{S}_a \}_{a=1}^{22}$.
The first column refers to the element in the basis. The second contains the pairs of irreducible representations corresponding to the states propagating in the $t_1$ and $t_2$ channel, labelled by SU(3) dimensions ($\textbf{1}$ being the singlet).
The last two columns show the associated Casimirs.
The $\textbf{10}$ stand for combinations of $\textbf{10}$ and $\overline{\textbf{10}}$, and the last two
structures denote two invariant tensors between $\textbf{10}, \overline{\textbf{10}}$ and the central gluon's
$\textbf{8}$ (for $N_c\geq 4$).}
\label{tab:Representations}
\end{table}

Therefore, we expand the hard function as
\begin{align} \label{eq: def H}
\mathcal{H}_5^{(\ell)} = \PT_1 \sum_{a=1}^{22} H_{a}^{(\ell)} \mathcal{S}_a\, ,
\end{align}
where we extract a factor of $\PT_1$ so that the coefficient functions $H_{a}^{(\ell)}$ are helicity-free. 
We show in Table~\ref{tab:Representations} our explicit choice for the basis of eigenvectors of the $t$-channel operators, $\{\mathcal{S}_a\}_{a=1}^{22}$.

This basis choice will greatly simplify the analysis of the Regge limit
since it is controlled by the quantum numbers flowing in the $t_i$ channels.
For example, $(\textbf{8}_a, \textbf{8}_a)$ is the only colour structure at tree-level: 
\begin{equation}
\begin{aligned} \label{eq: LL tree}
& h^{(0)}_{3} = 1\, ,\\
& h^{(0)}_{a}=0 \, , \quad \forall a\neq 3 \, .
\end{aligned}
\end{equation}
We will see that it is also the only structure at leading-logarithmic order (LL) to all loop orders. For convenience of the reader, we spell out in terms of single~\eqref{eq:single_traces} and double traces~\eqref{eq:double_traces} the colour structures which are of particular interest for this paper:
\begin{align} \label{eq: few S}
& \mathcal{S}_3 = \mathcal{T}_1 + \mathcal{T}_2 - \mathcal{T}_5 - \mathcal{T}_6 \, , \\ 
& \mathcal{S}_9 = \mathcal{T}_{1} - \mathcal{T}_{2} - \mathcal{T}_{5} + \mathcal{T}_{6} \, ,\\
& \mathcal{S}_{12} = \mathcal{T}_{1} - \mathcal{T}_{2} - \mathcal{T}_{5} + \mathcal{T}_{6} - (N_c-2) \left( \mathcal{T}_{7} - \mathcal{T}_{9} - \mathcal{T}_{10} + \mathcal{T}_{12} + \mathcal{T}_{13} + \mathcal{T}_{18} + \mathcal{T}_{20} + \mathcal{T}_{22} \right) , \\
& \mathcal{S}_{15} = \mathcal{T}_{1} - \mathcal{T}_{2} - \mathcal{T}_{5} + 
\mathcal{T}_{6} + (N_c+2) \left( \mathcal{T}_{7} - \mathcal{T}_{9} - \mathcal{T}_{10} + \mathcal{T}_{12} - \mathcal{T}_{13} - \mathcal{T}_{18} - \mathcal{T}_{20} - \mathcal{T}_{22} \right) \, , \\
& \mathcal{S}_{21} = \frac{N_c}{2} \left(-\mathcal{T}_{7} + \mathcal{T}_{9} - \mathcal{T}_{10} + \mathcal{T}_{12}\right) - \mathcal{T}_{13} + \mathcal{T}_{15} + \mathcal{T}_{16} + \mathcal{T}_{18} - \mathcal{T}_{19} - \mathcal{T}_{20} - \mathcal{T}_{21} + \mathcal{T}_{22} \, , \\
& \mathcal{S}_{22} = \mathcal{T}_{1} + \mathcal{T}_{2} - \mathcal{T}_{5} - \mathcal{T}_{6} + N_c \left(-\mathcal{T}_{13} + \mathcal{T}_{18} - \mathcal{T}_{20} + \mathcal{T}_{22}\right) \,.
\end{align}
We provide in an ancillary file the transformation matrix to the trace-based colour basis,
\begin{align}
\mathcal{S}_a = \sum_{b=1}^{22} E_{ab} \mathcal{T}_b \,.
\end{align}

\subsection{One-loop hard function}

Thanks to the uniform transcendental weight property of the amplitude, the one-loop hard function is given by a weight-two function. We organise it in powers of $\log(x)$,
\begin{align}
\label{eq:H1logx}
H_a^{(1)} = \sum_{k=0}^1 h^{(1)}_{a,k}\left(N_c, s,s_1,s_2,z,\zbar \right) \log^k(x) + o(1)\, ,
\end{align}
where the coefficient function $h^{(1)}_{a,k}$ has weight $(2-k)$. Note that, although the function has weight 2, the maximal power of $\log x$ is 1. The fact that there can appear at most one power of $\log x$ per loop order
is a well-known result of the BFKL formalism \cite{Lipatov:1976zz,Kuraev:1976ge}~(see e.g. Ref.~\cite{Caron-Huot:2017fxr} for a recent discussion in the scattering amplitudes context). It is related to the fact that boosting a projectile does not introduce any new collinear singularities.

Only one component has a non-vanishing contribution at order $\log x$,
\begin{equation}
\begin{aligned} \label{eq: LL 1loop}
& h^{(1)}_{3,1} = -2 N_c \left[2 \log\left(\frac{s}{s_1 s_2}\right) - \log\left((1-z)(1-\zbar) \right) - \log \left( z \zbar \right) \right] \, , \\
& h^{(1)}_{a,1}=0 \, , \quad \forall a\neq 3 \, .
\end{aligned}
\end{equation}
The non-trivial colour structure corresponds to the representations $(\textbf{8}_a, \textbf{8}_a)$, namely to an adjoint exchange in the $t_1$ and $t_2$ channels. This is expected. The leading logarithmic (LL) terms, i.e. those of order $(g^2)^{\ell} \log^{\ell} x$, all have the same colour structure as the tree-level amplitude, namely $(\textbf{8}_a, \textbf{8}_a)$, because at LL order a single elementary excitation (the reggeized gluon) propagates in each channel.
The other colour components are kinematically suppressed, either by logarithms or by powers of $x$~\cite{Lipatov:1976zz,Kuraev:1976ge}.
There are also selection rules at NLL order, where a (symmetrical) pair of adjoint excitations can also be exchanged.
Since such a pair cannot carry the colour representation $\textbf{10}$, many components vanish at order $\log^0 x$. From the explicit computation we find that
\begin{align}
\label{eq:h1a0}
h^{(1)}_{a,0} = 0 \, , \quad \forall a \in \{7, 10, 13, 16, 17, 18, 19, 20, 21, 22\} \, .
\end{align}
The remaining colour components are expressed in terms of the functions described in Section~\ref{sec:sYMfunctions}. All functions involved at one loop are manifestly single valued in the whole $s_{12}$ channel. We provide the complete expressions in both the upper and lower half of the complex $z$ plane in ancillary files.

\subsection{Two-loop hard function}
At two loops, the hard functions is given by a weight-4 function, which we expand in powers of $\log x$ as
\begin{align}
\label{eq:H2logx}
H_a^{(2)} = \sum_{k=0}^2 h^{(2)}_{a,k}\left(N_c, s,s_1,s_2,z,\zbar \right) \log^k(x) + o(1) \, ,
\end{align}
where the coefficient function $h^{(2)}_{a,k}$ has weight $4-k$.

Before discussing the different powers of $\log x$ separately, it is instructive to spell out some components. The two eigenvectors corresponding to the representation $(\textbf{10}, \textbf{10})$ are particularly simple. One, $H_{21}^{(2)}$, vanishes in the multi-Regge limit,
\begin{align}
\label{eq:H21}
H_{21}^{(2)} = \mathcal{O}(x) \, .
\end{align}
Interestingly, this can be seen by taking into account the behaviour of the rational functions~\eqref{eq:PTregge} alone.
The second eigenvector of the representation $(\textbf{10}, \textbf{10})$, $H_{22}^{(2)}$, is finite in the limit $x \to 0$
and has the explicit expression
\begin{equation}
\label{eq:H22}
\begin{aligned}
H_{22}^{(2)} = \, & 2 \pi^2 \biggl[ \log^2\left(\frac{s}{s_1 s_2}\right) - 2\log\left(\frac{s}{s_1 s_2}\right) \log\left(z \zbar (1 - z)(1 - \zbar)\right)+ \log^2(z \zbar)   \\
& + \log^2\left((1 - z)(1 - \zbar)\right) + \log\left((1 - z)(1 - \zbar)\right) \log(z\zbar)  \\
& - 2 \biggl({\rm Li}_2(z) - {\rm Li}_2(\zbar) - \frac{1}{2} \left(  \log(1 - z) - \log(1-\zbar)\right) \log(z\zbar) \biggr)\biggr] + \mathcal{O}(x)  \, .
\end{aligned}
\end{equation}
This function is manifestly single valued in the whole complex $z$ plane. In particular, in the last line we recognise the single-valued dilogarithm,
\begin{align}
\label{eq:Li2BlochWigner}
D_2(z,\zbar) = \text{Li}_2(z) - \text{Li}_2(\zbar) +\frac{1}{2} \biggl(\log(1-z)-\log(1-\zbar) \biggr) \log(z \zbar) \, .
\end{align}

Let us now turn to a discussion of the different orders in $\log x$. As expected, only one colour structure, $(\textbf{8}_a, \textbf{8}_a)$, contains the leading logarithm $\log^2 x$,
\begin{equation}
\begin{aligned} \label{eq: LL 2loop}
& h^{(2)}_{3,2} = 2 N_c^2 \left(2 \log\left(\frac{s}{s_1 s_2} \right) - \log\left(z \zbar\right) - \log\left( (1-z) (1-\zbar) \right) \right)^2\, , \\
& h^{(2)}_{a,2} = 0 \, , \quad  \forall a \neq 3 \, .
\end{aligned}
\end{equation}
The colour components which vanish at order $\log^0 x$ at one loop~\eqref{eq:h1a0}, vanish at order $\log x$ at two loops as well,
\begin{align}
h^{(2)}_{a,1} = 0 \qquad \forall a \in \{7,10,13,16,17,18,19,20,21,22 \} \, .
\end{align}
This is a general feature, as will be clear below:
only the colour representations which have non-trivial coefficients at the previous loop order can exhibit a logarithm of $x$.
The remaining non-vanishing components are very simple, as they contain just manifestly single-valued logarithms and the single-valued dilogarithm~\eqref{eq:Li2BlochWigner}. The non-trivial components, except $h^{(2)}_{3,1}$, are proportional to $i \pi$ and thus vanish at symbol level. $h^{(2)}_{3,1}$ is non-zero at symbol level. Still, it does not contain any genuine weight-3 function, but only products of logarithms. 

Finally, let us consider the finite terms, i.e. $h^{(2)}_{a,0}$. They have transcendental weight 4, but only $h^{(2)}_{3,0}$ does not vanish at symbol level. Nonetheless, the latter has a rather simple form, as it involves only logarithms and the single-valued dilogarithm~\eqref{eq:Li2BlochWigner}. The other components are proportional to either $i \pi$ or $\pi^2$. Three component vanish, $h^{(2)}_{a,0} = 0$ $\forall a = 10,18,21$. Some contain genuine weight-3 functions.

As we have seen in Section~\ref{sec:sYMfunctions}, the $\mathcal{N}=4$ super Yang-Mills hard function in the multi-Regge limit involves several functions which are not manifestly real analytic in the complex $z$ plane. It is worthwhile to highlight that they appear only in the order-$\log^0 x$ components $h^{(2)}_{a,0}$ with $a \in \{1,2,5,6,11,12,14,15\}$. Among these, $h^{(2)}_{12,0}$ and  $h^{(2)}_{15,0}$ are particularly simple, as they only involve --~of the functions which are not manifestly real analytic~-- the logarithms~\eqref{eq:funW1nsv}. For instance,
\begin{align}
h^{(2)}_{12,0} = \frac{i\pi}{6}\left( \left(g_{7}^{(1)}\right)^3-\left(g_{6}^{(1)}\right)^3\right) + \frac{2i\pi^3}{3}\left(g_{7}^{(1)}-g_{6}^{(1)}\right)
+ \text{(analytic)}\,. \label{eq: singular 12}
\end{align}

The other components, $h^{(2)}_{a,0}$ with $a \in \{1,2,5,6,11,14\}$, contain not only the logarithms~\eqref{eq:funW1nsv}, but also the weight-2 and 3 functions given by equations~\eqref{eq:funW2nsv} and~\eqref{eq:funW3nsv}. It is however interesting to note that the latter only appear in a specific combination,
\begin{align}
6 g^{(3)}_1 + g^{(3)}_2 - 2 g^{(3)}_3 + 2 g^{(3)}_4 + 2 g^{(3)}_5 + 2 g^{(3)}_6 + g^{(2)}_2 ( g^{(1)}_4 - g^{(1)}_5) \,.
\end{align}   
Nevertheless, we introduced all these functions separately, as they will enter the hard function in $\mathcal{N}=8$ supergravity.

It is interesting to note that, although the hard function is continuous throughout the whole complex $z$ plane, certain components are not analytic across the real axis. In particular, the second derivatives of certain components exhibit discontinuities:  at $\text{Re}[z]>0$ for components 1, 11 and 14; at $\text{Re}[z]<1$ for components 2, 5 and 6; and at $\text{Re}[z]<0$ and $\text{Re}[z]>1$ for components 12 and 15. Such non-analyticity of non-planar amplitudes in the multi-Regge limit, while absent in the planar limit,
it is not a new phenomenon. It was already observed, for instance, in the computation of the non-planar impact factor at one loop~\cite{Fadin:1999de}, as will be discussed in the next section.

We provide the complete expressions of the coefficient functions $h^{(2)}_{a,k}$ in both the upper and lower half of the complex $z$ plane in ancillary files. They are expressed in terms of the functions which we described in Section~\ref{sec:sYMfunctions}.

\section{Predictions of BFKL theory in $\mathcal{N}=4$ super Yang-Mills}
\label{sec:BFKL}

We now confront the preceding results with the effective description of the Regge limit based on
BFKL theory, where the degrees of freedom at different rapidity interact by exchange of (multiple) reggeized gluons.
The BFKL approach becomes particularly simple for certain colour structures, as explained shortly,
and therefore we will focus here on those simplest colour structures leaving a systematic analysis to the future. 
Interestingly, we will be able to express directly the $d\to 4$ limit of the hard functions in terms of an infrared $R$-operation,
by-passing the calculation of IR-divergent quantities.

\subsection{General considerations}

We begin by studying the bare (IR divergent) amplitude.
In multi-Regge kinematics, the degrees of freedom of the 5-point amplitude are split into three sets:
left-moving ($p_2,p_3$), central ($p_4$) and right-moving ($p_5,p_1$), which leads to the following factorization:
\begin{equation}
 \mathcal{A}_5 = \langle L | e^{-\eta_L H} a_{4} e^{-\eta_R H} |R\rangle \, , \label{A5 BFKL}
\end{equation}
where $\eta_L=\log |\frac{s_{34}}{-s_{23}}| - \tfrac{i\pi}{2}$ and
$\eta_R=\log |\frac{s_{45}}{-s_{15}}| - \tfrac{i\pi}{2}$ are convenient choices of rapidity variables (boost factors)
in the left and right channels, $H$ is the BFKL Hamiltonian (generator of boosts),
and $a_4$ stands for the annihilation operator of particle 4.
All the energy dependence is explicit in the exponent, while the remaining ingredients only produce functions of transverse momenta ${\bf p}_i$ (see Section~\ref{sec:MultiReggeLimit} for our kinematics).
Since the Hamiltonian commutes with colour projectors, this factorization shows that energy
logarithms can only dress colour components which were already present at a previous loop order.

The wavefunctions on the outside each represent a pair of fast-moving particles,
\begin{align}
\langle L| \equiv \langle 0| a_3 a_2^\dagger \, ,\qquad
| R\rangle \equiv a_5 a_1^\dagger |0\rangle\, ,
\end{align}
which are to be inserted inside time-ordered products.
To leading power, these
can be expanded over a basis of multi-reggeon states. These are closely related to null Wilson lines
as one might expect from an eikonal approximation. Briefly, the Wilson lines $U$ are exponential
of the reggeized gluon $W$:
$U_{\bf x} = e^{ig W^a_{\bf x} T^a}$, see Refs.~\cite{Caron-Huot:2013fea,Caron-Huot:2017fxr} (also~\cite{Kovner:2005qj})
for further detail.
The complete wavefunction contains an arbitrary number of Wilson lines and therefore also an arbitrary number of reggeons.
For our specific application below, we will exploit the following simplification of perturbation theory.
The leading contribution for a given number of reggeons takes a very simple form,
coming from expanding a single Wilson line:
\begin{equation}\begin{aligned} \label{eq: leading WF}
| R\rangle &=2p_5^- \int \frac{d^{2-2\epsilon}{\bf x}\ e^{i{\bf p}_5\cdot{\bf x}}}{(2\pi)^{2-2\epsilon}} 
\left( igT^a_{i_5i_1} |W^a_{\bf x}\rangle + \frac{(ig T)_{i_5i_1}^2}{2!}|W_{\bf x}W_{\bf x}\rangle + \frac{(igT)_{i_5i_1}^3}{3!}| W_{\bf x}W_{\bf x}W_{\bf x}\rangle+\ldots\right)
\\ & \quad + \mbox{(higher powers of $g^2$ for each $W$)}.
\end{aligned}\end{equation}
For notational convenience, we suppressed
the adjoint indices on the last two terms. (These are to be contracted between the $T^a$'s and $W^a$'s,
as written in the first term. The $W$'s commute so their ordering is immaterial.)
Notice that each additional reggeon comes with a factor $g$.
Notice also that in all the terms in Eq.~\eqref{eq: leading WF} the reggeons sit at the same transverse coordinate ${\bf x}$.

The general structure of loop corrections is as follows.
The coefficient of the $|W\rangle$ component is simply a function of the coupling and momentum,
known to three loops in a certain scheme \cite{Caron-Huot:2017fxr} (and known to all orders in the planar limit \cite{Beisert:2006ez}).
In the planar limit, this is the only contribution.
Loop corrections to multi-reggeon components are more complicated:
delocalized products such as $|W_{\bf x}W_{\bf y}\rangle$ appear at order $g^4$.
These are integrated against impact factors of the form $C^{(1)}({\bf x}-{\bf y})$\, ,
which are nontrivial functions of relative positions or momenta \cite{Fadin:1999de}.
This particular one-loop impact factor can contribute to certain colour structures in the two-loop amplitudes.

What will be important here is that the elementary reggeon $W^a$ is colour-adjoint, which restricts the possible colour structures
to which each term in Eq.~\eqref{eq: leading WF} can contribute.
Another restriction comes from the symmetry (signature) which interchanges legs $2$ and $3$ and reverses $W$.
For example, the state $|W\rangle$ contributes only to the antisymmetric adjoint, $\textbf{8}_a$,
producing the leading-log result shown in Eqs.~\eqref{eq: LL tree}, \eqref{eq: LL 1loop} and \eqref{eq: LL 2loop}.
The symmetric adjoint $\textbf{8}_s$ is not affected by single-$W$ exchange because of signature conservation:
the amplitude is invariant under the simultaneous crossing of legs $p_1$ and $p_5$ and reversal $W\mapsto -W$ in the $t_2$ channel. Tensor product rules together with signature conservation imply the following selection rules:
\begin{equation}\begin{aligned} \label{WW colors}
 | W\rangle &\mbox{ contributes to } \textbf{8}_a\, ,\\
 | WW\rangle &\mbox{ contributes to } \textbf{8}_s, \textbf{0}, \textbf{27}\,, \\
 | WWW\rangle &\mbox{ contributes to } \textbf{8}_a, \textbf{10}\, .
\end{aligned}\end{equation}
The two-loop amplitude in the $(r_1,r_2)=(\textbf{8}_a,\textbf{27})$ channel,
for instance, thus requires only the following ingredients: the gluon impact factors up to one-loop
(that is the coefficients of $|WW\rangle$ and of $\langle W|$ in Eq.~\eqref{eq: leading WF} and its analog on the left),
the regge trajectory and the central emission vertex to one-loop (respectively, the $H$ eigenvalue and overlap in Eq.~(\ref{A5 BFKL})).
All these ingredients are available and it will be interesting in future work to assemble them.

Below we focus on two simple non-planar colour components at two-loops: $(\textbf{10},\textbf{10})$,
where three reggeons must appear in each channel,
and the various $(WW,WW)$ components, where two reggeons must flow in each channel, where we will highlight a non-real analytic behaviour.

\subsection{Bare amplitude with maximal reggeon exchanges}

\begin{figure}[t]
  \begin{center}
    \def\svgwidth{0.2\columnwidth}%% Creator: Inkscape 1.0beta1 (32d4812, 2019-09-19), www.inkscape.org
%% Accompanies image file 'reggeons1.pdf' (pdf, eps, ps)
\begingroup%
  \makeatletter%
  \providecommand\color[2][]{%
    \errmessage{(Inkscape) Color is used for the text in Inkscape, but the package 'color.sty' is not loaded}%
    \renewcommand\color[2][]{}%
  }%
  \providecommand\transparent[1]{%
    \errmessage{(Inkscape) Transparency is used (non-zero) for the text in Inkscape, but the package 'transparent.sty' is not loaded}%
    \renewcommand\transparent[1]{}%
  }%
  \providecommand\rotatebox[2]{#2}%
  \newcommand*\fsize{\dimexpr\f@size pt\relax}%
  \newcommand*\lineheight[1]{\fontsize{\fsize}{#1\fsize}\selectfont}%
  \ifx\svgwidth\undefined%
    \setlength{\unitlength}{59.10582733bp}%
    \ifx\svgscale\undefined%
      \relax%
    \else%
      \setlength{\unitlength}{\unitlength * \real{\svgscale}}%
    \fi%
  \else%
    \setlength{\unitlength}{\svgwidth}%
  \fi%
  \global\let\svgwidth\undefined%
  \global\let\svgscale\undefined%
  \makeatother%
  \begin{picture}(1,1.15477199)%
    \lineheight{1}%
    \setlength\tabcolsep{0pt}%
    \put(0,0){\includegraphics[width=\unitlength,page=1]{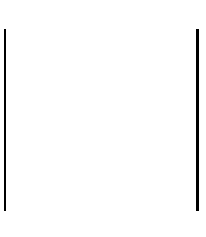}}%
    \put(0.92872796,0.00310412){\color[rgb]{0,0,0}\makebox(0,0)[lt]{\lineheight{1.25}\smash{\begin{tabular}[t]{l}1\end{tabular}}}}%
    \put(0.00079423,0){\color[rgb]{0,0,0}\makebox(0,0)[lt]{\lineheight{1.25}\smash{\begin{tabular}[t]{l}2\end{tabular}}}}%
    \put(-0.01034707,1.05700145){\color[rgb]{0,0,0}\makebox(0,0)[lt]{\lineheight{1.25}\smash{\begin{tabular}[t]{l}3\end{tabular}}}}%
    \put(0.45476002,1.0591638){\color[rgb]{0,0,0}\makebox(0,0)[lt]{\lineheight{1.25}\smash{\begin{tabular}[t]{l}4\end{tabular}}}}%
    \put(0,0){\includegraphics[width=\unitlength,page=2]{reggeons1.pdf}}%
    \put(0.928314,1.06251576){\color[rgb]{0,0,0}\makebox(0,0)[lt]{\lineheight{1.25}\smash{\begin{tabular}[t]{l}5\end{tabular}}}}%
    \put(0,0){\includegraphics[width=\unitlength,page=3]{reggeons1.pdf}}%
    \put(0.21079062,0.6559614){\makebox(0,0)[lt]{\lineheight{1.25}\smash{\begin{tabular}[t]{l}$t_1$\end{tabular}}}}%
    \put(0.67445408,0.65596101){\makebox(0,0)[lt]{\lineheight{1.25}\smash{\begin{tabular}[t]{l}$t_2$\end{tabular}}}}%
  \end{picture}%
\endgroup%
  \hspace{15mm}
    \def\svgwidth{0.2\columnwidth}%% Creator: Inkscape 1.0beta1 (32d4812, 2019-09-19), www.inkscape.org
%% Accompanies image file 'reggeons2.pdf' (pdf, eps, ps)
\begingroup%
  \makeatletter%
  \providecommand\color[2][]{%
    \errmessage{(Inkscape) Color is used for the text in Inkscape, but the package 'color.sty' is not loaded}%
    \renewcommand\color[2][]{}%
  }%
  \providecommand\transparent[1]{%
    \errmessage{(Inkscape) Transparency is used (non-zero) for the text in Inkscape, but the package 'transparent.sty' is not loaded}%
    \renewcommand\transparent[1]{}%
  }%
  \providecommand\rotatebox[2]{#2}%
  \newcommand*\fsize{\dimexpr\f@size pt\relax}%
  \newcommand*\lineheight[1]{\fontsize{\fsize}{#1\fsize}\selectfont}%
  \ifx\svgwidth\undefined%
    \setlength{\unitlength}{58.60423206bp}%
    \ifx\svgscale\undefined%
      \relax%
    \else%
      \setlength{\unitlength}{\unitlength * \real{\svgscale}}%
    \fi%
  \else%
    \setlength{\unitlength}{\svgwidth}%
  \fi%
  \global\let\svgwidth\undefined%
  \global\let\svgscale\undefined%
  \makeatother%
  \begin{picture}(1,1.06604835)%
    \lineheight{1}%
    \setlength\tabcolsep{0pt}%
    \put(0,0){\includegraphics[width=\unitlength,page=1]{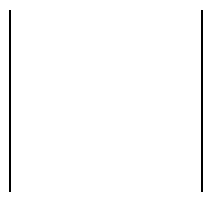}}%
    \put(0.01123666,0){\color[rgb]{0,0,0}\makebox(0,0)[lt]{\lineheight{1.25}\smash{\begin{tabular}[t]{l}   \end{tabular}}}}%
    \put(0,1.06604835){\color[rgb]{0,0,0}\makebox(0,0)[lt]{\lineheight{1.25}\smash{\begin{tabular}[t]{l}   \end{tabular}}}}%
    \put(0,0){\includegraphics[width=\unitlength,page=2]{reggeons2.pdf}}%
  \end{picture}%
\endgroup%
  \hspace{15mm}
    \def\svgwidth{0.2\columnwidth}%% Creator: Inkscape 1.0beta1 (32d4812, 2019-09-19), www.inkscape.org
%% Accompanies image file 'reggeons3.pdf' (pdf, eps, ps)
\begingroup%
  \makeatletter%
  \providecommand\color[2][]{%
    \errmessage{(Inkscape) Color is used for the text in Inkscape, but the package 'color.sty' is not loaded}%
    \renewcommand\color[2][]{}%
  }%
  \providecommand\transparent[1]{%
    \errmessage{(Inkscape) Transparency is used (non-zero) for the text in Inkscape, but the package 'transparent.sty' is not loaded}%
    \renewcommand\transparent[1]{}%
  }%
  \providecommand\rotatebox[2]{#2}%
  \newcommand*\fsize{\dimexpr\f@size pt\relax}%
  \newcommand*\lineheight[1]{\fontsize{\fsize}{#1\fsize}\selectfont}%
  \ifx\svgwidth\undefined%
    \setlength{\unitlength}{58.60423206bp}%
    \ifx\svgscale\undefined%
      \relax%
    \else%
      \setlength{\unitlength}{\unitlength * \real{\svgscale}}%
    \fi%
  \else%
    \setlength{\unitlength}{\svgwidth}%
  \fi%
  \global\let\svgwidth\undefined%
  \global\let\svgscale\undefined%
  \makeatother%
  \begin{picture}(1,1.06604835)%
    \lineheight{1}%
    \setlength\tabcolsep{0pt}%
    \put(0,0){\includegraphics[width=\unitlength,page=1]{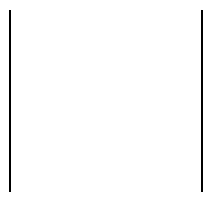}}%
    \put(0.01123666,0){\color[rgb]{0,0,0}\makebox(0,0)[lt]{\lineheight{1.25}\smash{\begin{tabular}[t]{l}   \end{tabular}}}}%
    \put(0,1.06604835){\color[rgb]{0,0,0}\makebox(0,0)[lt]{\lineheight{1.25}\smash{\begin{tabular}[t]{l}    \end{tabular}}}}%
    \put(0,0){\includegraphics[width=\unitlength,page=2]{reggeons3.pdf}}%
  \end{picture}%
\endgroup%

    \caption{Family of reggeon diagrams
    in which a maximal number of reggeons (horizontal lines) is exchanged in both the $t_1$ and $t_2$ channels.
    We also include all permutations of the insertions along the vertical (eikonal) lines.}
    \label{fig:max reggeons}
  \end{center}
\end{figure}

The two-loop $(\textbf{10},\textbf{10})$ amplitude is part of a family where
the maximal number of reggeons is forced to flow in each channel, as shown in Fig.~\ref{fig:max reggeons}.
(For colour-adjoint external states this family effectively terminates at two loops.
To force more reggeons one would have to scatter external particles in larger colour representations.)

At tree level, the analogous amplitude is $(\textbf{8}_a,\textbf{8}_a)$, where a single reggeon is exchanged, coming from the $|W\rangle$
term in Eq.~\eqref{eq: leading WF}.
The overlap in Eq.~\eqref{A5 BFKL} then gives the so-called Lipatov vertex, which gives for example for the external helicity configuration $1^-2^- 3^+4^+5^+$ (See Section 5.2 of Ref.~\cite{Caron-Huot:2013fea} for its computation in the present formalism, and for a review of the rules
used below):
\begin{equation}
\delta^{(8)}(Q) A_5^{(0)}\bigg|_{(\textbf{8}_a,\textbf{8}_a)}
\propto \frac{s_{12}}{{\bf \bar p}_{3} {\bf \bar p}_4 {\bf p}_{5}} \, , \label{A5 regge}
\end{equation}
which agrees with the Regge limit of the Parke-Taylor factor PT${}_1$ times the supersymmetric $\delta$-function
in Eq.~\eqref{eq: def A5}; our Regge kinematics is defined in Section~\ref{sec:MultiReggeLimit}.

The analogous loop amplitudes in Fig.~\ref{fig:max reggeons}(b) can be easily written down by means of the following rules.
Please note that these pictures are not Feynman diagrams in the full four-dimensional theory but rather they follow the rules
of a two-dimensional-like effective theory where only the transverse momenta are integrated over in loops.
Each horizontal propagator gives a factor $\frac{i}{|{\bf p}^2|}$, the central emission vertex
gives a factor $\frac{{\bf p}_L {\bf \bar{p}}_R}{{\bf \bar{p}}_4}$ (where ${\bf p}_{L/R}$ denote the transverse momenta
entering from the left and the right), and the vertical lines do nothing except to track group theory
factors $ig T^a$ inserted at each vertex (in accordance with the wavefunction in Eq.~\eqref{eq: leading WF})).
Converting the group theory factors to the trace basis,
these diagrams allow to predict the following colour components:
\begin{equation}\begin{aligned} 
 A^{(0)}_3 &= {\rm PT}_1\, , \\
 A^{(1)}_{\{9;12;15\}} &= {\rm PT}_1 \times -i\pi
 \left\{\frac{-N_c^3}{2(N_c^2-4)};\frac{1}{N_c-2};\frac{1}{N_c+2}\right\}
 I_{\rm tri}^{(1)}\, , \\
 A^{(2)}_{\{21; 22\}} &={\rm PT}_1 \times  2\pi^2 \left\{0 ; 1\right\} I_{\rm tri}^{(2)}\, , \label{regge integrand}
\end{aligned}\end{equation}
where the subscripts refer to the colour factors in Table \ref{tab:Representations}.
(For the reader's convenience, these are, in order of appearance:
$(\textbf{8}_a, \textbf{8}_a);(\textbf{8}_s, \textbf{8}_s); (\textbf{0}, \textbf{0}); (\textbf{27}_a, \textbf{27})$ and the two
$(\textbf{10},\textbf{10})$'s, defined explicitly in Eq.~\eqref{eq: few S}).
In particular, at two loops, just by doing colour algebra and without doing any integral, BFKL theory
predicts that one component vanishes: $A^{(2)}_{21}=0$, in agreement with the preceding section!

The nonzero components involve the following 2d triangle integrals:
\begin{equation}\begin{aligned} \label{eq: 2d integrals}
I_{\rm tri}^{(1)} &= \int \frac{d^{2-2\eps}{\bf \ell}}{\pi^{1-\eps} e^{-\eps\gamma}}
\ \frac{{\bf \bar p_3}({\bf p}_3-{\bf \ell})({\bf \bar p}_5+{\bf \bar \ell}) {\bf  p}_5}
 {(p_3-\ell)^2(p_5+\ell)^2\ell^2}\ ,
\\
I_{\rm tri}^{(2)} &= \int \frac{d^{2-2\eps}{\bf \ell}_1}{\pi^{1-\eps} e^{-\eps\gamma}} \int \frac{d^{2-2\eps}{\bf \ell}_2}{\pi^{1-\eps} e^{-\eps\gamma}}
 \ \frac{{\bf \bar p_3}({\bf p}_3-{\bf \ell}_1)({\bf \bar p}_5+{\bf \bar \ell}_1) {\bf  p}_5}
 {(p_3-\ell_1)^2(p_5+\ell_1)^2(\ell_1-\ell_2)^2\ell_2^2}\ .
\end{aligned}\end{equation}
The predictions of BFKL theory are thus phrased in terms of only 2d transverse-plane integrals,
which is one of the advantages of this effective theory.
The integrals may be computed as series in $\eps$, however we prefer to first
apply the infrared factorization procedure of section \ref{sec:IRsYM} and deal directly with the $\eps\to 0$ limit of the hard functions.

\subsection{Infrared-factorized amplitude}

To understand how the infrared singularities organize in the Regge limit, it is helpful to first analyze the Regge limit
of the dipole formula for the soft anomalous dimension (see Eq.~\eqref{eq:dipole}), following \cite{DelDuca:2011ae}:
\begin{equation}\begin{aligned} \label{eq: dipole Regge}
-\tfrac14\mathbf{\Gamma}_5^{(1)} &= 
\sum_{i<j} \left(\textbf{T}^i{\cdot}\textbf{T}^j\right) \log\left( \frac{-s_{ij}}{\mu^2}\right) 
= \textbf{T}^{2}{\cdot}\textbf{T}^3 \log \left( \frac{-t_1}{\mu^2} \right)
+ \textbf{T}^{1}{\cdot}\textbf{T}^5 \log \left( \frac{-t_2}{\mu^2}\right)
 \\
&\phantom{=}
+(\textbf{T}^2{+}\textbf{T}^3){\cdot}\textbf{T}^{4}L_{34}
+\textbf{T}^{4}{\cdot}(\textbf{T}^1{+}\textbf{T}^5)L_{45}
+(\textbf{T}^2{+}\textbf{T}^3){\cdot}(\textbf{T}^1{+}\textbf{T}^5) L_{12}
\\
&\phantom{=}
+\tfrac{i\pi}{2}
 (\textbf{T}^{2}{-}\textbf{T}^{3}){\cdot}\textbf{T}^{4}
+\tfrac{i\pi}{2}
\textbf{T}^{4}{\cdot}(\textbf{T}^{1}{-}\textbf{T}^{5})
-\tfrac{i\pi}{2}
(\textbf{T}^{2}{-}\textbf{T}^3){\cdot}(\textbf{T}^1{-}\textbf{T}^5)\, ,
\end{aligned}\end{equation}
where $L_{ij} = \log \left( \frac{|s_{ij}|}{\mu^2}\right) -\tfrac{i\pi}{2}$ is a signature-even combination.
Here we have restricted to the kinematics where particles $3,4,5$ are outgoing.

We have organised the terms in Eq.~\eqref{eq: dipole Regge} according to their signature in the $t_1$ and $t_2$ channels, 
that is, their properties under swapping particles 2 and 3,  or 1 and 5, respectively.
(For example, the last two terms are odd under the $t_2$ signature while all others are even.
This means that,  if we were to flip the energies of particles $1$ and $5$ to make 5 incoming, all that would change is that
those two terms would flip sign.)

The key point to observe is that all the terms in the first two lines commute with the Casimirs ${\bf T}_{t_1}^2$
and ${\bf T}_{t_2}^2$: they do not modify the colour representation which is exchanged in the $t_1$ and $t_2$-channel.
Thus, when working in the colour basis of Table.~\ref{tab:Representations}, the renormalization operator $\textbf{Z}_5$
is essentially diagonal, except for the three terms on the last line of Eq.~\eqref{eq: dipole Regge}.
This dramatically simplifies the study of infrared factorisation in the Regge limit.

In particular, when focusing on the maximal-reggeon colour structures of the preceding subsection,
only the \emph{last term} in Eq.~\eqref{eq: dipole Regge} needs to be kept: all the others project
onto colour structures which vanish at the lower loop order!  Denoting this term as
\begin{align}
\mathbf{\Gamma}_{\rm max} = 2i\pi(\textbf{T}^{2}{-}\textbf{T}^3){\cdot}(\textbf{T}^1{-}\textbf{T}^5)\, ,
\end{align}
the Regge limit of the maximum-reggeon colour component of the hard function  is
\begin{align}
 H_{\rm max} =
 \lim_{\eps\to 0}
 \left[1-\frac{a}{2\eps}\mathbf{\Gamma}_{\rm max} + \frac{a^2}{8\eps^2}\left(\mathbf{\Gamma}_{\rm max}\right)^2\right] \frac{A_{\rm max}}{{\rm PT}_1}\,, \label{eq: regge renormalization}
\end{align}
where $H$ is defined in Eq.~\eqref{eq: def H} and the right-hand-side involves the amplitudes given
in Eq.~\eqref{regge integrand}.
The action of the colour operators can be worked out by expressing the bare amplitude $A$ in the trace basis,
where the $\textbf{T}$ operators can be computed using standard $SU(N_c)$ trace identities,
and converting back to the colour-flow basis. In this way we find:
\begin{equation}\begin{aligned}
(\textbf{T}^{2}{-}\textbf{T}^3){\cdot}(\textbf{T}^1{-}\textbf{T}^5) \circ \mathcal{S}_3 &= 
\frac{-N_c^3}{2(N_c^2-4)}\mathcal{S}_9 + \frac{1}{N_c-2}\mathcal{S}_{12} + \frac{1}{N_c+2}\mathcal{S}_{15}\, ,\\
\left((\textbf{T}^{2}{-}\textbf{T}^3){\cdot}(\textbf{T}^1{-}\textbf{T}^5)\right)^2 \circ \mathcal{S}_3 &= 
\frac{N_c^2+16}{4} \mathcal{S}_3 + \mathcal{S}_7+\mathcal{S}_{17} -4\mathcal{S}_{22}\,.
\end{aligned}\end{equation}
Notice that the first line
involves the same combination as in Eq.~\eqref{regge integrand}, and that $\mathcal{S}_{21}$ is absent from the second line.
We thus find, for the above colour components of the hard functions:
\begin{equation}\begin{aligned} 
H^{(0)}_3 &= 1\, , \\
H^{(1)}_{\{9;12;15\}} &= -i\pi \left\{\frac{-N_c^3}{2(N_c^2-4)};\frac{1}{N_c-2};\frac{1}{N_c+2}\right\}
 \times \left( I_{\rm tri}^{(1)} +\frac{1}{\epsilon}\right)\, ,\\
H^{(2)}_{\{21;22\}} &= 2\pi^2 \left\{0;1\right\}\times\left(I_{\rm tri}^{(2)} +\frac{2}{\eps}I_{\rm tri}^{(1)} +
\frac{1}{\eps^2}\right)\, . \label{regge integrand R}
\end{aligned}\end{equation}
Before presenting the analogous two-loop result, it is useful to observe the following:
the parentheses are simply the result of the
infrared $R$-operation \cite{Chetyrkin:1982nn,Chetyrkin:1984xa,Larin:2002sc,Herzog:2017bjx} applied to the integrand!

The IR $R$-operation is conceptually similar to the perhaps more
familiar ultraviolet $R$-operation, which recursively collapses divergent subgraphs of a Feynman integral and replaces them with local counterterms.  The infrared $R$-operation recursively subtracts infrared subdivergences by \emph{erasing} (as opposed to collapsing) IR-divergent sub-integrals, replacing them with the infrared divergent part of corresponding integrals. (The IR and UV $R$-operations commute, and are often combined into a single $R^*$.)
The IR $R$-operation enables a clean distinction between IR and UV divergences and has been used for
higher-loop calculations of the QCD $\beta$-functions and UV divergences of on-shell scattering amplitudes \cite{Bern:2010tq}.
Applying it to the integrands in Eq.~(\ref{eq: 2d integrals}), we find
\begin{equation}\begin{aligned}
 R\left[I_{\rm tri}^{(1)}\right] &= I_{\rm tri}^{(1)} + \frac{1}{\epsilon}\, , \\
 R\left[I_{\rm tri}^{(2)}\right] &= I_{\rm tri}^{(2)} + \frac{2}{\epsilon} I_{\rm tri}^{(1)} + \frac{1}{\epsilon^2}\, . \label{R operation}
\end{aligned}\end{equation}
These are precisely the combinations appearing in Eq.~\eqref{regge integrand R}.

Physically, it is perhaps not surprising that the infrared $R$-operation correctly predicts the hard function,
because its combinatorics are similar to those which underly infrared factorization. We thus expect the
$R$ operation (or a suitable generalisation which can deal with collinear divergences) to correctly predict
the hard function even away from the Regge limit. It could be interesting to study this more, notably because
the $R$-operation may be computable graph by graph.

A convenient way to evaluate the one-loop triangle integral is to expand the numerator
over a standard basis of (parity even and odd)
bubble integrals and a dimension-shifted scalar triangle (the triangle with numerator $\ell_\perp^2$ where $\ell_\perp$ denotes
the $(-2\eps)$-dimensional components of $\ell$, which we take to be orthogonal to ${\bf p}_3$ and ${\bf p}_5$).
We omit intermediate steps and outline the calculation.
Since the latter integrates to $O(\eps)$, it can be dropped, as well as the odd bubbles, and the surviving bubbles all come with coefficient $\pm \tfrac12$. We obtain the following simple result:
\begin{align}
 R[I_{\rm tri}^{(1)}] =\log\left(\frac{\big|{\bf p}_3^2{\bf p}_5^2\big|}{\mu^2\big|{\bf p}_4^2\big|}\right)\ .
\end{align}
We stress that this calculation is entirely algebraic and we did need to make use of integration-by-parts identities.
Now for the two-loop triangle, we can first integrate out $\ell_2$, which leaves a one-loop triangle with a non-standard propagator $1/\left(\eps(\ell^2)^{1+\eps}\right)$. While the same algebraic reorganisation carries through,
the result would now seem much more complicated since some of the odd bubbles are now still triangles, and the odd one does not vanish. However, thanks to the $R$-operation, we find that the result neatly reduces to bubbles plus the dimension-shifted triangle:
\begin{align}
 R[I_{\rm tri}^{(2)}] =\log^2 \left( \frac{\big|{\bf p}_3^2{\bf p}_5^2\big|}{\mu^2\big|{\bf p}_4^2\big|}\right)
 -\log \left( \frac{\big|{\bf p}_3^2\big|}{\big|{\bf p}_4^2\big|}\right) \log \left( \frac{\big|{\bf p}_5^2\big|}{\big|{\bf p}_4^2\big|} \right)
 -2D_2\left(-\frac{{\bf p}_3}{{\bf p}_4},-\frac{\bar{\bf p}_3}{\bar{\bf p}_4}\right)\,, 
\label{R triangles} \end{align}
where the last term, coming from the dimension-shifted triangle,
is the Bloch-Wigner dilogarithm in Eq.~\eqref{eq:Li2BlochWigner}.

It is important to stress that the $R$-subtracted two-loop integral is \emph{not} simply
the ``finite term" in the $\eps\to0$ Laurent series of the integral, which would be much more complicated. In particular, functions with the letter $(z-\zbar)$ in their symbol cancel between the two-loop integral and the $O(\epsilon)$ term at one loop.
This parallels the cancellation of the letter $W_{31}$ observed for the hard functions in Section~\ref{sec:HardFunctions}.

Converting to the $g$ functions from Section \ref{sec:sYMfunctions}, where $z=-{\bf p}_3/{\bf p}_4$,
the result in Eq.~\eqref{regge integrand R} can thus be written as
\begin{equation}\begin{aligned}
& H^{(1)}_{\{9;12;15\}} = i\pi  
 \left\{\frac{-N_c^3}{2(N_c^2-4)};\frac{1}{N_c-2};\frac{1}{N_c+2}\right\} \left(g^{(1)}_3-g^{(1)}_4-g^{(1)}_5\right) \, , \\
& H^{(2)}_{\{21; 22\}} = 2 \pi^2 \left\{0 ; 1\right\} \left(
\left(g^{(1)}_{3}-g^{(1)}_{4}-g^{(1)}_{5}\right)^2 - g^{(1)}_{4}g^{(1)}_{5}- 2 g^{(2)}_1 \right)\,, \label{bfkl H}
\end{aligned}\end{equation}
in complete agreement with the results of the direct computation discussed in Section~\ref{sec:sYM}.

\subsection{Examples of non real-analytic amplitudes}
\label{sec: bfkl sing}

We now analyse the effect of the one-loop impact factors.
These do not affect the two-loop components recorded in Eq.~\eqref{sec:sYM}, but
appear in all colour structures where two-reggeon states can propagate ($\textbf{8}_s, \textbf{0}$ and $\textbf{27}$ according to
the second line of Eq.~\eqref{WW colors}).

A striking feature of the impact factors computed in \cite{Fadin:1999de}
is that they contain terms which are not real-analytic as a function of angles: their derivatives are discontinuous.
We extract the $\mathcal{N}=4$ super Yang-Mills result by taking the leading transcendental part
of their formulas.
The unique non-real-analytic contribution comes from $\frac{2}{\eps}K_2$ in $I_4$ in their Eq.~(5.27).
It gives, in our notation (using also Eq.~(3.12) there), the following $O(a)$ wavefunction in Eq.~\eqref{eq: leading WF}:
\begin{align} \label{WW NLO}
  |R\rangle\Big|_{WW} = -g^2p_5^+ \int_{\ell}
 |W^a_{{\bf p}_5+\ell}W^b_{-\bf \ell}\rangle\left( (T^aT^b)_{i_5i_1} + \tilde{a}\left[(T^c T^d)_{a i_1}(T^c T^d)_{i_5b}
 \frac{4K_2}{\eps} +\mbox{(analytic)}\right]\right),
\end{align}
where $\tilde{a}=a\frac{\Gamma(1-\eps)\Gamma^2(1+\eps)}{\Gamma(1+2\eps)e^{\eps \gamma_E}}$ is a factor in their equation,
and
\begin{equation}\begin{aligned}
 \frac{K_2}{\eps} &\equiv \frac{1}{\eps}\int_0^1 \frac{d\beta}{\beta(1-\beta)}\left(
 \left|\beta({\bf p}_5+\ell)-(1-\beta)\ell\right|^{2\eps} - (1-\beta)^{2\eps} |\ell|^{2\eps}- \beta^{2\eps} |{\bf p}_5+\ell|^{2\eps}\right)
 \\ &= -\frac{1}{\eps^2} -\frac{1}{\eps}\log|\ell({\bf p}_5+\ell)| + \frac{2\pi^2}{3} -2\log|\ell|\log|{\bf p}_5+\ell| -\phi^2.
\end{aligned}\end{equation}
The variable $\beta$ represents the energy fraction of a projectile parton.
The offending term is the last one:
$\phi= \arccos\left(\tfrac{\ell{\cdot}(\ell+{\bf p}_5)}{|\ell(\ell+{\bf p}_5)|}\right)\in [0,\pi]$,
whose first derivative jumps when the two reggeons are parallel.

Notice that a nonplanar colour factor multiplies $K_2$ in Eq.~\eqref{WW NLO}.
This is in agreement with the absence of such contributions in the planar impact factor, first discussed in \cite{Lipatov:2010ad}.

Since a 2d integral over a real-analytic function is real-analytic, we expect all non-analyticity in the final answer
to originate from impact factor corrections.
These include corrections to the vertex for emission of ${\bf p}_4$, see for example \cite{Kozlov:2012zz}.
However, for colour structures with $WW$ in each channel, we may argue that these should be real-analytic.
(There is both a loop correction to the vertex at the center of Fig.~\ref{fig:max reggeons}, which is a planar ingredient hence safe,
as well as a connected correction involving all reggeons, which should be safe being at this order
the expansion of a single Wilson line as in Eq.~\eqref{eq: leading WF}.)

Working out the colour algebra, we thus find that the impact factor in Eq.~(\ref{WW NLO}) predicts
\begin{align} \label{H2 WW}
H^{(2)}_{\{9;12;15\}} = -2\pi i\left\{0,1,1\right\} \times \left( I^{(1)'}_{\rm tri} + (z{\leftrightarrow}1-\zbar)\right) + \mbox{(analytic)} \, ,
\end{align}
where the symmetrisation accounts for the left- and right- impact factors.
The triangle integral is as before but with an extra $-\phi^2$ inserted at the vertex, which can also be written as follows:
\begin{equation}\label{Itri prime}\begin{aligned}
I^{(1)'}_{\rm tri} &=
\int \frac{d^{2-2\eps}{\bf \ell}}{\pi^{1-\eps} e^{-\eps\gamma}}
\ \frac{{\bf \bar p_3}({\bf p}_3-{\bf \ell})({\bf \bar p}_5+{\bf \bar \ell}) {\bf  p}_5}
 {\big|({\bf p}_3-{\bf \ell})^2({\bf p}_5+{\bf \ell})^2\ell^2\big|} \times
 \frac14 \left[\log\left(\frac{\ell}{\ell+{\bf p}_5}\right)-\log\left(\frac{\bar\ell}{\bar\ell+\bar{\bf p}_5}\right)\right]^2
\\ &= \frac1{12} (\log(z)-\log(\zbar))^3  + \frac{\pi^2}{3} (\log(z)-\log(\zbar))^2 +\mbox{(analytic)}.
\end{aligned}\end{equation}
This is non real analytic for ${\rm Re}[z]<0$: the second derivative of the amplitude jumps when the transverse momenta of
particles ${\bf p}_3$ and ${\bf p}_4$ become parallel.
Eqs.~\eqref{H2 WW} and \eqref{Itri prime} are in precise agreement with the explicit four-dimensional calculation in
Eq.~\eqref{eq: singular 12}.
The present calculation allows us to track this effect to integration regions
involving highly energetic collinear particles inside the loop.

In summary, we have found precise agreement between the methods: both for the non-analytic terms for two-reggeon exchange,
and for the three-reggeon exchanges in the preceding subsection.
We view this as a highly nontrivial check of the full five-point amplitude, including the integration constants,
as well as a nontrivial confirmation of the BFKL framework, including the impact factor of \cite{Fadin:1999de}.
It is interesting that non-planar amplitudes appear much richer than their planar limit,
where the five-particle amplitude is essentially trivial \cite{DelDuca:2008jg} and only single-valued polylogarithms
are believed to appear for any number of legs \cite{DelDuca:2016lad}.

\section{The multi-Regge limit of the $\mathcal{N}=8$ supergravity amplitude}
\label{sec:SUGRA}

The analysis of the multi-Regge limit of the two-loop five-graviton amplitude in $\mathcal{N}=8$ supergravity was initiated in Ref.~\cite{Chicherin:2019xeg}. There, the symbol of the hard function in the multi-Regge limit was provided, up to and including the finite terms as $x\to 0$. Moreover, the leading logarithmic part of the hard function, of order $\log^4 x$, was provided analytically. Here we lift the entire analysis to function level.

The multi-Regge limit of the $\mathcal{N}=8$ supergravity amplitude presents several novel features with respect to the $\mathcal{N}=4$ super Yang-Mills one. 
First of all, the supergravity amplitude exhibits a much more complicated structure of the kinematic factors. In order to see this, we first need to extract an overall factor to cancel the helicity. We choose to normalise the amplitudes and the hard functions by $\PT_1^2$,
\begin{align}
\label{eq:SUGRAnormalisation}
\widetilde{\mathcal{M}}_5^{(\ell)} = \frac{\mathcal{M}_5^{(\ell)}}{\PT_1^2} \,, \qquad \widetilde{\mathcal{F}}_5^{(\ell)} = \frac{\mathcal{F}_5^{(\ell)}}{\PT_1^2} \,.
\end{align}
With this normalisation, the tree-level five-graviton amplitude given by Eq.~\eqref{eq:M0sugra} is finite in the multi-Regge limit,
\begin{align}
\widetilde{\cal M}^{(0)}_5 = -\frac{s_1^2 s_2^2 }{s^2} z(1-\bar z)(z-\bar z) + {\cal O}(x)\,.
\end{align}
However, the rational functions in the supergravity amplitudes diverge in general as $1/x^2$ at one loop and as $1/x^4$ at two loops. This has two important consequences. First, the expansion of the hard function is a double series $x^{-m}\log^k(x)$. Second, in order to determine the leading contributions in the limit $x\to 0$ we need to take into account the power corrections of the pentagon function, given by the gauge transformation matrix in Eq.~\eqref{eq:AsymptoticSolution}. The presence of the power corrections imply a drop in transcendentality. In other words, the asymptotics of the hard function is not of uniform and maximal weight, but contains lower weight functions and rational terms as well. 

One might suspect that this feature is an artifact of the chosen normalisation, but it is in fact inevitable. We want a common normalisation factor at all loop orders, but the dimensionality of the gravitational coupling constant implies that the rational functions have different dimensions at each loop order. As a result, the rational functions diverge as $1/x^4$ at two loops and as $1/x^2$ at one loop with the normalisation~\eqref{eq:SUGRAnormalisation}. Any correction to the latter which would make the two-loop rational functions finite in the limit $x\to 0$ would also make the one-loop ones vanish, and would not cure the three-loop ones.

On the other hand it is true that the normalisation~\eqref{eq:SUGRAnormalisation} is completely arbitrary. It is however motivated. 
We want to cancel the helicity with the same normalisation at all loop orders. The minimal choice of normalisation is given by a product of two Parke-Taylor factors~\eqref{eq:PTdefinition}. It turns out that the choice made in Eq.~\eqref{eq:SUGRAnormalisation} is a particularly good one. Most other products of Parke-Taylor factors develop higher poles in the limit. No product of two Parke-Taylor factors can improve the behaviour at $x\to 0$ of the rational functions with respect to Eq.~\eqref{eq:SUGRAnormalisation}. 

Another new feature related to the rational functions is that some of them are singular at $z = \zbar$ in the multi-Regge limit. This follows from the necessity of taking into account power corrections in the pentagon functions. As we discussed at the end of Section~\ref{sec:PentagonMR}, certain non-planar integrals have discontinuities at $\eps_5 = 0$, corresponding to $z = \zbar$ in the multi-Regge kinematics. The five-particle hard function in $\mathcal{N}=4$ super Yang-Mills moreover is not real analytic across $\text{Im}[z]=0$: its second derivatives are discontinuous. If we were to take into account subleading power corrections, therefore, we would encounter rational functions singular at $z = \zbar$ also in the $\mathcal{N}=4$ super Yang-Mills case. Again, this cannot be fixed by changing the normalisation Eq.~\eqref{eq:SUGRAnormalisation}. We checked that the only way to cancel the poles at $z = \zbar$ in our result is the trivial one, namely to multiply it by some power of $(z-\zbar)$. If we were to include higher power corrections in the result, however, higher and higher poles at $z=\zbar$ whould show up, so that there is no overall fixed power of $(z-\zbar)$ that would remove this divergence entirely.

In the next sections we discuss the multi-Regge asymptotics of the hard function at one and two loops. We carry out the computation separately in the upper and in the lower half of the complex plane, as described in Section~\ref{sec:PentagonMR}. We provide the explicit expressions in both regions in ancillary files, written in terms of the transcendental functions classified in Sections~\ref{sec:sYMfunctions} and~\ref{sec:SUGRAfunctions}.

\subsection{One-loop hard function}

The one-loop leading singularities $r^{(1)}_i$, normalised by $\PT_1^2$ to cancel the helicity, are in general divergent in the multi-Regge limit as
\begin{align}
\frac{r^{(1)}_i}{\PT_1^2} \underset{ x \to 0 }{\sim} \mathcal{O}\left(\frac{1}{x^2}\right)\, .
\end{align}
As a consequence, we need to take into account power corrections up to $x^2$ in the asymptotic expansion of the pentagon functions~\eqref{eq:AsymptoticSolution}.

The multi-Regge limit of the normalised hard function at one loop then takes to form
\begin{align}
\widetilde{{\cal F}}^{(1)}_5   =  \sum_{m=0}^2 \sum_{k=0}^2 \frac{1}{x^m}\log^k(x)\, F^{(1)}_{m,k}(s_1,s_2,s,z,\zbar) + o(1)  \, , \label{eq:1loopHardsugra}
\end{align}
where we omit the infinitesimal terms, i.e. those of order $x^m \log^k(x)$ with $m>0$ and $k \geq 0$. The coefficient functions $F^{(1)}_{m,k}$ contain a mixture of rational functions of $s, s_1, s_2, z$ and $\zbar$, and transcendental functions with up to weight 2. Interestingly, the latter involve only the single-valued logarithms given by Eq.~\eqref{eq:funW1sv}. As a consequence, the expression for the multi-Regge limit of the one-loop hard function is straightforwardly valid in the whole complex plane $z\in \mathbb{C}$. 
For instance, the leading power term, $F_{2,0}^{(1)}$, is given by
\begin{align}
\label{eq:F120}
F_{2,0}^{(1)} = 2 i \pi \frac{ s_1^2 s_2^2 z (1-\zbar) }{s} \left[ g^{(1)}_5+ z \left(g^{(1)}_3 - g^{(1)}_5 \right) + \zbar \left(g^{(1)}_4-g^{(1)}_3\right)\right] \, .
\end{align}
We comment on the other terms proceeding by powers of $\log x$.

Unlike the gauge theory case, the one-loop hard function in supergravity has a $\log^2 x$ term. It appears at power-suppressed order ($\mathcal{O}(x^0)$), so $F^{(1)}_{2,2}=0$ and $F^{(1)}_{1,2}=0$, and is multiplied by the rational function
\begin{align}
F^{(1)}_{0,2} = \frac{2 s_1^3 s_2^3 z (1 - \zbar) (3 z - 3 z^2 - \zbar + 4 z^2 \zbar + \zbar^2 - 4 z \zbar^2)}{s^3}\, .
\end{align}
Also $\log^1 x$ is present only at order $\mathcal{O}(x^0)$, namely $F^{(1)}_{2,1}=0$ and $F^{(1)}_{1,1}=0$. Its coefficient, $F^{(1)}_{0,1}$, contains both a rational and a weight-1 term. 
The order-$\log^0 x$ part of the hard function is the most complicated. We have already shown in Eq.~\eqref{eq:F120} the leading power term, $F^{(1)}_{2,0}$. $F^{(1)}_{0,0}$ and $F^{(1)}_{1,0}$ contain a mixture of terms of different transcendentalities, from 0 to 2. 
We find it useful to summarise the transcendentality structure of the result as
\begin{equation}
\label{eq:ReggeSugra1}
\begin{aligned}
F^{(1)}_{2,2} &= 0\, , & F^{(1)}_{1,2} &= 0\,, & F^{(1)}_{0,2} &=  {\cal Q}_{0,2}^{(0)} \,, \\
F^{(1)}_{2,1} &= 0 \,, & F^{(1)}_{1,1} &= 0\,, & F^{(1)}_{0,1} &=  {\cal Q}_{0,1}^{(1)} + i\pi {\cal Q}_{0,1}^{(0)} + {\cal Q}_{0,1}^{(0)'} \,, \\
F^{(1)}_{2,0} &= i\pi {\cal Q}^{(1)}_{2,0} \, , & F^{(1)}_{1,0} &= i\pi {\cal Q}_{1,0}^{(1)} + i\pi {\cal Q}_{1,0}^{(0)} \,  , & F^{(1)}_{0,0} &= {\cal Q}_{0,0}^{(2)} + i\pi {\cal Q}_{0,0}^{(1)} + {\cal Q}_{0,0}^{(1)'} + i\pi {\cal Q}_{0,0}^{(0)} + {\cal Q}_{0,0}^{(0)'} \,,
\end{aligned}
\end{equation}
where $\mathcal{Q}^{(w)}$ schematically denotes a uniform weight-$w$ function containing rational functions of $s, s_1, s_2, z$ and $\zbar$.  Such equations thus show which components vanish and what transcendental weight the others have.
We provide the explicit expressions in ancillary files.

Finally, it is worth noting that the one-loop hard function is singular on the real axis $z=\zbar$ due to its rational factors. In particular, $F^{(1)}_{1,0}$ and $F^{(1)}_{0,0}$ are divergent. $\mathcal{Q}_{1,0}^{(1)}$ and $\mathcal{Q}_{0,0}^{(1)}$ have a simple pole, and $\mathcal{Q}_{0,0}^{(1)}$ has a third-order pole at $z=\zbar$. These poles are not canceled upon series expansion of the transcendental functions around $z = \zbar$.

\subsection{Two-loop hard function}

Similarly to the one-loop case, the two-loop rational factors $r^{(2)}_i$, given by Eqs.~\eqref{eq:r2_1} and~\eqref{eq:r2_41}, are singular in the multi-Regge limit,
\begin{align}
\frac{r^{(2)}_i}{\PT_1^2} \underset{ x \to 0 }{\sim} \mathcal{O}\left(\frac{1}{x^4}\right)\, .
\end{align}
The soft factor $\sigma_5$ is singular as well,
\begin{align}
\sigma_5  = - \frac{2 i \pi s}{x^2} - \frac{2 s_1 s_2}{s} (1- z- \zbar + 2 z \zbar) \log(x) + \mathcal{O}(1)\, .
\end{align}
Therefore, in order to determine the multi-Regge limit of the two-loop hard function up to infinitesimal contributions we need to expand the two-loop leading singularities $r^{(2)}_i$ up to order $x^0$, the one-loop leading singularities $r^{(1)}_i$ and the soft factor $\sigma_5$ up to order $x^2$. As for the pentagon functions, we need to keep into account the power corrections in Eq.~\eqref{eq:AsymptoticSolution} up to order $x^4$ at both one and two loops.

We organise the multi-Regge limit of the normalised hard function at two loops as
\begin{align}
\widetilde{{\cal F}}^{(2)}_5   =  \sum_{m=0}^4 \sum_{k=0}^4 \frac{1}{x^m}\log^k(x)\, F^{(2)}_{m,k}(s_1,s_2,s,z,\zbar) + o(1) \, . \label{eq:2loopHardsugra}
\end{align}
The coefficients functions ${\cal F}^{(2)}_5$ have mixed transcendentality up to weight  $(4-k)$.
Following the discussion of the one-loop hard function, we first present the leading power contribution in Eq.~\eqref{eq:2loopHardsugra},
\begin{equation}
\label{eq:F2_40}
\begin{aligned}
F^{(2)}_{4,0} = \, & -2 \pi^2 s_1^2 s_2^2 z (-1 + \zbar) \biggl(-\zbar \left(g^{(1)}_3 - g^{(1)}_4\right)^2 + 
   z \left(g^{(1)}_3 - g^{(1)}_5\right)^2 + 2 g^{(1)}_3 g^{(1)}_5 \\
   & - g^{(1)}_4 g^{(1)}_5 - 
   \left(g^{(1)}_5\right)^2 + 2 g^{(2)}_1 \biggr) \, ,
\end{aligned} 
\end{equation}
and then comment on the remaining terms order by order in $\log x$. Since most expressions are rather bulky, we present the results adopting the same schematic notation of Eq.~\eqref{eq:ReggeSugra1}, in order to highlight their transcendental weight structure. Explicit expressions are provided in ancillary files.

The order-$\log^4(x)$ part of the hard function is the simplest. As at one loop, the leading logarithm (LL) is present only at order $x^0$,
\begin{align}
F^{(2)}_{4,4} = 0 \,, \qquad F^{(2)}_{3,4} = 0  \,, \qquad
F^{(2)}_{2,4} = 0  \,, \qquad F^{(2)}_{1,4} =0  \,, \qquad
F^{(2)}_{0,4} = {\cal Q}_{0,4}^{(0)} \, .
\end{align}
where ${\cal Q}_{0,4}^{(0)}$ is a rational function.
Note that we do not need to take into account the power expansion of the pentagon functions and of the soft factor $\sigma_5$ in order to determine the LL contributions.

At order $\log^3(x)$ we find
\begin{align}
F^{(2)}_{4,3} = 0  \,, \quad F^{(2)}_{3,3} = 0  \,, \quad
F^{(2)}_{2,3} = i\pi {\cal Q}_{2,3}^{(0)}  \,, \quad F^{(2)}_{1,3} =  i\pi {\cal Q}_{1,3}^{(0)}  \,, \quad
F^{(2)}_{0,3} ={\cal Q}_{0,3}^{(1)} +i\pi{\cal Q}_{0,3}^{(0)} + {\cal Q}_{0,3}^{(0)'} \,.
\end{align}
The transcendental part of ${\cal Q}_{0,3}^{(1)}$ involves only the manifestly single-valued logarithms~\eqref{eq:funW1sv}. The presence of the rational term ${\cal Q}_{0,3}^{(0)'}$ in $F^{(2)}_{0,3}$ is a manifestation of the transcendentality drop. Some rational functions, $\mathcal{Q}_{1,3}^{(0)}$ and $\mathcal{Q}_{0,3}^{(0)}$, have poles at $z=\zbar$, of order 1 and 3, respectively.

At order $\log^2(x)$ we have
\begin{equation}
\begin{aligned}
& F^{(2)}_{4,2} = 0 \,, \\
& F^{(2)}_{3,2} = 0 \,, \\
& F^{(2)}_{2,2} = i\pi {\cal Q}_{2,2}^{(1)} + \pi^2  {\cal Q}_{2,2}^{(0)} + i\pi {\cal Q}_{2,2}^{(0)'} \,,\\
& F^{(2)}_{1,2} =  i\pi {\cal Q}_{1,2}^{(1)} + \pi^2  {\cal Q}_{1,2}^{(0)} + i\pi {\cal Q}_{1,2}^{(0)'}\,, \\
& F^{(2)}_{0,2} ={\cal Q}_{0,2}^{(2)} +i\pi{\cal Q}_{0,2}^{(1)} + \pi^2{\cal Q}_{0,2}^{(0)} + {\cal Q}_{0,2}^{(1)'} + i \pi {\cal Q}_{0,2}^{(0)'} + {\cal Q}_{0,2}^{(0)''}  \, .
\end{aligned}
\end{equation}
The transcendental part of the previous expressions, including the weight-2 functions in ${\cal Q}^{(2)}_{0,2}$, are given by (products of) the logarithms~\eqref{eq:funW1sv} only. Some rational functions have up to poles at $z= \zbar$ with up to order three.

At order $\log(x)$ we find
\begin{equation}
\begin{aligned}
& F^{(2)}_{4,1} = 0 \,,\\
& F^{(2)}_{3,1} = 0 \,, \\
& F^{(2)}_{2,1} = i\pi {\cal Q}_{2,1}^{(2)} + \pi^2  {\cal Q}_{2,1}^{(1)} + i\pi^3 {\cal Q}_{2,1}^{(0)}  + i \pi {\cal Q}_{2,1}^{(1)'} + \pi^2 {\cal Q}_{2,1}^{(0)'} + i \pi {\cal Q}_{2,1}^{(0)''} \,, \\
& F^{(2)}_{1,1} = i\pi {\cal Q}_{1,1}^{(2)} + \pi^2  {\cal Q}_{1,1}^{(1)} + i\pi^3 {\cal Q}_{1,1}^{(0)} + i\pi {\cal Q}_{1,1}^{(1)'} + \pi^2 {\cal Q}_{1,1}^{(0)'} + i\pi {\cal Q}_{1,1}^{(0)''} \,, \\
& F^{(2)}_{0,1} = {\cal Q}_{0,1}^{(3)} + i \pi {\cal Q}_{0,1}^{(2)} + \pi^2 {\cal Q}_{0,1}^{(1)} + i \pi^3 {\cal Q}_{0,1}^{(0)} + \zeta_3 {\cal Q}_{0,1}^{(0)'} + {\cal Q}_{0,1}^{(2)'}  + i\pi {\cal Q}_{0,1}^{(1)'} \\
& \quad \qquad  + \pi^2 {\cal Q}_{0,1}^{(0)''}  + {\cal Q}_{0,1}^{(1)''}  + i\pi {\cal Q}_{0,1}^{(0)'''} + {\cal Q}_{0,1}^{(0)''''}\,.
\end{aligned}
\end{equation}
In these expressions almost all weight-1 and 2 functions involve only the logarithms~\eqref{eq:funW1sv} and the single-valued dilogarithm~\eqref{eq:funW2sv}. Only the coefficient $F^{(2)}_{0,1}$ has a more involved functional structure, more precisely in its weight-2 and 3 parts, i.e. ${\cal Q}_{0,1}^{(2)'}$ and ${\cal Q}_{0,1}^{(3)}  + \pi^2  {\cal Q}_{0,1}^{(1)}$. The latter contain more complicated functions, which we described thoroughly in Section~\ref{sec:SUGRAfunctions}.

Finally, at order $\log^0(x)$ we have
\begin{equation}
\begin{aligned}
& F^{(2)}_{4,0} = \pi^2 {\cal Q}_{4,0}^{(2)} \, \\
& F^{(2)}_{3,0} = \pi^2 {\cal Q}_{3,0}^{(2)} + \pi^2 {\cal Q}_{3,0}^{(1)}  \,, \\
& F^{(2)}_{2,0} = i\pi {\cal Q}_{2,0}^{(3)} + \pi^2 {\cal Q}_{2,0}^{(2)} + i\pi^3 {\cal Q}_{2,0}^{(1)} + i \pi \zeta_3 {\cal Q}_{2,0}^{(0)} + \pi^4 {\cal Q}_{2,0}^{(0)'} + \pi^2 {\cal Q}_{2,0}^{(1)'} \\
& \qquad \quad  + i \pi^3 {\cal Q}_{2,0}^{(0)''} +i \pi {\cal Q}_{2,0}^{(1)''} + \pi^2 {\cal Q}_{2,0}^{(0)'''} + i \pi {\cal Q}_{2,0}^{(0)''''} \,, \\ 
& F^{(2)}_{1,0} =  i\pi {\cal Q}_{1,0}^{(3)} + \pi^2 {\cal Q}_{1,0}^{(2)} + i\pi^3 {\cal Q}_{1,0}^{(1)} + i \pi \zeta_3 {\cal Q}_{1,0}^{(0)} + \pi^4 {\cal Q}_{1,0}^{(0)'} + i\pi {\cal Q}_{1,0}^{(2)'} \\
& \qquad \quad + \pi^2 {\cal Q}_{1,0}^{(1)'} + i \pi^3 {\cal Q}_{1,0}^{(0)''} +i \pi {\cal Q}_{1,0}^{(1)''} + \pi^2 {\cal Q}_{1,0}^{(0)'''} + i \pi {\cal Q}_{1,0}^{(0)''''} \,, \\  
& F^{(2)}_{0,0} = {\cal Q}_{0,0}^{(4)} + i \pi {\cal Q}_{0,0}^{(3)} + \pi^2 {\cal Q}_{0,0}^{(2)} + i \pi^3 {\cal Q}_{0,0}^{(1)} + \zeta_3 {\cal Q}_{0,0}^{(1)'} + \pi^4 {\cal Q}_{0,0}^{(0)} + i\pi\zeta_3 {\cal Q}_{0,0}^{(0)'}  \\
& \quad \qquad  + {\cal Q}_{0,0}^{(3)'} + i\pi {\cal Q}_{0,0}^{(2)'} + \pi^2 {\cal Q}_{0,0}^{(1)''} + i \pi^3 {\cal Q}_{0,0}^{(0)''} + \zeta_3 {\cal Q}_{0,0}^{(0)'''}  +  {\cal Q}_{0,0}^{(2)''} \\
& \quad \qquad + i \pi  {\cal Q}_{0,0}^{(1)'''} + \pi^2  {\cal Q}_{0,0}^{(0)''''} + {\cal Q}_{0,0}^{(1)''''}  +  i \pi {\cal Q}_{0,0}^{(0)'''''} + {\cal Q}_{0,0}^{(0)''''''} \,. 
\end{aligned}
\end{equation}
We have given $F_{4,0}^{(2)}$ explicitly in Eq.~\eqref{eq:F2_40}. Both $F_{4,0}^{(2)}$ and $F_{3,0}^{(2)}$, non-vanishing only at this order in $\log x$, are rather simple. They involve only the logarithms~\eqref{eq:funW1sv} and the single-valued dilogarithm~\eqref{eq:funW2sv}. 
The coefficients $F^{(2)}_{2,0}$, $F^{(2)}_{1,0}$ and $F^{(2)}_{0,0}$ have a more complicated structure. They involve the functions described in Section~\ref{sec:SUGRAfunctions}. In particular, $F^{(2)}_{2,0}$ and $F^{(2)}_{1,0}$ contain functions with up to genuine weight 3, whereas in $F^{(2)}_{0,0}$ also genuine weight-4 functions appear.

In general, it is worth noting that all the transcendental functions appearing in the multi-Regge asymptotics of the one and two-loop hard functions are smooth (real-analytic) in both the upper and lower half of the complex $z$ plane. In particular, let us stress that the hard function is real analytic across $\text{Re}[z]=1/2$. The rational functions however bring in divergences on the real axis $\text{Im}[z] = 0$. This makes it impossible to check the continuity across the real axis of the whole hard function. 
The coefficients which are not singular at $z=\zbar$, namely $F^{(2)}_{0,4}$, $F^{(2)}_{2,3}$, $F^{(2)}_{2,2}$, $F^{(2)}_{2,1}$ and $F^{(2)}_{4,0}$, do match at $\text{Im}[z]=0$, and only their derivatives jump.

The coefficients $F^{(2)}_{1,3}$, $F^{(2)}_{0,3}$, $F^{(2)}_{1,2}$, $F^{(2)}_{0,2}$, $F^{(2)}_{1,1}$, $F^{(2)}_{3,0}$ are instead singular at $z=\zbar$, but they involve only functions which are single-valued in the whole complex plane. In particular, note that $F^{(2)}_{1,3}$ is purely rational.
$F^{(2)}_{0,1}$ diverges at $z=\zbar$, and involves both single-valued functions and functions which are defined separately in the upper and in the lower half of the complex plane. The singular rational factors multiply only single-valued functions, and the part which stays finite at $z = \zbar$ can be shown to be continuous across $\text{Im}[z]=0$.
This separation is however impossible for $F^{(2)}_{0,0}$ and $F^{(2)}_{1,0}$, in which non-single-valued functions appear alongside with rational factors singular at $z=\zbar$. This coefficient is the most complicated and it contains the highest pole at $z=\zbar$, of order 7.

\section{Conclusion and outlook}
\label{sec:Conclusions}

In this paper we studied the two-loop five-particle amplitudes in $\mathcal{N}=4$ super Yang-Mills theory and $\mathcal{N}=8$ supergravity. Previously, these amplitudes were computed at symbol level~\cite{Abreu:2018aqd,Chicherin:2018yne,Chicherin:2019xeg,Abreu:2019rpt}. The analysis of their multi-Regge limit was also initiated at symbol level~\cite{Chicherin:2018yne,Chicherin:2019xeg}.
Recently, all massless five-point integrals were computed at two-loop order in the $s_{12}$ scattering channel~\cite{Badger:2019djh}.  

In this paper, we used the knowledge of the integrals to lift the previous results to function level. We obtained expressions for the two amplitudes in terms of rational functions and pentagon functions~\cite{Chicherin:2017dob,Chicherin:2020oor}. The amplitudes have manifestly uniform transcendental weight at all orders in the dimensional regularisation parameter $\epsilon$. 
We verified the infrared structure and extracted the hard functions, written in terms of weight-4 pentagon functions. The hard functions can therefore be studied analytically and numerically.
We provided a numerical reference point for the purpose of future cross-checks. 

As the main focus of this paper, we computed the multi-Regge limit of the two-loop five-particle amplitudes in $\mathcal{N}=4$ super Yang-Mills theory and $\mathcal{N}=8$ supergravity. Interestingly, the space of pentagon functions vastly simplifies in the multi-Regge kinematics. For any massless two-loop five-particle amplitude, it reduces to that of two-dimensional harmonic polylogarithms. This enabled us to compute relatively compact expressions for the multi-Regge limit of the two amplitudes. Our results are written in terms of a small number of transcendental functions, whose analytic properties are transparent, as discussed in Section~\ref{sec:MultiReggeLimit}. We provided the multi-Regge limit of the hard functions in ancillary files. We kept up to and including the finite terms in the limit.

We also carried out a completely independent calculation of the multi-Regge limit of the $\mathcal{N}=4$ super Yang-Mills amplitude using the BFKL effective theory. We focused on certain colour structures, for which this approach dramatically simplifies the computation. It allowed us to target directly the infrared-finite hard function, this way by-passing the calculation of infrared-divergent quantities. The relationship between divergent and finite quantities follows neatly from an infrared $R$-operation. 
We compared the results obtained with the two approaches and found complete agreement. This provided a highly non-trivial check of our computation. 
 
It is interesting to comment some more on the structure of the results.
The multi-Regge asymptotics of the $\mathcal{N}=4$ super Yang-Mills hard function is significantly simpler than its supergravity counterpart. The rational functions appearing in it have an extremely simple behaviour, and the result maintains the uniform transcendentality in the limit. Moreover, the transcendental functions are very simple: they involve only classical polylogarithms of at most weight three. Despite this simplicity, the result features a very non-trivial analytic property, related to the presence of the pseudo-scalar invariant $\eps_5$. The kinematics constrains $\eps_5$ to be pure imaginary, and its sign defines two copies of the scattering region separated by the hypersurface $\eps_5 = 0$. The Regge asymptotics of the $\mathcal{N}=4$ super Yang-Mills hard function is continuous across this surface, but the second derivatives of certain colour structures are not. For convenience of the reader, we illustrated how this non-trivial analytic property stems from certain non-planar integrals.
As an additional cross-check, we independently reproduced this non-analyticity in section \ref{sec: bfkl sing} 
using known ingredients from BFKL theory.

The $\mathcal{N}=8$ supergravity hard function appears to have a richer structure.
In this case we do not have an underlying effective theory as developed as BFKL
and we opted to simply present our results as-is.
Like in the $\mathcal{N}=4$ super Yang-Mills case, we kept up to and including the finite terms in the limit. However, since the rational functions develop poles in the limit, we needed to include power corrections of the integrals. As a consequence, the result contains terms of different transcendental weight (up to four), and a plethora of rational functions. Unlike the $\mathcal{N}=4$ super Yang-Mills case, some rational functions are singular on the hypersurface $\epsilon_5=0$. Furthermore, the transcendental functions are more complicated: they involve genuine weight-four functions and multiple polylogarithms.

It would be interesting to extend the BFKL analysis to other colour structures.
We believe that the explicit amplitudes presented here are sufficient to determine the multi-Regge limit of two-loop amplitudes of any multiplicity. A systematic analysis would be very interesting.
The multi-Regge limit of $\mathcal{N}=8$ supergravity is more involved, but is of great relevance as well~\cite{Bartels:2012ra,SabioVera:2019edr,DiVecchia:2019kta}. It would be interesting if the rich expressions we obtained by direct computation could be explained through some effective theory approach.
Note that, although we restricted our analysis of the multi-Regge limit to the finite terms, the setup we presented is general. This means that further terms can be obtained, e.g. the power-suppressed corrections of the $\mathcal{N}=4$ super Yang-Mills amplitude, if desired.

\section*{Acknowledgments}
We thank Simon Badger, Thomas Gehrmann, Gudrun Heinrich, Tiziano Peraro, Ben Ruijl and Pascal Wasser for useful discussions. This research received funding from the European Research Council (ERC) under the European Union's Horizon 2020 research and innovation programme, {\it Novel structures in scattering amplitudes} (grant agreement No 725110),
from the National Science and Engineering Council of Canada,
the Canada Research Chair program, and the Fonds de Recherche du Qu\'ebec - Nature et Technologies.
The research is also supported by the NSF of China through Grant No. 11947301.

\bibliography{5point_refs_Regge.bib}
\bibliographystyle{JHEP}

\end{document}